\documentclass[fleqn,usenatbib]{mnras}

\usepackage{subfig}
\usepackage{comment}
\usepackage{longtable}
\usepackage{caption}
\usepackage{threeparttable}
\usepackage{booktabs}
\usepackage[figuresright]{rotating}
\newcommand{\be}{\begin{equation}}
\newcommand{\ee}{\end{equation}}
\newcommand{\bea}{\begin{eqnarray}}
\newcommand{\eea}{\end{eqnarray}}
\newcommand{\hunit}{$\rm{km \ s^{-1} \ Mpc^{-1}}$}
\newcommand{\lcdm}{$\Lambda$CDM}
\newcommand{\pcdm}{$\phi$CDM}
\usepackage{array}
\makeatletter
\newcommand{\thickhline}{%
    \noalign {\ifnum 0=`}\fi \hrule height 1pt
    \futurelet \reserved@a \@xhline
}
\newcolumntype{"}{@{\hskip\tabcolsep\vrule width 1pt\hskip\tabcolsep}}
\makeatother

\newcommand{\hiig}{H\,\textsc{ii}G}
\newcommand{\hii}{H\,\textsc{ii}}
\newcommand{\om}{$\Omega_{\rm m_0}$}
\newcommand{\ok}{$\Omega_{\rm k_0}$}

\usepackage[T1]{fontenc}

\DeclareRobustCommand{\VAN}[3]{#2}
\let\VANthebibliography\thebibliography
\def\thebibliography{\DeclareRobustCommand{\VAN}[3]{##3}\VANthebibliography}

\usepackage{graphicx}	
\usepackage{amsmath}	
\usepackage{amssymb}	

\title[Cosmological constraints from observations]{Cosmological constraints from higher-redshift gamma-ray burst, \hii\ starburst galaxy, and quasar (and other) data}

\author[Cao et al.]{
Shulei Cao,$^{1}$\thanks{E-mail: shulei@phys.ksu.edu}
Joseph Ryan,$^{1}$\thanks{E-mail: jwryan@phys.ksu.edu}
Narayan Khadka,$^{1}$\thanks{E-mail: nkhadka@phys.ksu.edu}
Bharat Ratra$^{1}$\thanks{E-mail: ratra@phys.ksu.edu}
\\
$^{1}$Department of Physics, Kansas State University, 116 Cardwell Hall, Manhattan, KS 66502, USA
}

\date{Accepted XXX. Received YYY; in original form ZZZ}

\pubyear{2020}

\begin{document}
\label{firstpage}
\pagerange{\pageref{firstpage}--\pageref{lastpage}}
\maketitle

\begin{abstract}
We use higher-redshift gamma-ray burst (GRB), \hii\ starburst galaxy (\hiig), and quasar angular size (QSO-AS) measurements to constrain six spatially flat and non-flat cosmological models. These three sets of cosmological constraints are mutually consistent. Cosmological constraints from a joint analysis of these data sets are largely consistent with currently-accelerating cosmological expansion as well as with cosmological constraints derived from a combined analysis of Hubble parameter ($H(z)$) and baryon acoustic oscillation (BAO, with Planck-determined baryonic matter density) measurements. A joint analysis of the $H(z)$ + BAO + QSO-AS + \hiig\ + GRB data provides fairly model-independent determinations of the non-relativistic matter density parameter $\Omega_{\rm m_0}=0.313\pm0.013$ and the Hubble constant $H_0=69.3\pm1.2\ \rm{km \ s^{-1} \ Mpc^{-1}}$. These data are consistent with the dark energy being a cosmological constant and with spatial hypersurfaces being flat, but they do not rule out mild dark energy dynamics or a little spatial curvature. We also investigate the effect of including quasar flux measurements in the mix and find no novel conclusions.
\end{abstract}




\section{Introduction} \label{sec:intro}

There is a large body of evidence indicating that the Universe recently transitioned from a decelerated to an accelerated phase of expansion (at redshift $z \sim 3/4$; see e.g. \citealp{Farooq_Ranjeet_Crandall_Ratra_2017}) and has been undergoing accelerated expansion ever since (for reviews, see e.g. \citealp{Ratra_Vogeley,Martin,Coley_Ellis}). In the standard model of cosmology, called the \lcdm\ model \citep{peeb84}, the accelerated expansion is powered by a constant dark energy density (the cosmological constant, $\Lambda$). This model also assumes that spatial hypersurfaces are flat on cosmological scales, and that the majority of non-relativistic matter in the Universe consists of cold dark matter (CDM). 

Out of all the models that have been devised to explain the observed accelerated expansion of the Universe, the \lcdm\ model is currently the most highly favored in terms of both observational data and theoretical parsimony (see e.g. \citealp{Farooq_Ranjeet_Crandall_Ratra_2017,scolnic_et_al_2018,planck2018b,eBOSS_2020}). In spite of these virtues, however, there are some indications that the \lcdm\ model may not tell the whole story. On the observational side, some workers have found evidence of discrepancies between the \lcdm\ model and cosmological observations (\citealp{riess_2019, martinelli_tutusaus_2019}) and on the theoretical side, the origin of $\Lambda$ has yet to be explained in fundamental terms (e.g., \citealp{Martin}). One way to pin down the nature of dark energy is by studying its dynamics phenomenologically. It is possible that the dark energy density may evolve in time (\citealp{peebrat88}), and many dark energy models exhibiting this behavior have been proposed.

Cosmological models have largely been tested in the redshift range $0 \lesssim z \lesssim 2.3$, with baryon acoustic oscillation (BAO\footnote{In our BAO data analyses in this paper the sound horizon computation assumes a value for the current baryonic matter physical density parameter $\Omega_{\rm b_0} h^2$, appropriate for the model under study, computed from Planck CMB anisotropy data.}) measurements probing the upper end of this range, and at $z\sim1100$, using cosmic microwave background (CMB) anisotropy data. To determine the accuracy of our cosmological models, we also need to test them in the redshift range $2.3 \lesssim z \lesssim 1100$. Quasar angular size (QSO-AS), \hii\ starburst galaxy (\hiig), quasar X-ray and UV flux (QSO-Flux), and gamma-ray burst (GRB) measurements are some of the handful of data available in this range. The main goal of this paper is, therefore, to examine the effect that QSO-AS, \hiig, and GRB data have on cosmological model parameter constraints, in combination with each other, and in combination with more well-known probes.\footnote{We relegate the analysis of QSO-Flux data to an appendix, the reasons for which are discussed there.} 

Gamma-ray bursts are promising cosmological probes for two reasons. First, it is believed that they can be used as standardizable candles \citep{Lamb2000, Lamb2001, Amati2002, Amati2008, Amati2009, Ghirlanda2004, Demianski2011, Fyan2015}. Second, they cover a redshift range that is wider than most other commonly-used cosmological probes, having been observed up to $z \sim 8.2$ \citep{Amati2008, Amati2009, Amati2019, samushia_ratra_2010, Demianski2011, Wang_2016, Demianski_2017a, Demianski_2019, Dirirsa2019, Khadka_2020c}. In particular, the $z\sim 2.7$--8.2 part of the Universe is primarily accessed by GRBs,\footnote{Though QSO-Flux measurements can reach up to $z \sim 5.1$.} so if GRBs can be standardized, they could provide useful information about a large, mostly unexplored, part of the Universe.

QSO-AS data currently reach to $z\sim 2.7$. These data, consisting of measurements of the angular size of astrophysical radio sources, furnish a standard ruler that is independent of that provided by the BAO sound horizon scale. The intrinsic linear size $l_m$ of intermediate luminosity QSOs has recently been accurately determined by \cite{Cao_et_al2017b}, opening the way for QSOs to, like GRBs, test cosmological models in a little-explored region of redshift space.\footnote{The use of QSO-AS measurements to constrain cosmological models dates back to near the turn of the century (e.g. \citealp{gurvits_kellermann_frey_1999, vishwakarma_2001, lima_alcaniz_2002, zhu_fujimoto_2002, Chen_Ratra_2003}), but, as discussed in \cite{Ryan_2}, these earlier results are suspect, because they are based on an inaccurate determination of $l_m$.}

\hiig\ data reach to $z\sim 2.4$, just beyond the range of current BAO data. Measurements of the luminosities of the Balmer lines in \hii\ galaxies can be correlated with the velocity dispersion of the radiating gas, making \hii\ galaxies a standard candle that can complement both GRBs and lower-redshift standard candles like supernovae (\citealp{Siegel_2005,Plionis_2009,Mania_2012,Chavez_2014,G-M_2019}).

Current QSO-Flux measurements reach to $z\sim 5.1$, but they favor a higher value of the current (denoted by the subscript ``0'') non-relativistic matter density parameter ($\Omega_{\rm m_0}$) than what is currently thought to be reasonable. The $\Omega_{\rm m_0}$ values obtained using QSO-Flux data, in a number of cosmological models, are in nearly 2$\sigma$ tension with the values obtained by using other well-established cosmological probes like CMB, BAO, and Type Ia supernovae (\citealp{RisalitiandLusso_2019, Yang_2019, Wei_Melia_2020, Khadka_2020b}). Techniques for standardizing QSO-Flux measurements are still under development, so it might be too early to draw strong conclusions about the cosmological constraints obtained from QSO-Flux measurements. Therefore, in this paper, we use QSO-Flux data alone and in combination with other data to constrain cosmological parameters in four different models, and record these results in Appendix \ref{sec:appendix}.

We find that the GRB, \hiig, and QSO-AS constraints are largely mutually consistent, and that their joint constraints are consistent with those from more widely used, and more restrictive, BAO and Hubble parameter ($H(z)$) data. When used jointly with the $H(z)$ + BAO data, these higher-$z$ data tighten the $H(z)$ + BAO constraints.

This paper is organized as follows. In Section \ref{sec:data} we introduce the data we use. Section \ref{sec:model} describes the models we analyze, with a description of our analysis method in Section \ref{sec:analysis}. Our results are in Section \ref{sec:results}, and we provide our conclusions in Section \ref{sec:conclusion}. Additionally, we discuss our results for QSO-Flux measurements in Appendix \ref{sec:appendix}.

\section{Data}
\label{sec:data}
We use QSO-AS, \hiig, QSO-Flux, and GRB data to obtain constraints on the cosmological models we study. The QSO-AS data, comprising 120 measurements compiled by \cite{Cao_et_al2017b} (listed in Table 1 of that paper) and spanning the redshift range $0.462 \leq z \leq 2.73$, are also used in \cite{Ryan_2}; see these papers for descriptions. The \hiig\ data, comprising 107 low redshift ($0.0088 \leq z \leq 0.16417$) \hiig\ measurements, used in \cite{Chavez_2014} (recalibrated by \citealp{G-M_2019}), and 46 high redshift ($0.636427 \leq z \leq 2.42935$) \hiig\ measurements, used in \cite{G-M_2019}, are also used in \cite{Caoetal_2020}. The GRB data, spanning the redshift range $0.48 \leq z \leq 8.2$, are collected from \cite{Dirirsa2019} (25 from Table 2 of that paper (F10), and the remaining 94 from Table 5 of the same, which are a subset of those compiled by \citealp{Wang_2016}) and also used in \cite{Khadka_2020c}. We also add 1598 QSO-Flux measurements spanning the redshift range $0.036 \leq z \leq 5.1003$, from \cite{RisalitiandLusso_2019}. These data are used in \cite{Khadka_2020b}; see that paper for details. Results related to these QSO-Flux data are discussed in Appendix \ref{sec:appendix}. 

In order to be useful as cosmological probes, GRBs need to be standardized, and many phenomenological relations have been proposed for this purpose (\citealp{Amati2002}, \citealp{Ghirlanda2004}, \citealp{Liang2005}, \citealp{Muccino_2020}, and references therein). As in \cite{Khadka_2020c}, we use the Amati relation (\citealp{Amati2002}), which is an observed correlation between the peak photon energy $E_{\rm p}$ and the isotropic-equivalent radiated energy $E_{\rm iso}$ of long-duration GRBs, to standardize GRB measurements. There have been many attempts to standardize GRBs using the Amati relation. Some analyses assume a fixed value of $\Omega_{\rm m_0}$ to calibrate the Amati relation, so they favor a relatively reasonable value of $\Omega_{\rm m_0}$. Others use supernovae data to calibrate the Amati relation, while some use $H(z)$ data to calibrate it. This means that most previous GRB analyses are affected by some non-GRB external factors. In some cases this leads to a circularity problem, in which the models to be constrained by using the Amati relation are also used to calibrate the Amati relation itself (\citealp{Liu_Wei_2015, Demianski_2017a, Demianski_2019, Dirirsa2019}). In other cases, the data used in the calibration process dominate the analysis results. To overcome these problems, we fit the parameters of the Amati relation simultaneously with the parameters of the cosmological models we study (as done in \citealp{Khadka_2020c}; also see \citealp{Wang_2016}).

The isotropic radiated energy $E_{\rm iso}$ of a source in its rest frame at a luminosity distance $D_L$ is
\be
\label{Eiso}
    E_{\rm iso}=\frac{4\pi D_L^2}{1+z}S_{\rm bolo},
\ee
where $S_{\rm bolo}$ is the bolometric fluence, and $D_L$ (defined below) depends on $z$ and on the parameters of our cosmological models. $E_{\rm iso}$ is connected to the source's peak energy output $E_{\rm p}$ via the Amati relation \citep{Amati2008, Amati2009}
\begin{equation}
    \label{eq:Amati}
    \log E_{\rm iso} = a  + b\log E_{\rm p},
\end{equation}
where $a$ and $b$ are free parameters that we vary in our model fits.\footnote{$\log=\log_{10}$ is implied hereinafter.} Note here that the peak energy $E_{\rm p} = (1+z)E_{\rm p, obs}$ where $E_{\rm p, obs}$ is the observed peak energy.

The correlation between \hiig\ luminosity ($L$) and velocity dispersion ($\sigma$) is:
\begin{equation}
\label{eq:logL}
    \log L = \beta \log \sigma + \gamma,
\end{equation}
where $\beta$ is the slope and $\gamma$ is the intercept. As in \cite{Caoetal_2020} (see that paper for details), we use the values
\begin{equation}
    \label{eq:Gordon_beta}
    \beta = 5.022 \pm 0.058,
\end{equation}
and
\begin{equation}
    \label{eq:Gordon_gamma}
    \gamma = 33.268 \pm 0.083.
\end{equation}
One can test a cosmological model with parameters $\textbf{p}$ by using it to compute a theoretical distance modulus
\begin{equation}
\label{eq:mu_th}
    \mu_{\rm th}\left(\textbf{p}, z\right) = 5\log D_{L}\left(\textbf{p}, z\right) + 25,
\end{equation}
and comparing this prediction to the distance modulus computed from observational \hiig\ luminosity and flux ($f$) data
\begin{equation}
\label{eq:mu_obs}
    \mu_{\rm obs} = 2.5\log L - 2.5\log f - 100.2,
\end{equation}
(\citealp{Terlevich_2015, G-M_2019}).

QSO-AS data can be used to test cosmological models by comparing the theoretical angular size of the QSO
\be
\label{eq:theta_th}
\theta_{\rm th} = \frac{l_m}{D_{A}}
\ee
with its observed angular size $\theta_{\rm obs}$. In equation \eqref{eq:theta_th}, $l_m$ is the characteristic linear size of the QSO,\footnote{For the data sample we use, this quantity is equal to $11.03 \pm 0.25$ pc; see \cite{Cao_et_al2017b}.} and $D_{A}$ (defined below) is its angular size distance.

Underestimated systematic uncertainties for both \hiig\ and QSO-AS data might be responsible for the large reduced $\chi^2$ (described in Sec. \ref{subsec:comparison}).

The transverse comoving distance $D_M(\textbf{p}, z)$ is related to the luminosity distance $D_L(\textbf{p}, z)$ and the angular size distance $D_A(\textbf{p}, z)$ through $D_M(\textbf{p}, z)=D_L(\textbf{p}, z)/(1+z)=(1+z)D_A(\textbf{p}, z)$, and is a function of $z$ and the parameters $\textbf{p}$:
\be
\label{eq:DL}
\resizebox{0.44\textwidth}{!}{%
  $D_M(\textbf{p}, z) = 
    \begin{cases}
    \vspace{1mm}
    D_C(\textbf{p}, z) & \text{if}\ \Omega_{\rm k_0} = 0,\\
    \vspace{1mm}
    \frac{c}{H_0\sqrt{\Omega_{\rm k_0}}}\sinh\left[\sqrt{\Omega_{\rm k_0}}H_0D_C(\textbf{p}, z)/c\right] & \text{if}\ \Omega_{\rm k_0} > 0, \\
    \vspace{1mm}
    \frac{c}{H_0\sqrt{|\Omega_{\rm k_0}|}}\sin\left[\sqrt{|\Omega_{\rm k_0}|}H_0D_C(\textbf{p}, z)/c\right] & \text{if}\ \Omega_{\rm k_0} < 0.
    \end{cases}$%
    }
\ee
In the preceding equation,
\be
\label{eq:DC}
    D_C(\textbf{p}, z) \equiv c\int^z_0 \frac{dz'}{H(\textbf{p}, z')},
\ee
$H_0$ is the Hubble constant, $\Omega_{\rm k_0}$ is the current value of the spatial curvature energy density parameter, and $c$ is the speed of light (\citealp{Hogg}).

We also use $H(z)$ and BAO measurements to constrain cosmological parameters. The $H(z)$ data, 31 measurements spanning the redshift range $0.070 \leq z \leq 1.965$, are compiled in Table 2 of \cite{Ryan_1}. The BAO data, 11 measurements spanning the redshift range $0.38 \leq z \leq 2.34$, are listed in Table 1 of \cite{Caoetal_2020}. 

Systematic errors that affect $H(z)$ measurements include assumptions about the stellar metallicity of the galaxies in which cosmic chronometers are found, progenitor bias, the presence of a population of young stars in these galaxies, and assumptions about stellar population synthesis models. These effects were studied in \cite{moresco_et_al_2012, moresco_et_al_2016, moresco_et_al_2018, moresco_et_al_2020}. \cite{moresco_et_al_2020} found that the dominant contribution to the systematic error budget comes from the choice of stellar population synthesis model, which introduces an average systematic error of $\sim8.9$\% (though the authors say that this can be reduced to $\sim4.5$\% by removing an outlier model from the analysis). The impacts of a population of young stars and of the progenitor bias were found to be negligible in \cite{moresco_et_al_2018, moresco_et_al_2012}, and \cite{moresco_et_al_2020} found that the impact of a $\sim$ 5--10\% uncertainty in the metallicity estimates produces a $\sim$ 4--9\% systematic error in the $H(z)$ measurements.

The systematic uncertainties of BAO from \cite{Alam_et_al_2017} (described in Sec. 7) are included in their covariance matrix. The BAO data from \cite{Carter_2018} is the combined result of the 6dF Galaxy Survey1 (6dFGS) and the SDSS DR7 MGS, where the systematic effects are described in detail in \cite{Jones_et_al_2009} and \cite{Ross_et_al_2015} (negligible), respectively. As described in \cite{DES_2019b}, the BAO systematic uncertainty is 15\% of their statistical uncertainty and thus negligible. The same negligible systematic effect applies to the BAO measurement from \cite{3}. \cite{Agathe} added polynomial terms to the correlation function, so as to test the sensitivity of the slowly-varying part of the correlation function to systematic effects. They found that this shifted the BAO peak position by less than $1\sigma$ relative to its position in their fiducial model.

\section{Cosmological models}
\label{sec:model}

In this paper we consider three pairs of flat and non-flat cosmological models, with non-dynamical and dynamical dark energy density.\footnote{Observational constraints on non-flat models are discussed in \cite{Farooq_Mania_Ratra_2015}, \cite{Chen_et_al_2016}, \cite{yu_wang_2016}, \cite{rana_jain_mahajan_mukherjee_2017}, \cite{ooba_etal_2018a, ooba_etal_2018b, ooba_etal_2018c}, \cite{yu_etal_2018}, \cite{park_ratra_2018, park_ratra_2019a, park_ratra_2019b, park_ratra_2019c, park_ratra_2020}, \cite{wei_2018}, \cite{DES_2019}, \cite{coley_2019}, \cite{jesus_etal_2019}, \cite{handley_2019}, \cite{zhai_etal_2020}, \cite{li_etal_2020}, \cite{geng_etal_2020}, \cite{kumar_etal_2020}, \cite{efstathiou_gratton_2020}, \cite{divalentino_etal_2020}, \cite{gao_etal_2020}, \cite{Yang_2020}, \cite{Agudelo_Ruiz_2020}, \cite{Velasquez-Toribio_2020}, and references therein.} Since the data we use are at low redshift, we neglect the contribution that radiation makes to the cosmological energy budget.

In the \lcdm\ model the Hubble parameter is
\be\label{Hz1}
H(z)=H_0\sqrt{\Omega_{\rm m_0}(1+z)^3+\Omega_{\rm k_0}(1+z)^2+\Omega_{\Lambda}},
\ee
where $H_0$, \om, and the cosmological constant dark energy density parameter $\Omega_{\Lambda}$ are the parameters to be constrained, and $\Omega_{\rm k_0}$ obeys $\Omega_{\rm k_0} = 1 - \Omega_{\rm m_0} - \Omega_{\Lambda}$. When $\Omega_{\rm k_0} = 0$ (flat \lcdm), we only constrain $H_0$ and \om, as the value of $\Omega_{\Lambda}$ is fixed by $\Omega_{\Lambda} = 1 - \Omega_{\rm m_0}$.

The XCDM parametrization is an extension of the \lcdm\ model in which the dark energy equation of state parameter, $w_{\rm X} = p_{\rm X}/\rho_{\rm X}$, is allowed to take values different from $-1$, where $p_{\rm X}$ and $\rho_{\rm X}$ are the pressure and energy density, respectively, of the dark energy, treated in this case as an ideal, spatially homogeneous X-fluid.\footnote{Unlike the \lcdm\ and \pcdm\ models, the XCDM parametrization is physically incomplete because it cannot sensibly describe the evolution of spatial inhomogeneities. The XCDM parametrization can be made sensible by allowing for an additional free parameter $c^2_{s, {\rm X}} = dp_{\rm X}/d\rho_{\rm X}$ and requiring $c^2_{s, {\rm X}} > 0$.} In the XCDM parametrization the Hubble parameter takes the form
\be\label{Hz2}
\resizebox{0.47\textwidth}{!}{%
    $H(z)=H_0\sqrt{\Omega_{\rm m_0}(1+z)^3+\Omega_{\rm k_0}(1+z)^2+\Omega_{\rm X_0}(1+z)^{3(1+w_{\rm X})}},$%
    }
\ee
where $\Omega_{\rm X_0}$ is the current value of the X-fluid energy density parameter (its constraints are not reported in this paper) subject to $\Omega_{\rm X_0}=1-\Omega_{\rm m_0}-\Omega_{\rm k_0}$. When $w_{\rm X} = -1$ XCDM reduces to \lcdm. In the general, non-flat case, the model parameters to be constrained are $H_0$, \om, \ok, and $w_{\rm X}$. When $\Omega_{\rm k_0} = 0$ (flat XCDM), we only constrain $H_0$, \om, and $w_{\rm X}$, as the value of $\Omega_{\rm X_0}$ is fixed by $\Omega_{\rm X_0} = 1 - \Omega_{\rm m_0} - \Omega_{k_0}$.

In the $\phi$CDM model, a dynamical scalar field $\phi$, whose stress-energy tensor acts like that of a time-variable $\Lambda$, characterizes the dark energy, and has a potential energy density
\be\label{PE}
V(\phi)=\frac{1}{2}\kappa m_p^2\phi^{-\alpha}.
\ee
Here $m_p$ is the Planck mass, $\alpha \geq 0$, and 
\be\label{kp}
\kappa=\frac{8}{3m_p^2}\bigg(\frac{\alpha+4}{\alpha+2}\bigg)
\bigg[\frac{2}{3}\alpha(\alpha+2)\bigg]^{\alpha/2}
\ee
(\citealp{peebrat88, ratpeeb88, pavlov13}).\footnote{Observational constraints on the \pcdm\ model are discussed in, e.g., \cite{Chen_Ratra_2004}, \cite{samushia_et_al_2007}, \cite{yashar_et_al_2009}, \cite{Samushia_2010}, \cite{chen_ratra_2011b}, \cite{Campanelli_etal_2012}, \cite{Farooq_Ratra_2013}, \cite{farooq_crandall_ratra_2013}, \cite{Avsajanishvili_2015}, \cite{Sola_etal_2017}, \cite{zhai_blanton_slosar_tinker_2017}, \cite{sangwan_tripathi_jassal_2018}, \cite{Sola_perez_gomez_2018,sola_gomez_perez_2019}, \cite{ooba_etal_2019}, \cite{singh_etal_2019}, \cite{Ryan_2}, \cite{Khadka_2020a},\cite{Urena-Lopez_2020}.} For $\alpha=0$ the \pcdm\ models reduce to the \lcdm\ models. 

In this paper we make the approximation, valid for our purposes, that the scalar field is spatially homogeneous. When $\phi$ is approximated in this way, two coupled non-linear ordinary differential equations control its dynamics. The first is its equation of motion
\be\label{em}
\ddot{\phi}+3\bigg(\frac{\dot{a}}{a}\bigg)\dot{\phi}-\frac{1}{2}\alpha\kappa m_p^2\phi^{-\alpha-1}=0,
\ee
and the second is the Friedmann equation
\be\label{fe}
\bigg(\frac{\dot{a}}{a}\bigg)^2=\frac{8\pi}{3m_p^2}(\rho_{\rm m}+\rho_{\phi})-\frac{k}{a^2},
\ee
where $a$ is the scale factor and an overdot denotes a time derivative. In equation \eqref{fe}, ${-k}/{a^2}$ is the spatial curvature term (with $\Omega_{\rm k_0} = 0$, $>0$, $<0$ corresponding to $k = 0$, $-1$, $+1$, respectively), and $\rho_{\rm m}$ and $\rho_{\phi}$ are the non-relativistic matter and scalar field energy densities, respectively, where
\be\label{rp}
\rho_{\phi}=\frac{m_p^2}{32\pi}\bigg(\dot{\phi}^2+\kappa m_p^2\phi^{-\alpha}\bigg).
\ee
It follows that the Hubble parameter in $\phi$CDM is
\be\label{Hz3}
H(z)=H_0\sqrt{\Omega_{\rm m_0}(1+z)^3+\Omega_{\rm k_0}(1+z)^2+\Omega_{\phi}(z,\alpha)},
\ee
where the scalar field energy density parameter
\be\label{op}
\Omega_{\phi}(z,\alpha)=\frac{1}{12H_0^2}\bigg(\dot{\phi}^2+\kappa m_p^2\phi^{-\alpha}\bigg).
\ee
In the general, non-flat case, the parameters to be constrained are $H_0$, \om, \ok, and $\alpha$. In the special case that $\Omega_{\rm k_0} = 0$ (flat \pcdm), we only constrain $H_0$, \om, and $\alpha$.

\section{Data Analysis Methodology}
\label{sec:analysis}

By using the \textsc{python} module \textsc{emcee} \citep{2013PASP..125..306F}, we perform a Markov chain Monte Carlo (MCMC) analysis to maximize the likelihood function, $\mathcal{L}$, and thereby determine the best-fitting values of the free parameters. The flat cosmological parameter priors are the same as those used in \cite{Caoetal_2020} and the flat priors of the parameters of the Amati relation are non-zero over $0\leq\sigma_{\rm ext}\leq10$ (described below), $40\leq a\leq60$, and $0\leq b\leq5$. 

The likelihood functions associated with $H(z)$, BAO, \hiig, and QSO-AS data are described in \cite{Caoetal_2020}. For GRB data, the natural log of its likelihood function \citep{D'Agostini_2005} is
\be
\label{eq:LH_GRB}
    \ln\mathcal{L}_{\rm GRB}= -\frac{1}{2}\Bigg[\chi^2_{\rm GRB}+\sum^{119}_{i=1}\ln\left(2\pi(\sigma_{\rm ext}^2+\sigma_{{y_i}}^2+b^2\sigma_{{x_i}}^2)\right)\Bigg],
\ee
where
\be
\label{eq:chi2_GRB}
    \chi^2_{\rm GRB} = \sum^{119}_{i=1}\bigg[\frac{(y_i-b x_i-a)^2}{(\sigma_{\rm ext}^2+\sigma_{{y_i}}^2+b^2\sigma_{{x_i}}^2)}\bigg],
\ee
$x=\log\frac{E_{\rm p}}{\rm keV}$, $\sigma_{x}=\frac{\sigma_{E_{\rm p}}}{E_{\rm p}\ln 10}$, $y=\log\frac{E_{\rm iso}}{\rm erg}$, and $\sigma_{\rm ext}$ is the extrinsic scatter parameter, which contains the unknown systematic uncertainty. For the GRB with $\sigma_z$ uncertainty in $z$,
\be
\label{eq:err_y2}
    \sigma^2_{y}=\left(\frac{\sigma_{S_{\rm bolo}}}{S_{\rm bolo}\ln 10}\right)^2+\left(\frac{2(1+z)\frac{\partial D_M}{\partial z}+D_M}{(1+z)D_M\ln 10}\sigma_z\right)^2,
\ee
and for those without $z$ uncertainties $\sigma_z=0$ (the non-zero $\sigma_z$ has a negligible effect on our results).

The Akaike Information Criterion ($AIC$) and the Bayesian Information Criterion ($BIC$) are used to compare the goodness of fit of models with different numbers of parameters, where
\be
\label{AIC}
    AIC=-2\ln \mathcal{L}_{\rm max} + 2n,
\ee
and
\be
\label{BIC}
    BIC=-2\ln \mathcal{L}_{\rm max} + n\ln N.
\ee
In these equations, $\mathcal{L}_{\rm max}$ is the maximum value of the relevant likelihood function, $n$ is the number of free parameters of the model under consideration, and $N$ is the number of data points (e.g., for GRB $N=119$).

\section{Results}
\label{sec:results}
\subsection{\hiig, QSO-AS, and GRB constraints, individually}
\label{subsec:GRB}

\begin{figure*}
\centering
  \subfloat[All parameters]{%
    \includegraphics[width=3.5in,height=3.5in]{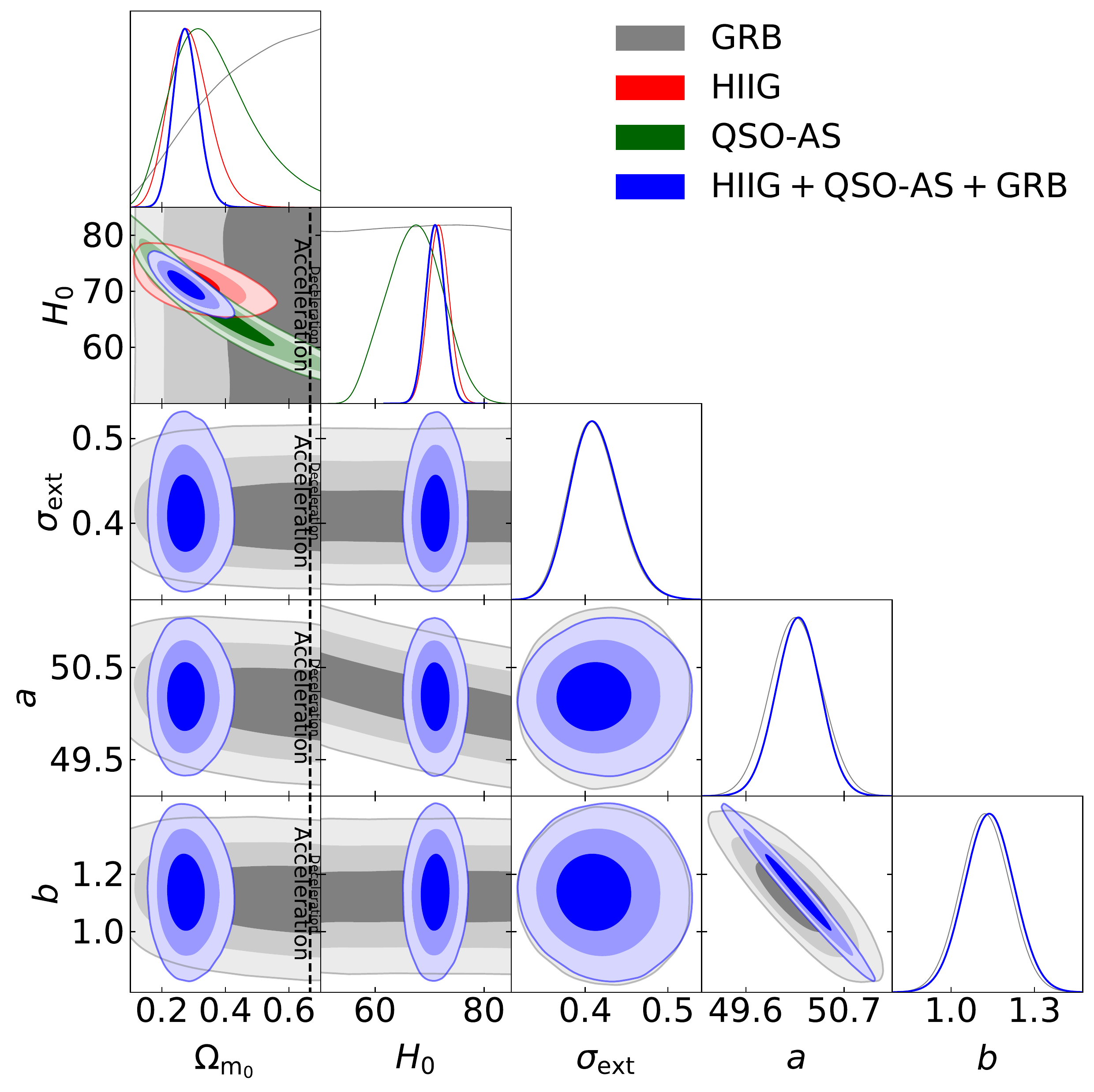}}
  \subfloat[Cosmological parameters zoom in]{%
    \includegraphics[width=3.5in,height=3.5in]{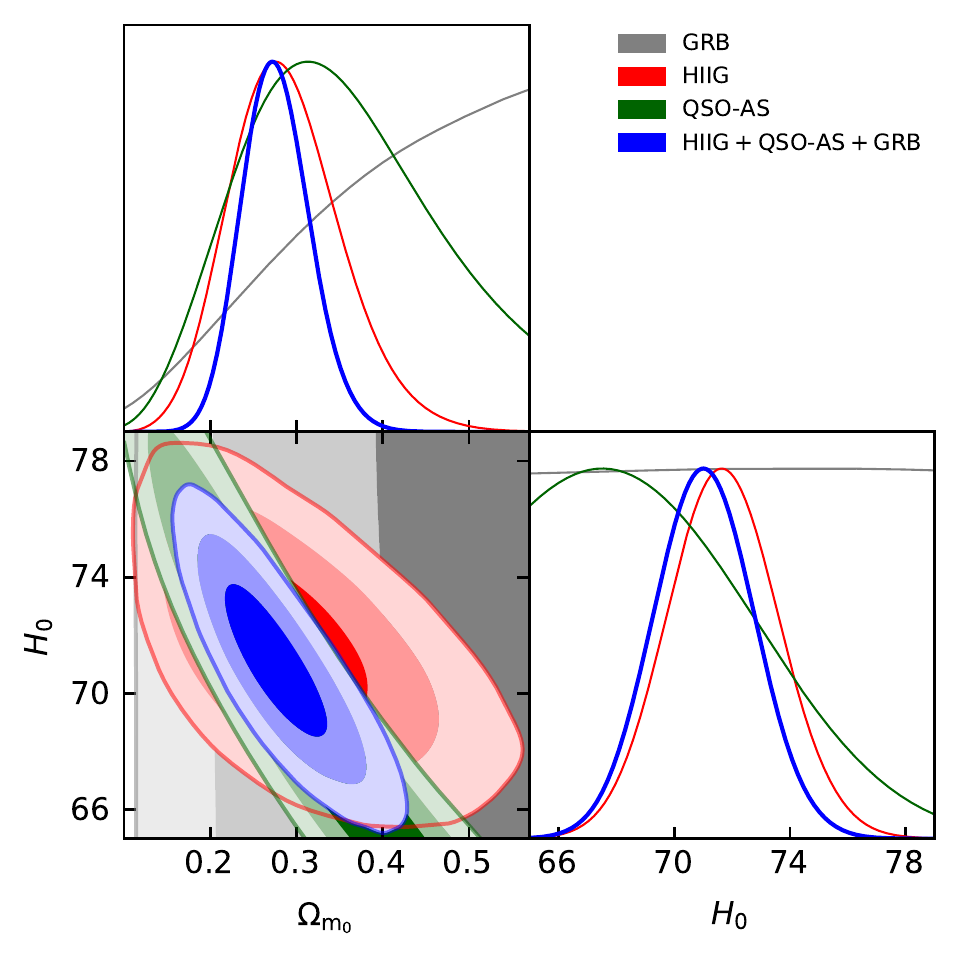}}\\
\caption{1$\sigma$, 2$\sigma$, and 3$\sigma$ confidence contours for flat \lcdm, where the right panel is the cosmological parameters comparison zoomed in. The black dotted lines in the left sub-panels of the left panel are the zero-acceleration lines, which divide the parameter space into regions associated with currently-accelerating (left) and currently-decelerating (right) cosmological expansion.}
\label{fig1}
\end{figure*}

\begin{figure*}
\centering
  \subfloat[All parameters]{%
    \includegraphics[width=3.5in,height=3.5in]{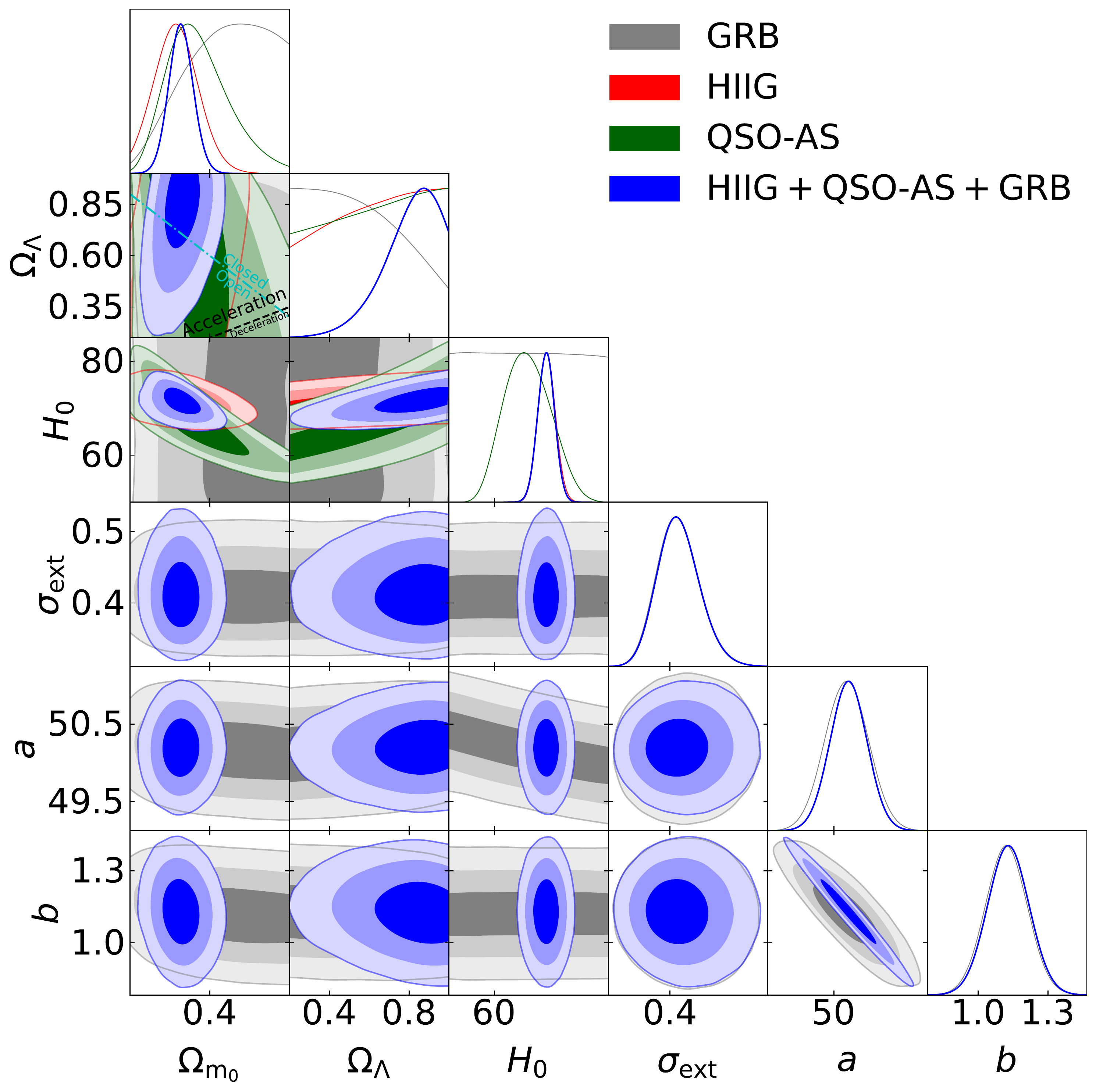}}
  \subfloat[Cosmological parameters zoom in]{%
    \includegraphics[width=3.5in,height=3.5in]{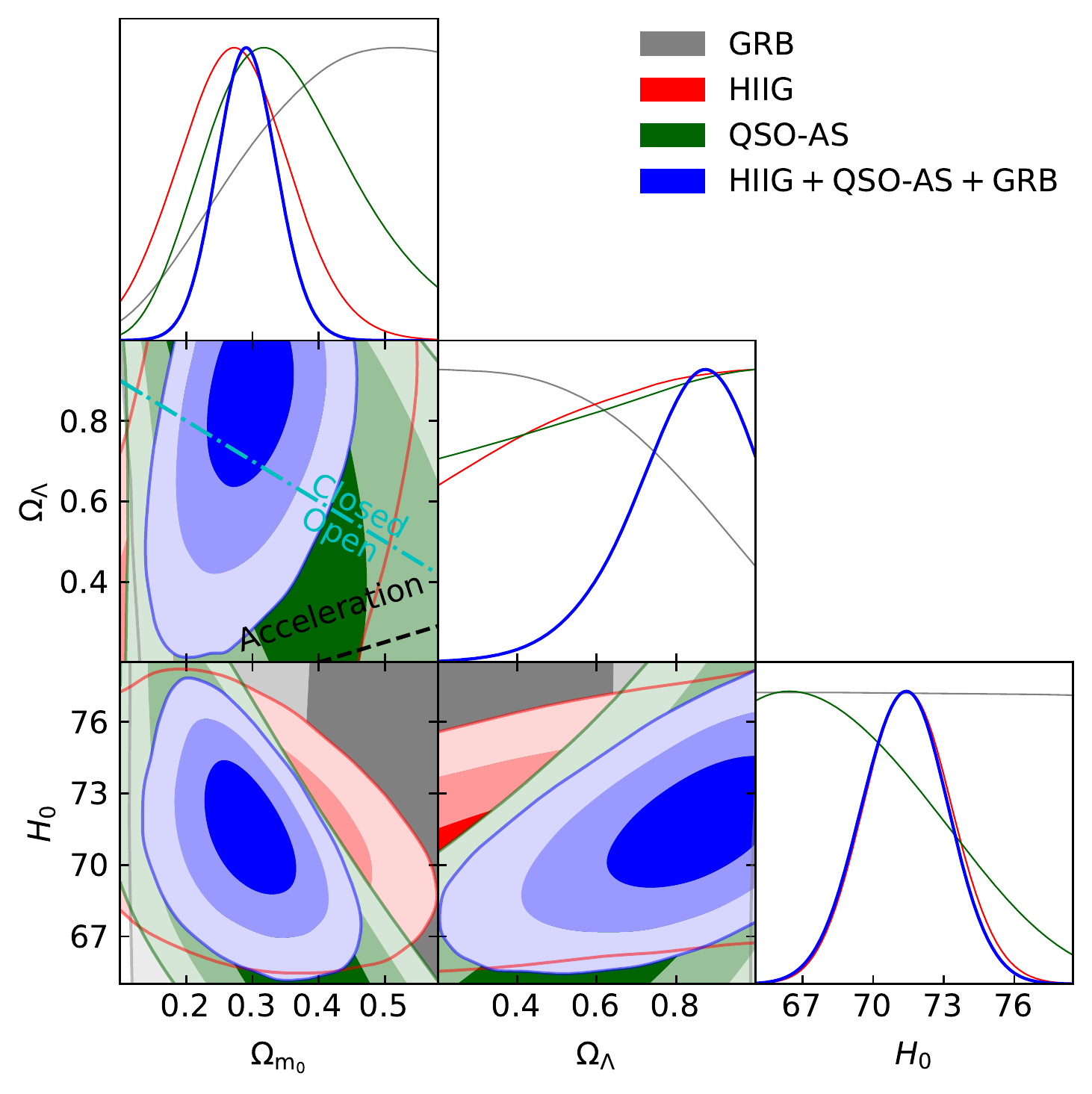}}\\
\caption{Same as Fig. \ref{fig1} but for non-flat \lcdm. The cyan dash-dot line represents the flat \lcdm\ case, with closed spatial hypersurfaces to the upper right. The black dotted line is the zero-acceleration line, which divides the parameter space into regions associated with currently-accelerating (above left) and currently-decelerating (below right) cosmological expansion.}
\label{fig2}
\end{figure*}

\begin{figure*}
\centering
  \subfloat[All parameters]{%
    \includegraphics[width=3.5in,height=3.5in]{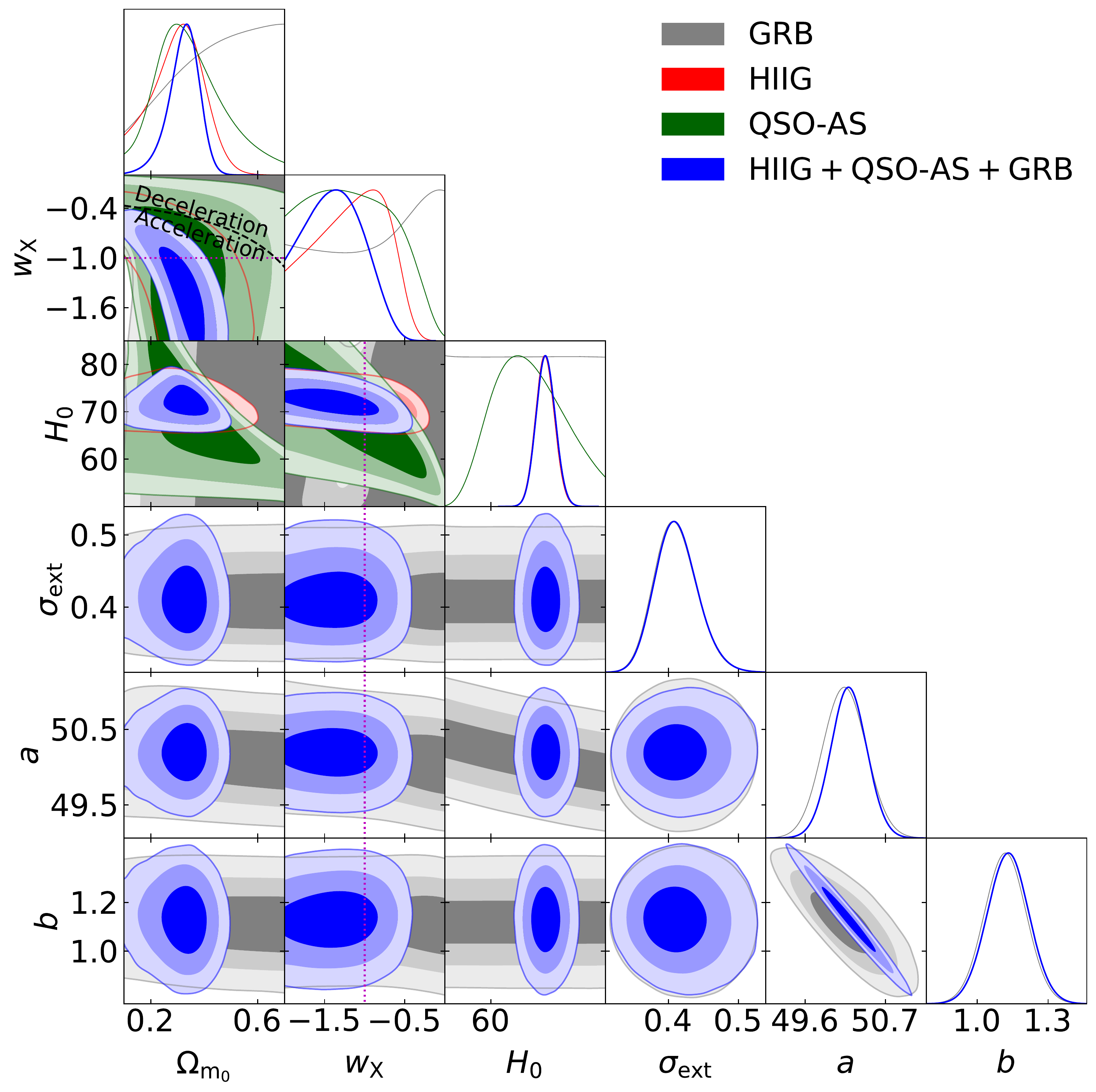}}
  \subfloat[Cosmological parameters zoom in]{%
    \includegraphics[width=3.5in,height=3.5in]{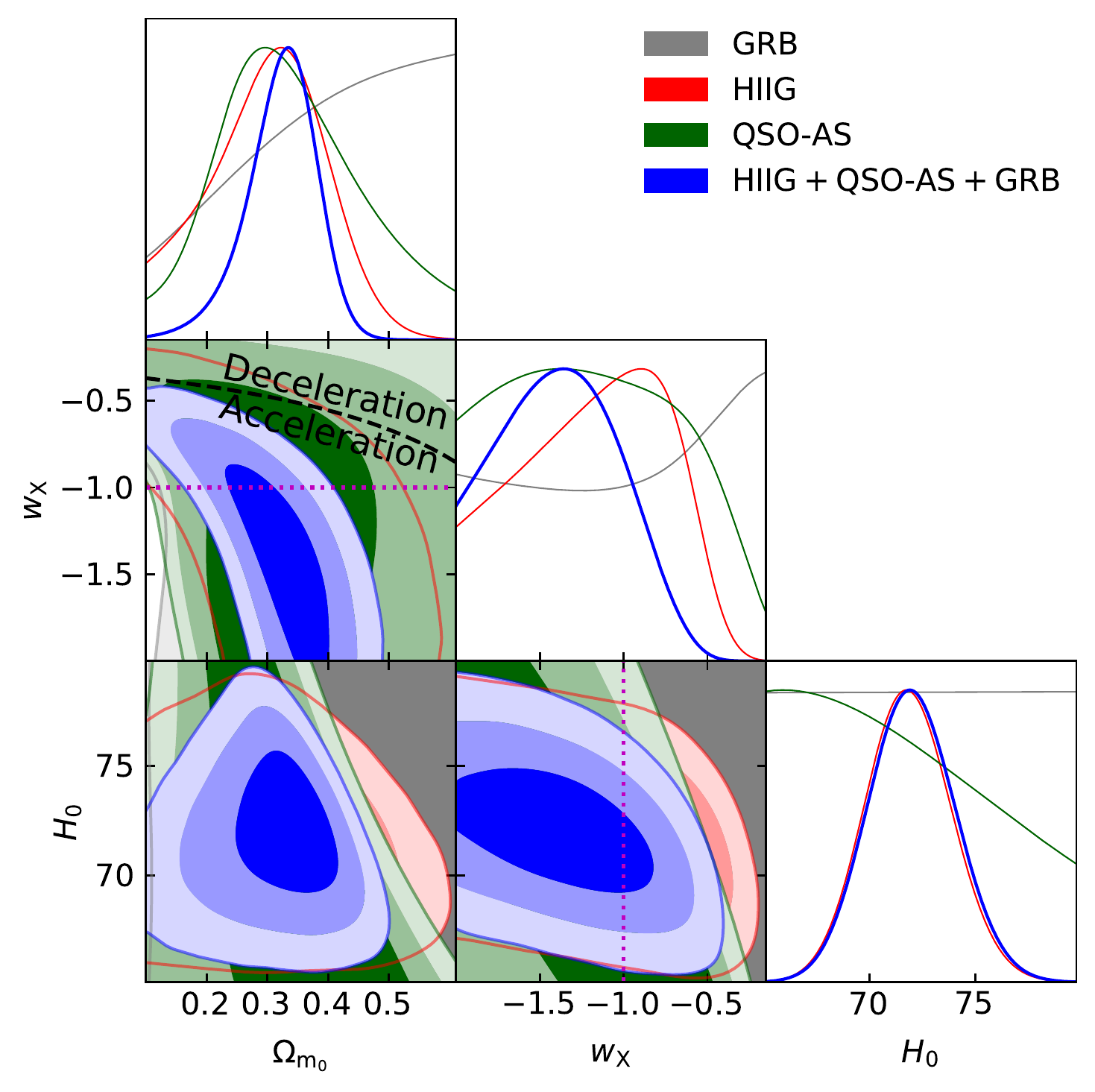}}\\
\caption{1$\sigma$, 2$\sigma$, and 3$\sigma$ confidence contours for flat XCDM. The black dotted line is the zero-acceleration line, which divides the parameter space into regions associated with currently-accelerating (below left) and currently-decelerating (above right) cosmological expansion. The magenta lines denote $w_{\rm X}=-1$, i.e. the flat \lcdm\ model.}
\label{fig3}
\end{figure*}

\begin{figure*}
\centering
  \subfloat[All parameters]{%
    \includegraphics[width=3.5in,height=3.5in]{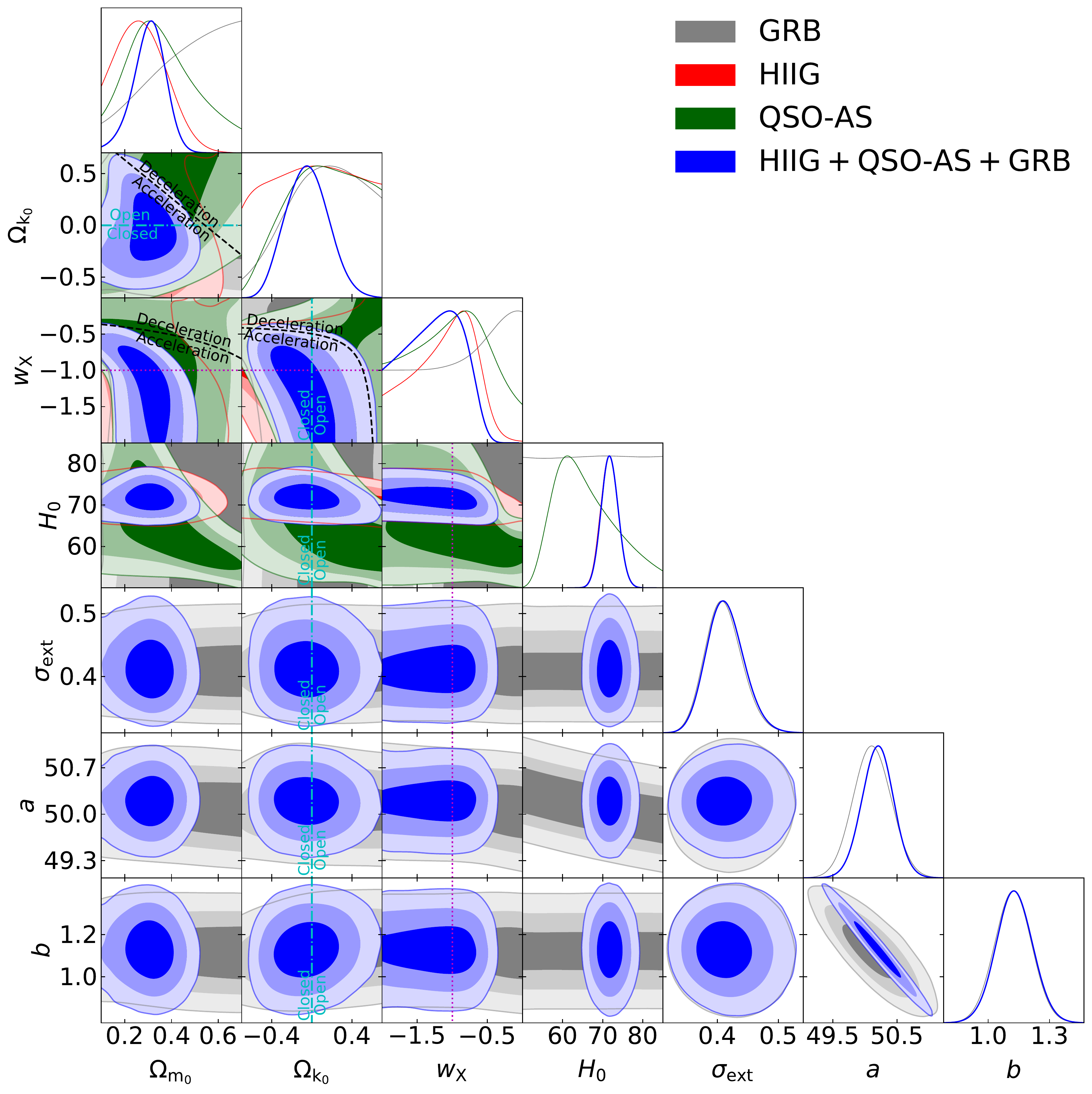}}
  \subfloat[Cosmological parameters zoom in]{%
    \includegraphics[width=3.5in,height=3.5in]{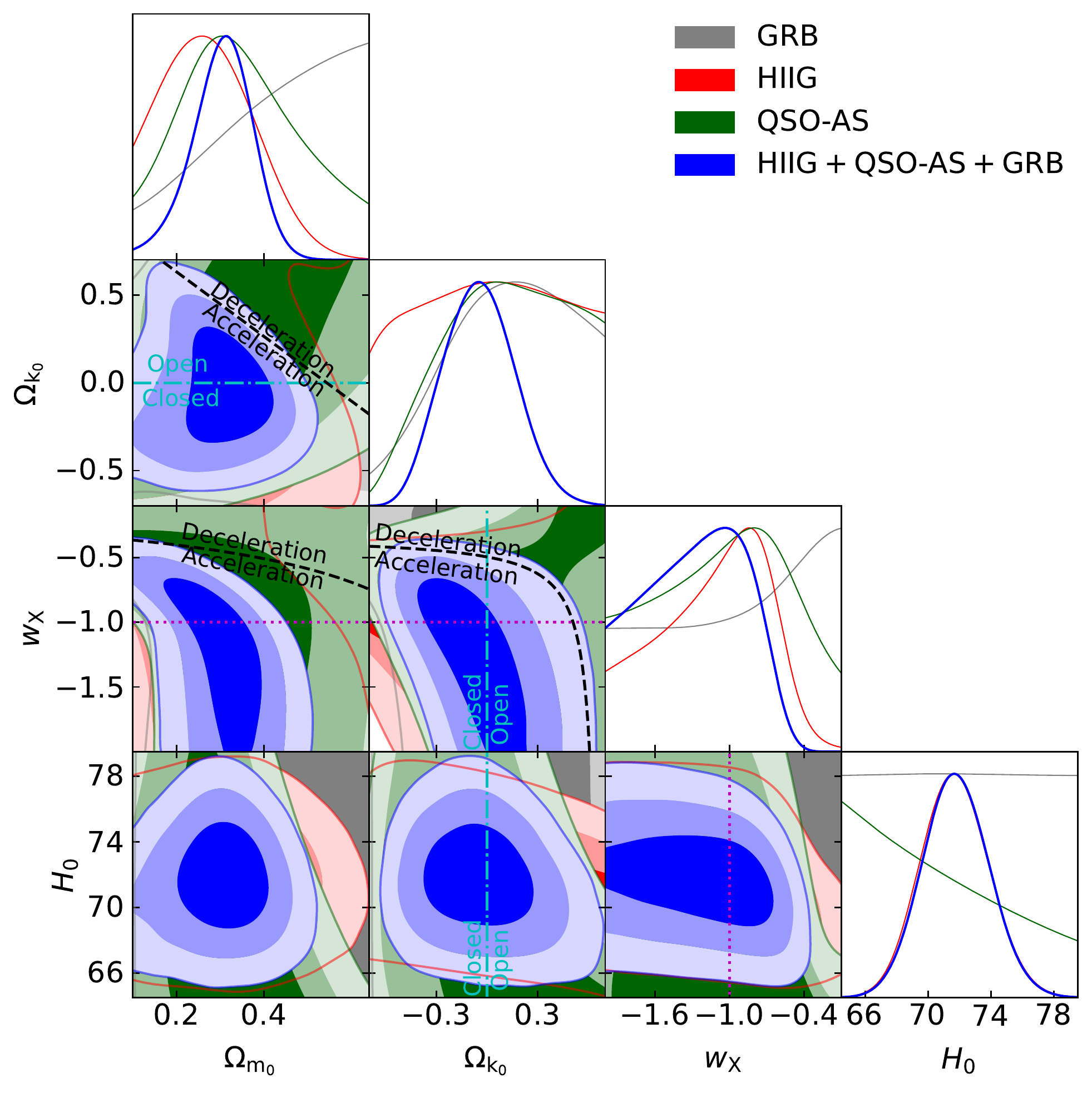}}\\
\caption{Same as Fig. \ref{fig3} but for non-flat XCDM, where the zero acceleration lines in each of the three subpanels are computed for the third cosmological parameter set to the $H(z)$ + BAO data best-fitting values listed in Table \ref{tab:BFP}. Currently-accelerating cosmological expansion occurs below these lines. The cyan dash-dot lines represent the flat XCDM case, with closed spatial hypersurfaces either below or to the left. The magenta lines indicate $w_{\rm X} = -1$, i.e. the non-flat \lcdm\ model.}
\label{fig4}
\end{figure*}

We present the posterior one-dimensional (1D) probability distributions and two-dimensional (2D) confidence regions of the cosmological and Amati relation parameters for the six flat and non-flat models in Figs. \ref{fig1}--\ref{fig6}, in gray (GRB), red (\hiig), and green (QSO-AS). The unmarginalized best-fitting parameter values are listed in Table \ref{tab:BFP}, along with the corresponding $\chi^2$, $-2\ln\mathcal{L}_{\rm max}$, $AIC$, $BIC$, and degrees of freedom $\nu$ (where $\nu \equiv N - n$).\footnote{Note that the $\chi^2$ values listed in Tables \ref{tab:BFP} and \ref{tab:BFP2} are computed from the best-fitting parameter values and are not necessarily the minimum (especially when including GRB and QSO-Flux data).} The values of $\Delta\chi^2$, $\Delta AIC$, and $\Delta BIC$ reported in Table \ref{tab:BFP} are discussed in Section \ref{subsec:comparison}, where we define $\Delta \chi^2$, $\Delta AIC$, and $\Delta BIC$, respectively, as the differences between the values of the $\chi^2$, $AIC$, and $BIC$ associated with a given model and their corresponding minimum values among all models. The marginalized best-fitting parameter values and uncertainties ($\pm 1\sigma$ error bars or $2\sigma$ limits) are given in Table \ref{tab:1d_BFP}.\footnote{We use the \textsc{python} package \textsc{getdist} \citep{Lewis_2019} to plot these figures and compute the central values (posterior means) and uncertainties of the free parameters listed in Table \ref{tab:1d_BFP}.}
From Table \ref{tab:1d_BFP} we find that the QSO-AS constraints on \om\ are consistent with other results within a 1$\sigma$ range but with large error bars, ranging from a low of $0.329^{+0.086}_{-0.171}$ (flat \pcdm) to a high of $0.364^{+0.083}_{-0.150}$ (flat \lcdm). 

The QSO-AS constraints on $H_0$ are between $H_0=61.91^{+2.83}_{-4.92}$ \hunit (non-flat \pcdm) and $H_0=68.39^{+6.14}_{-8.98}$ \hunit (flat XCDM), with large error bars and relatively low values for non-flat XCDM and the \pcdm\ models. 

The non-flat models mildly favor open geometry, but are also consistent, given the large error bars, with spatially-flat hypersurfaces (except for non-flat \pcdm, where the open case is favored at $2.76\sigma$). For non-flat \lcdm, non-flat XCDM, and non-flat \pcdm, we find $\Omega_{\rm k_0}=0.017^{+0.184}_{-0.277}$, $\Omega_{\rm k_0}=0.115^{+0.466}_{-0.293}$, and $\Omega_{\rm k_0}=0.254^{+0.304}_{-0.092}$, respectively.\footnote{From Table \ref{tab:1d_BFP} we see that GRB data are also consistent with flat spatial geometry in the non-flat \lcdm\ and XCDM cases, but also favor, at $2.92\sigma$, open spatial geometry in the case of non-flat \pcdm.}

The fits to the QSO-AS data favor dark energy being a cosmological constant but do not strongly disfavor dark energy dynamics. For flat (non-flat) XCDM, $w_{\rm X}=-1.161^{+0.430}_{-0.679}$ ($w_{\rm X}=-1.030^{+0.593}_{-0.548}$), and for flat (non-flat) \pcdm, $2\sigma$ upper limits of $\alpha$ are $\alpha<2.841$ ($\alpha<4.752$). In the former case, both results are within 1$\sigma$ of $w_{\rm X}=-1$, and in the latter case, both 1D likelihoods peak at $\alpha=0$. 

Constraints on cosmological model parameters derived solely from \hiig\ data are discussed in Sec. 5.1 of \cite{Caoetal_2020}, while those derived from GRB data are described in Sec. 5.1 of \cite{Khadka_2020a} (though there are slight differences coming from the different treatments of $H_0$ and the different ranges of flat priors used there and here); both are listed in Table \ref{tab:1d_BFP} here. In contrast to the \hiig\ and QSO-AS data sets, the GRB data alone cannot constrain $H_0$ because there is a degeneracy between the intercept parameter ($a$) of the Amati relation and $H_0$; for consistency with the analyses of the \hiig\ and QSO-AS data, we treat $H_0$ as a free parameter in the GRB data analysis here.

Cosmological constraints obtained using the \hiig, QSO-AS, and GRB data sets are mutually consistent, and are also consistent with those obtained from most other cosmological probes. This is partially a consequence of the larger \hiig, QSO-AS, and GRB data error bars, which lead to relatively weaker constraints on cosmological parameters when each of these data sets is used alone (see Table \ref{tab:1d_BFP}). However, because the \hiig, QSO-AS, and GRB constraints are mutually consistent, we may jointly analyze these data. Their combined cosmological constraints will therefore be more restrictive than when they are analyzed individually.

We note, from Figs. \ref{fig1}--\ref{fig6}, that a significant part of the likelihood of each of these three data sets lies in the parameter space part with currently-accelerating cosmological expansion.

\subsection{\hiig, QSO-AS, and GRB (HQASG) joint constraints}
\label{subsec:HQASG}

Because the \hiig, QSO-AS, and GRB contours are mutually consistent for all six of the models we study, we jointly analyze these data to obtain HQASG constraints.

\begin{figure*}
\centering
  \subfloat[All parameters]{%
    \includegraphics[width=3.5in,height=3.5in]{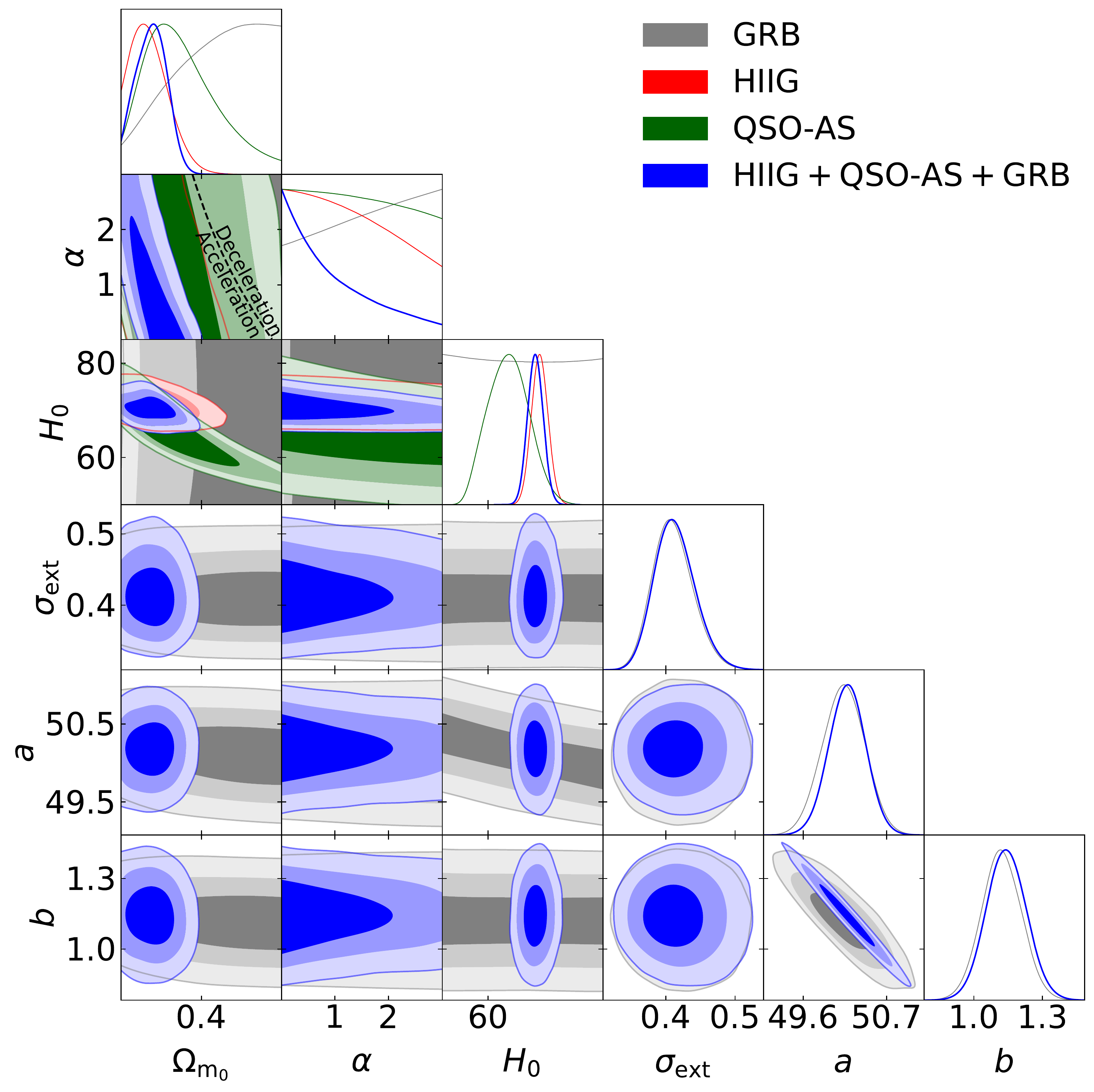}}
  \subfloat[Cosmological parameters zoom in]{%
    \includegraphics[width=3.5in,height=3.5in]{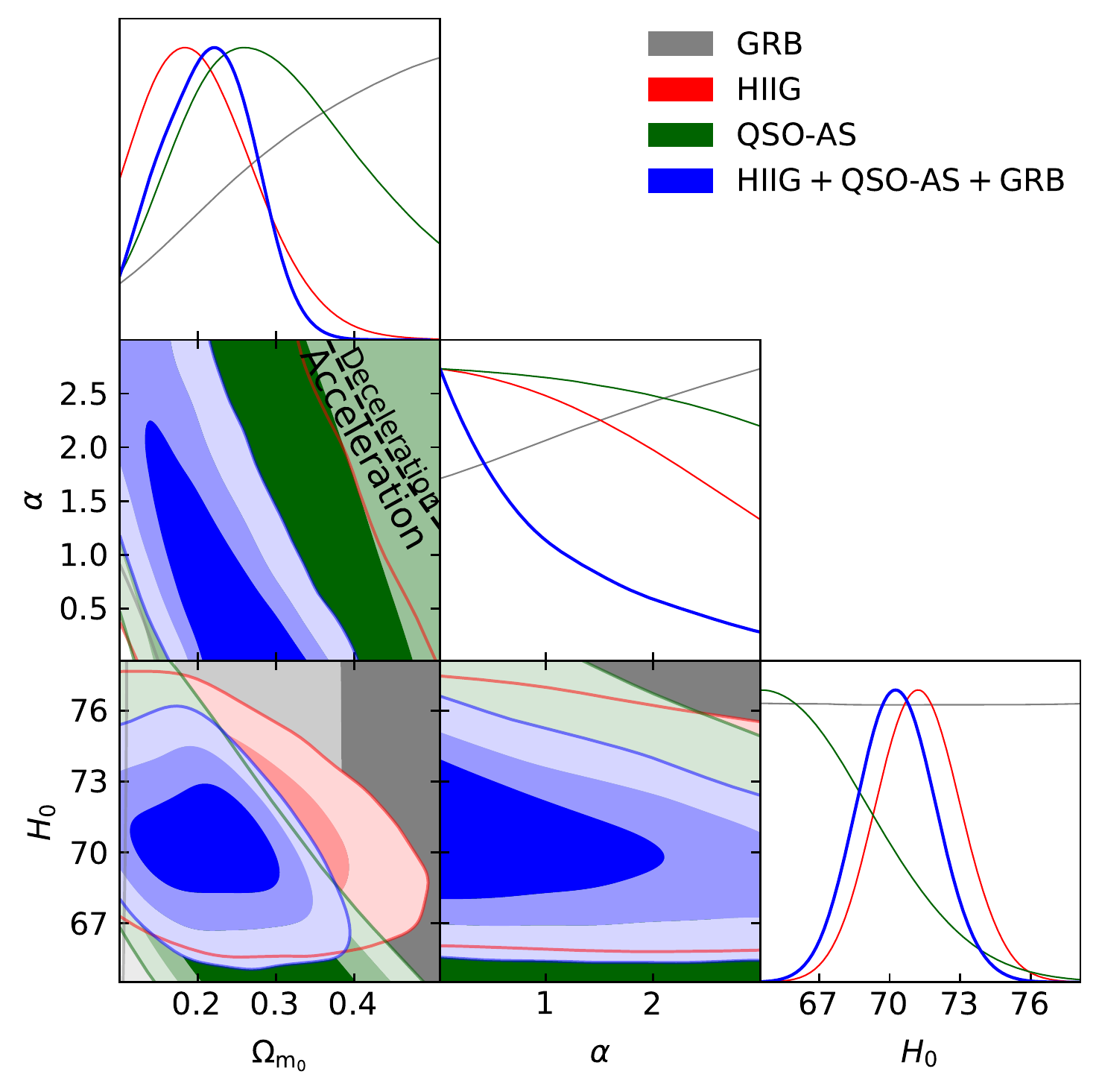}}\\
\caption{1$\sigma$, 2$\sigma$, and 3$\sigma$ confidence contours for flat \pcdm. The black dotted zero-acceleration line splits the parameter space into regions of currently-accelerating (below left) and currently-decelerating (above right) cosmological expansion. The $\alpha = 0$ axis is the flat \lcdm\ model.}
\label{fig5}
\end{figure*}

\begin{figure*}
\centering
  \subfloat[All parameters]{%
    \includegraphics[width=3.5in,height=3.5in]{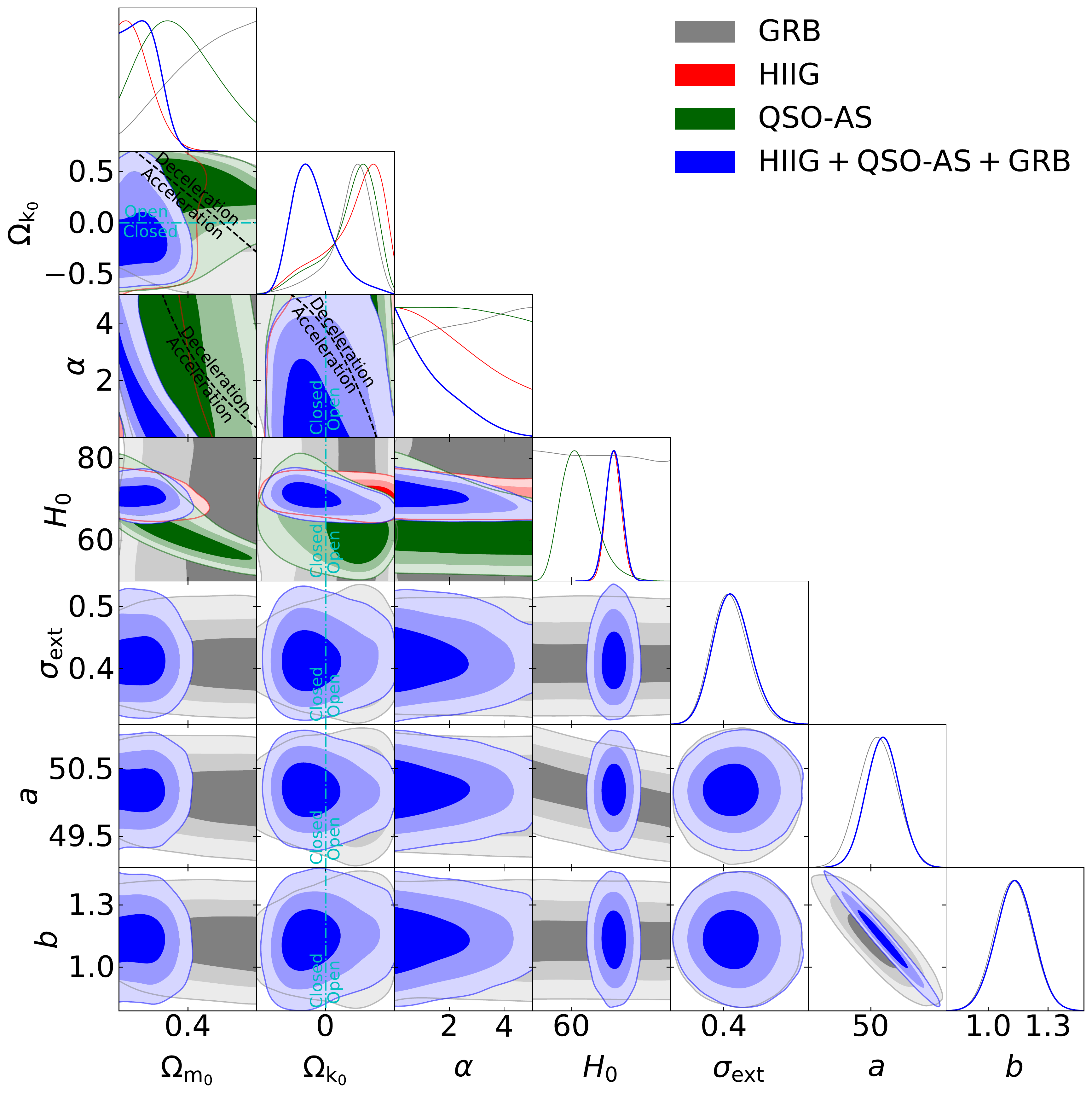}}
  \subfloat[Cosmological parameters zoom in]{%
    \includegraphics[width=3.5in,height=3.5in]{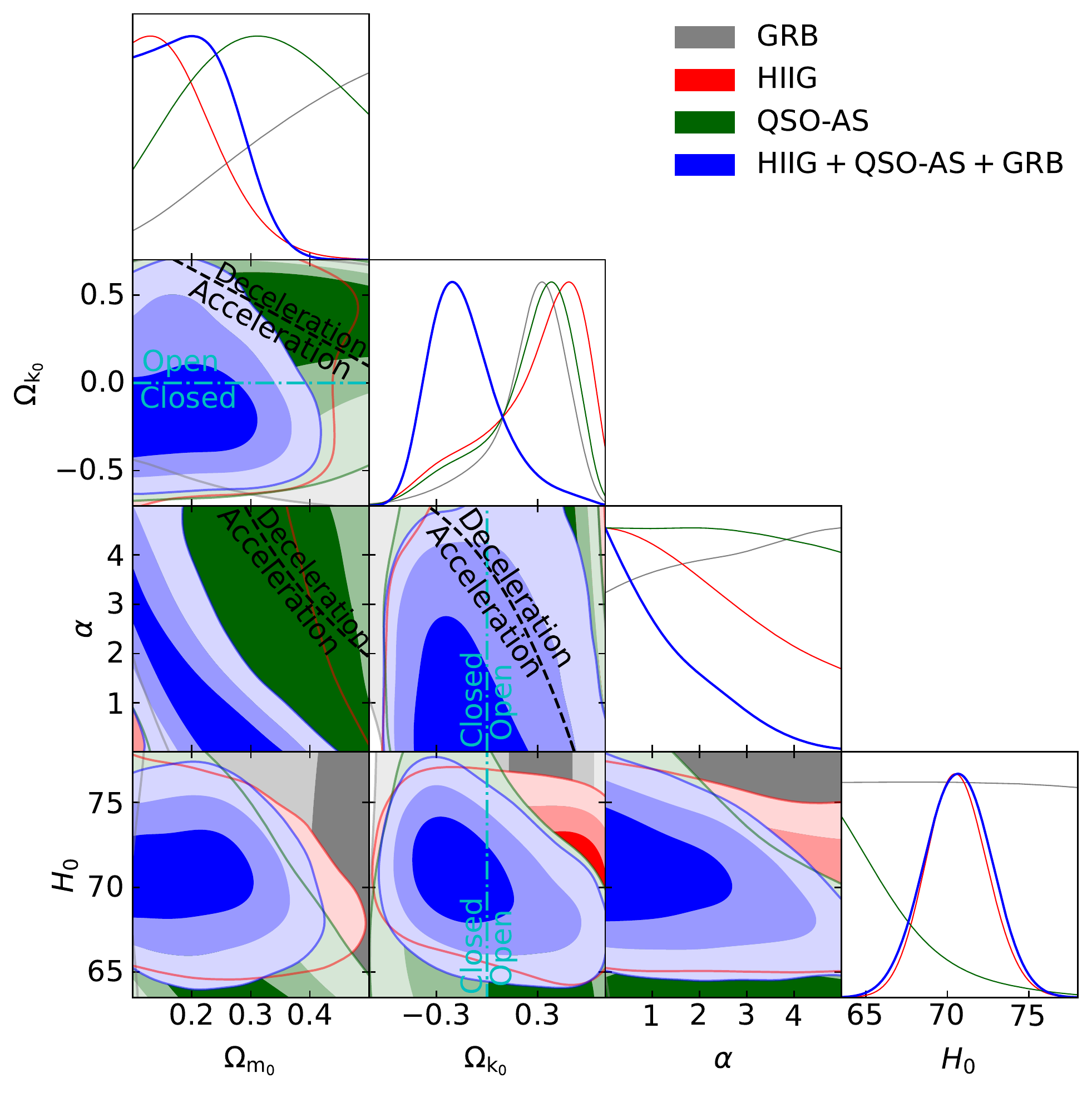}}\\
\caption{Same as Fig. \ref{fig5} but for non-flat \pcdm, where the zero-acceleration lines in each of the sub-panels are computed for the third cosmological parameter set to the $H(z)$ + BAO data best-fitting values listed in Table \ref{tab:BFP}. Currently-accelerating cosmological expansion occurs below these lines. The cyan dash-dot lines represent the flat \pcdm\ case, with closed spatial geometry either below or to the left. The $\alpha = 0$ axis is the non-flat \lcdm\ model.}
\label{fig6}
\end{figure*}

\begin{figure*}
\centering
  \subfloat[All parameters]{%
    \includegraphics[width=3.5in,height=3.5in]{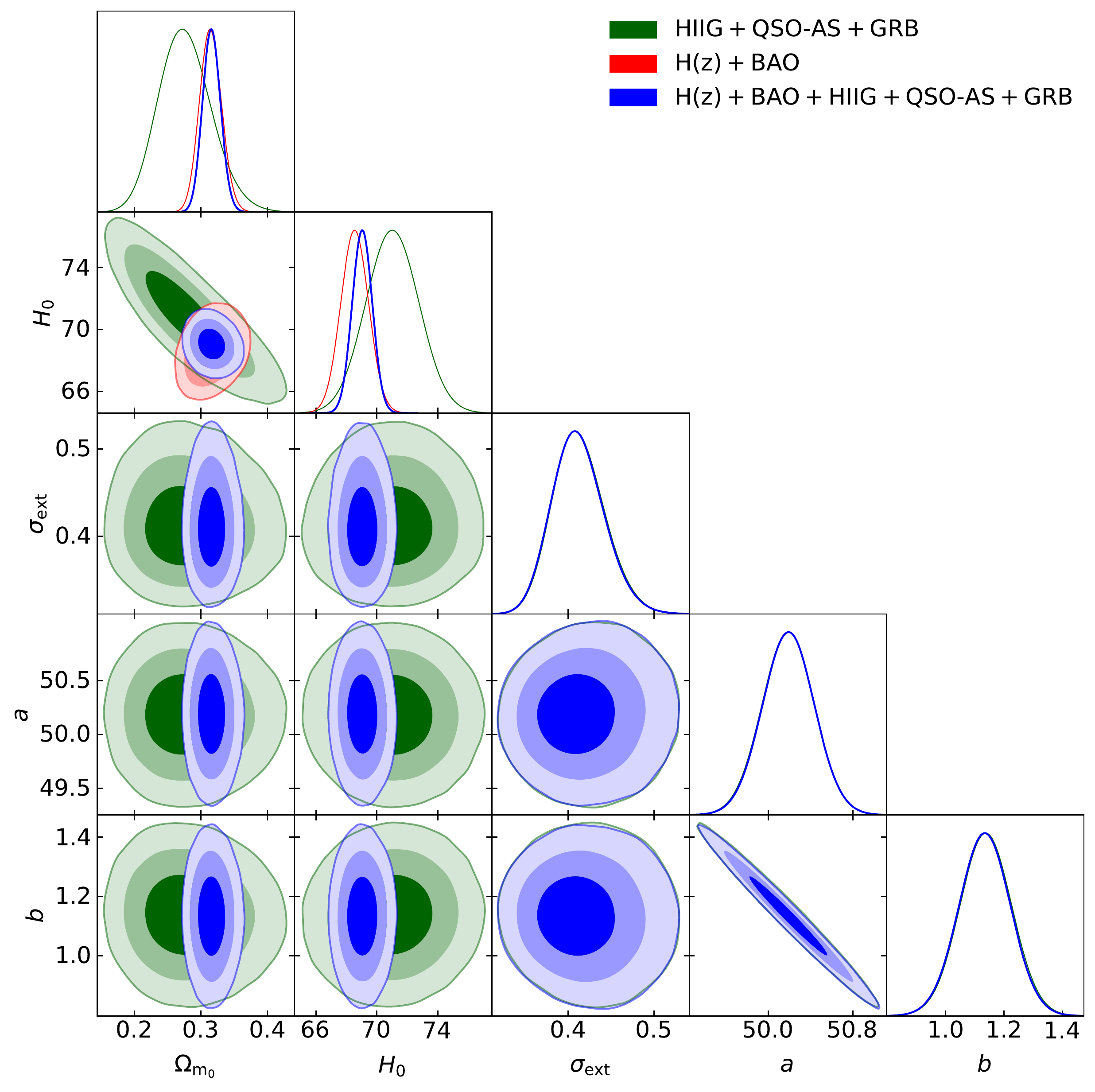}}
  \subfloat[Cosmological parameters zoom in]{%
    \includegraphics[width=3.5in,height=3.5in]{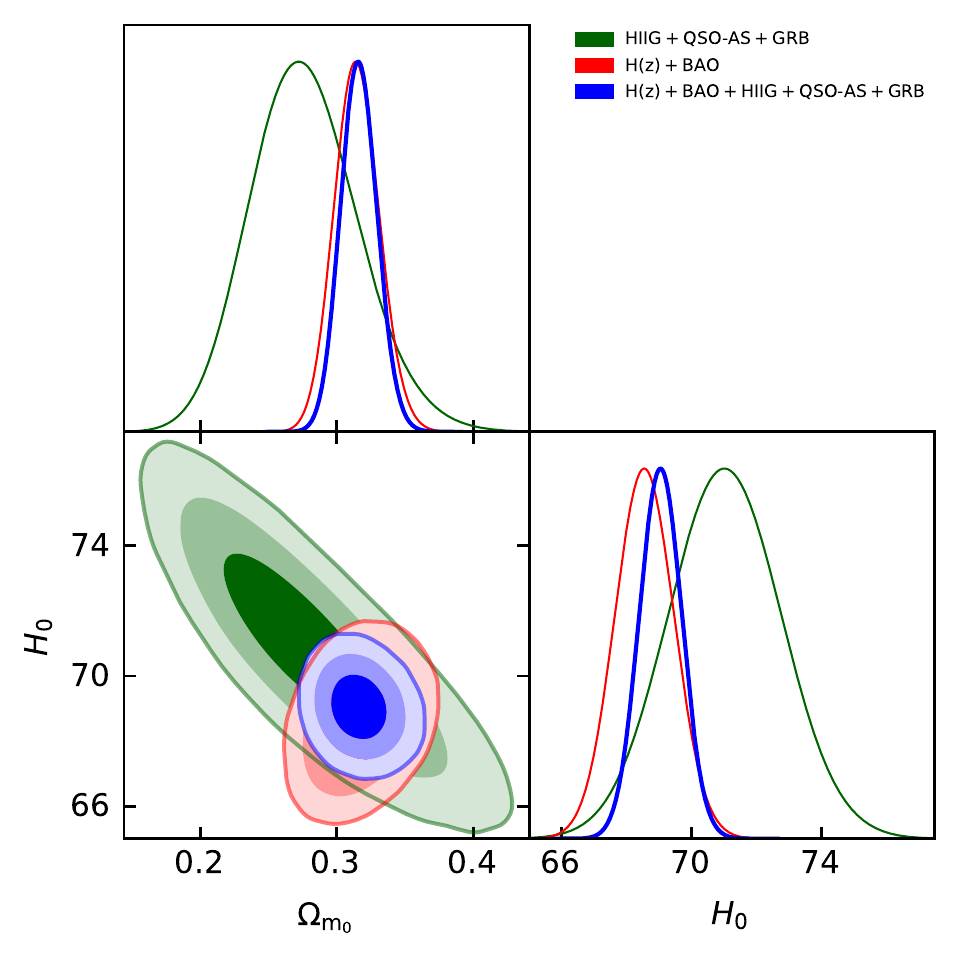}}\\
\caption{Same as Fig. \ref{fig1} (flat \lcdm) but for different combinations of data.}
\label{fig7}
\end{figure*}

\begin{figure*}
\centering
  \subfloat[All parameters]{%
    \includegraphics[width=3.5in,height=3.5in]{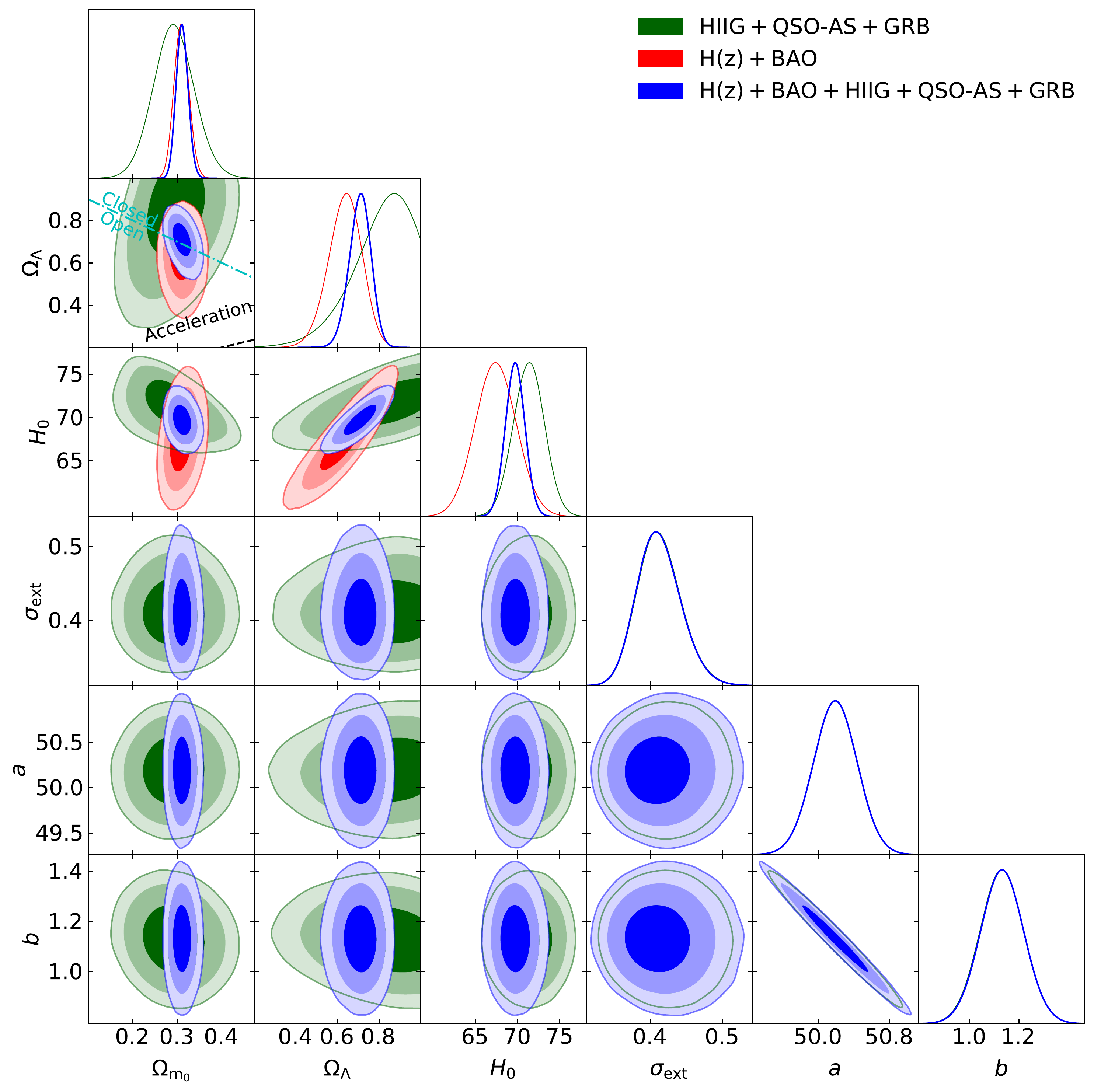}}
  \subfloat[Cosmological parameters zoom in]{%
    \includegraphics[width=3.5in,height=3.5in]{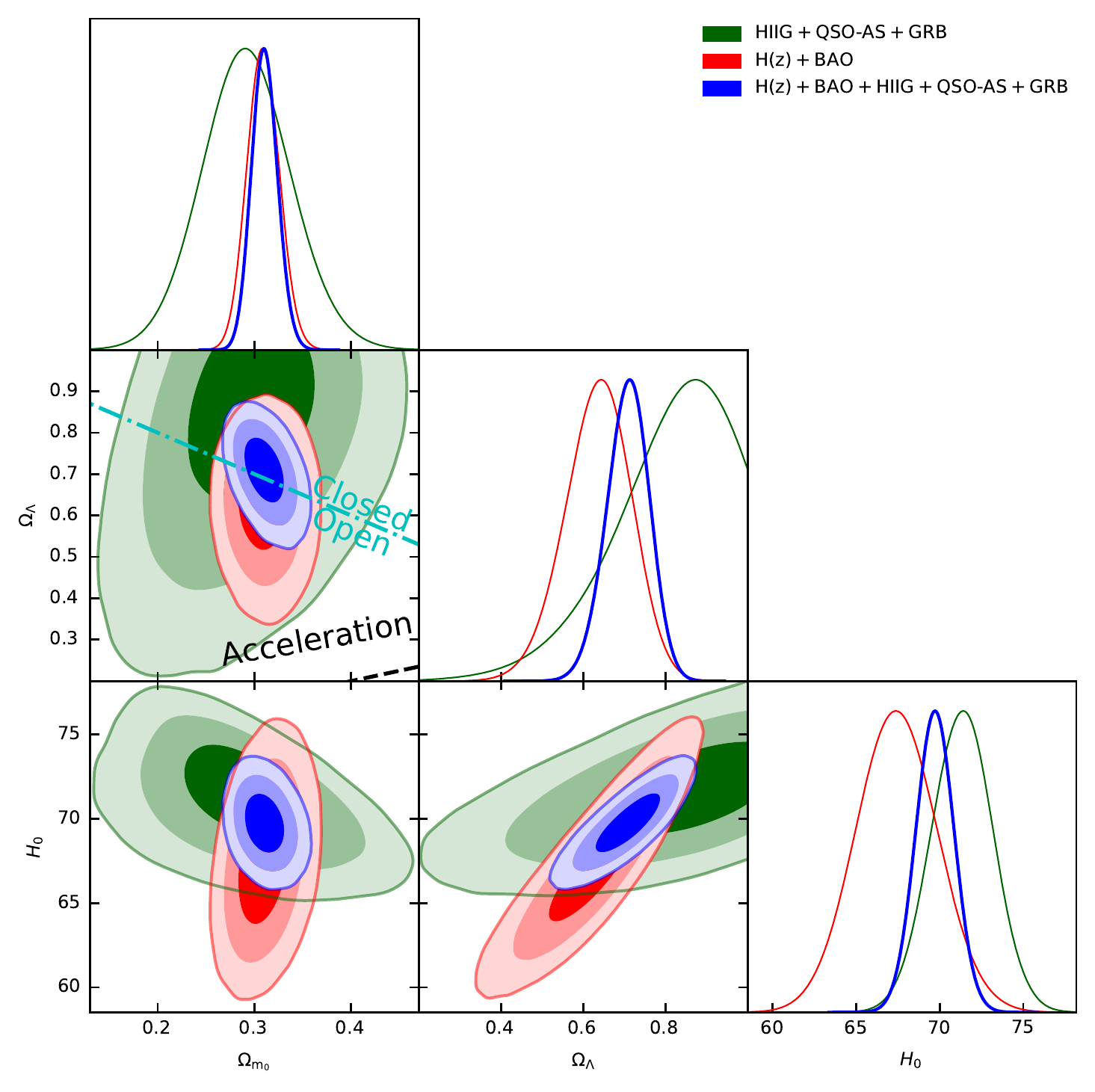}}\\
\caption{Same as Fig. \ref{fig2} (non-flat \lcdm) but for different combinations of data.}
\label{fig8}
\end{figure*}

\begin{figure*}
\centering
  \subfloat[All parameters]{%
    \includegraphics[width=3.5in,height=3.5in]{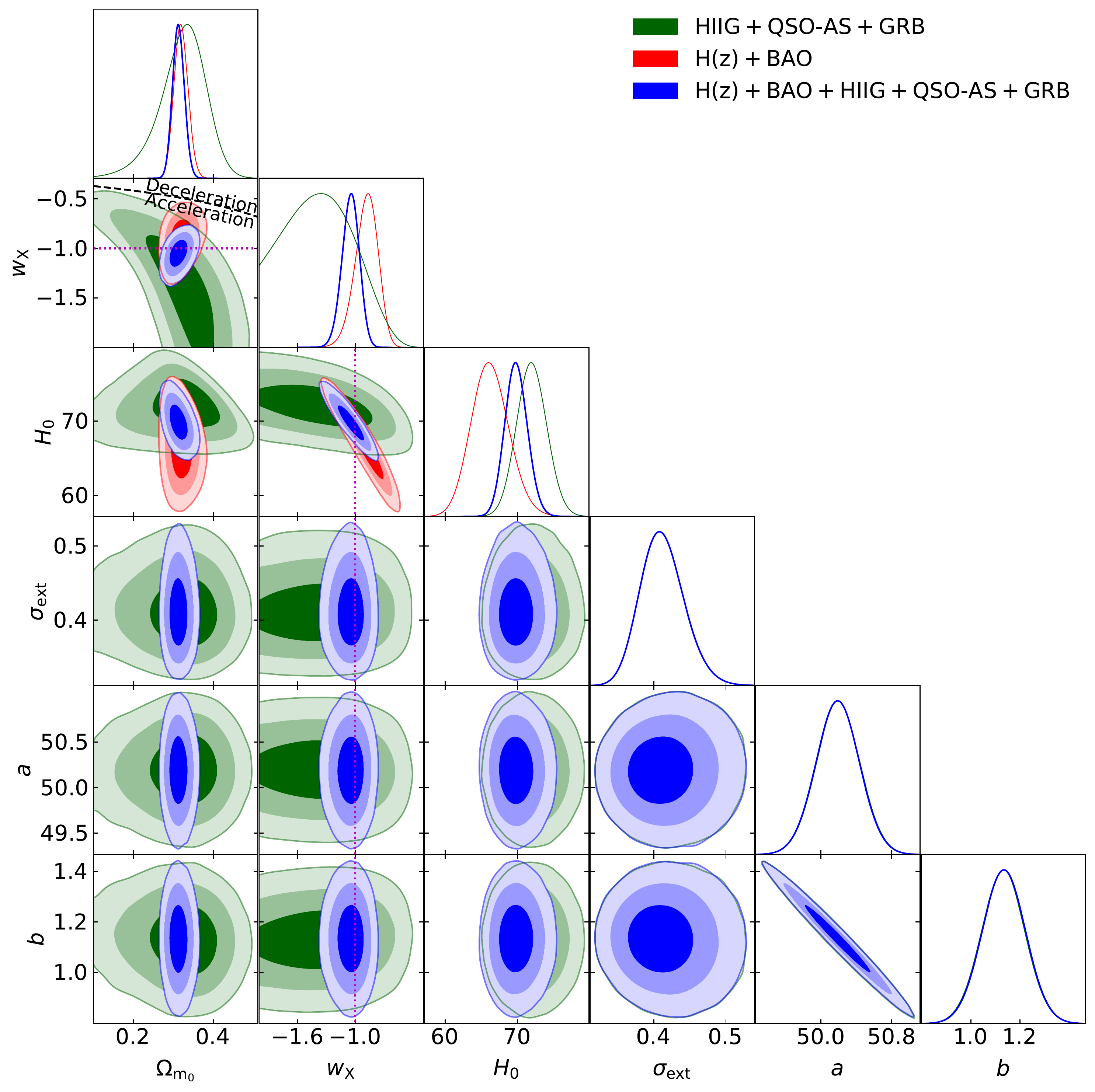}}
  \subfloat[Cosmological parameters zoom in]{%
    \includegraphics[width=3.5in,height=3.5in]{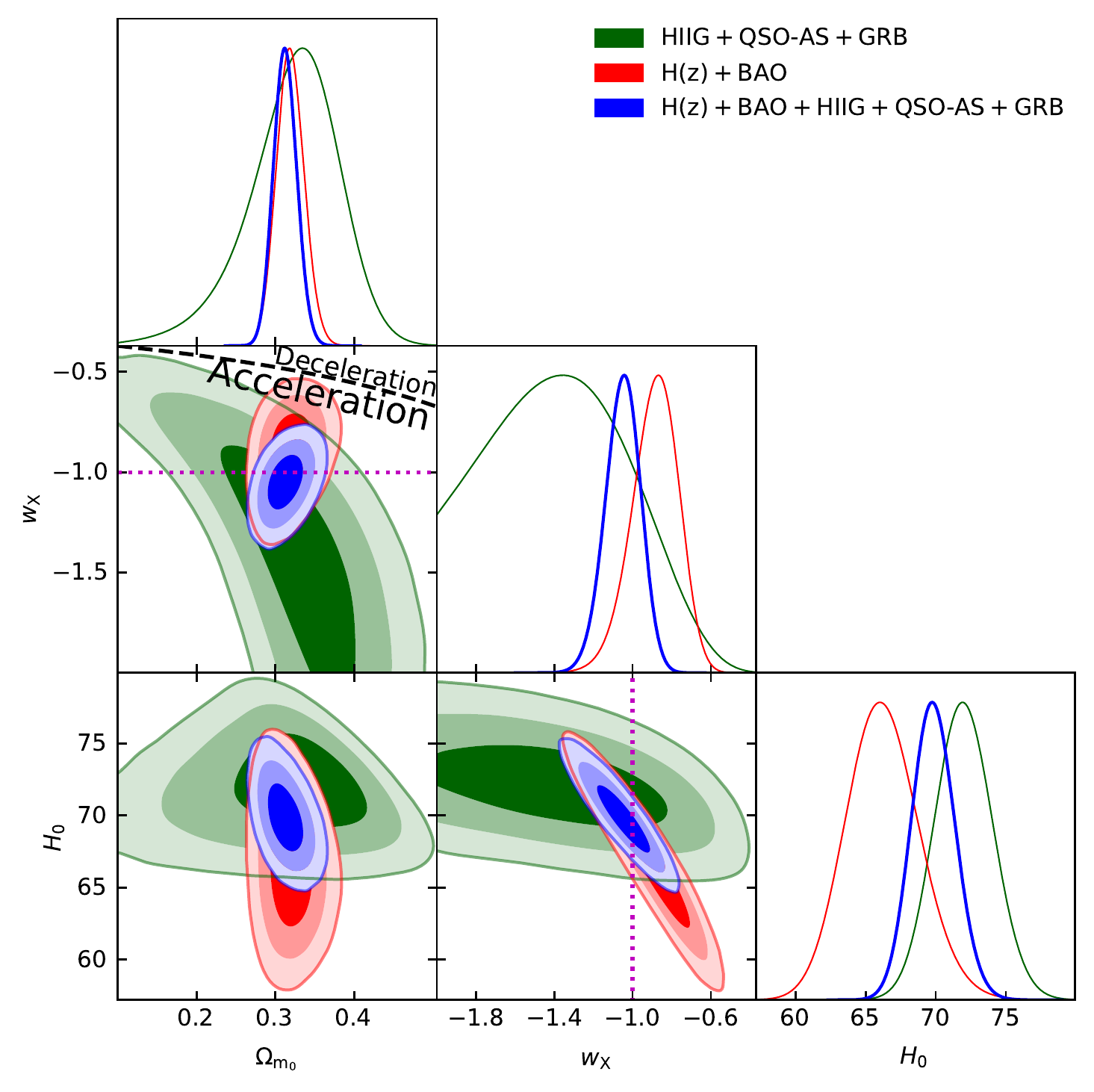}}\\
\caption{Same as Fig. \ref{fig3} (flat XCDM) but for different combinations of data.}
\label{fig9}
\end{figure*}

\begin{figure*}
\centering
  \subfloat[All parameters]{%
    \includegraphics[width=3.5in,height=3.5in]{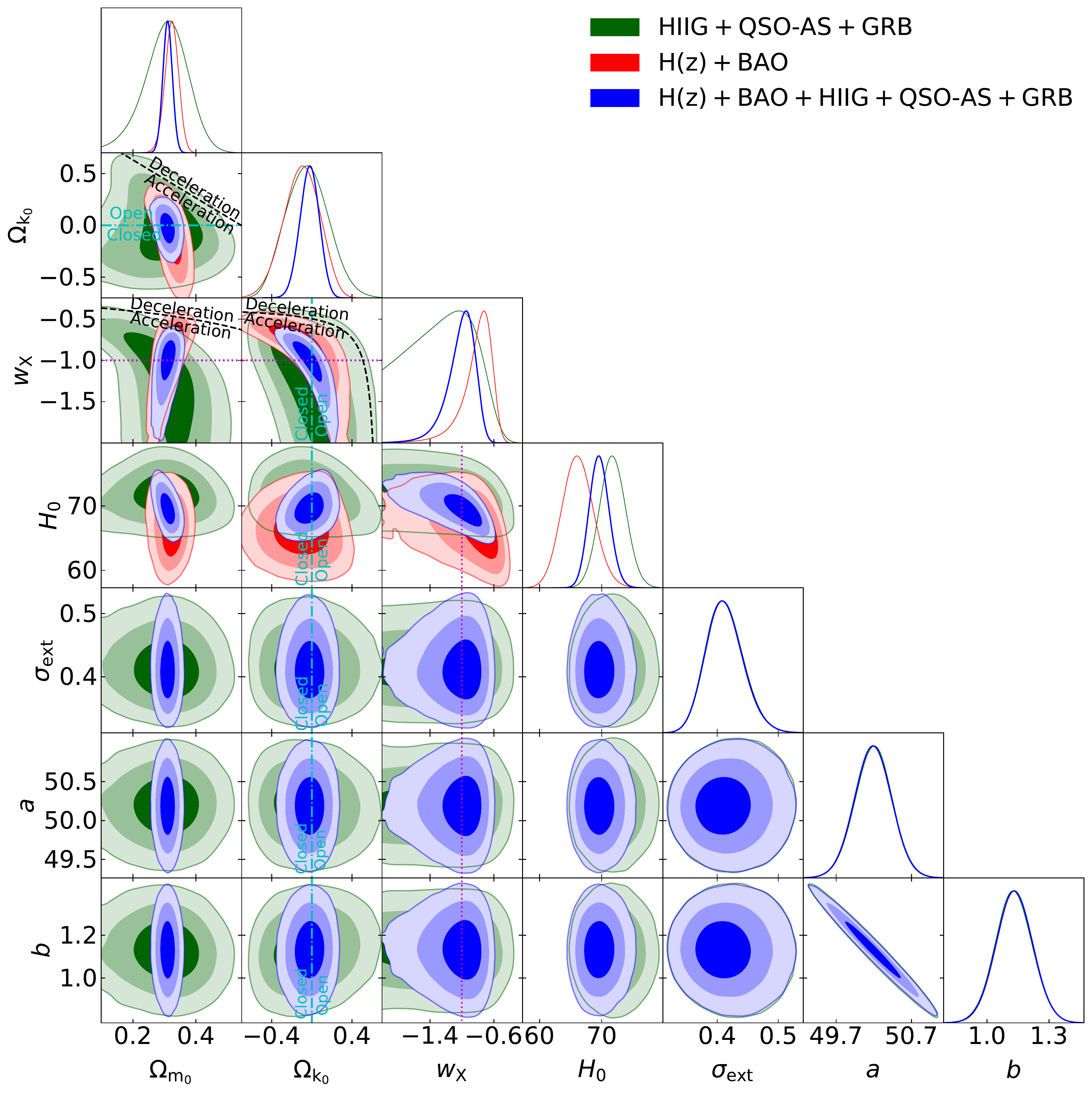}}
  \subfloat[Cosmological parameters zoom in]{%
    \includegraphics[width=3.5in,height=3.5in]{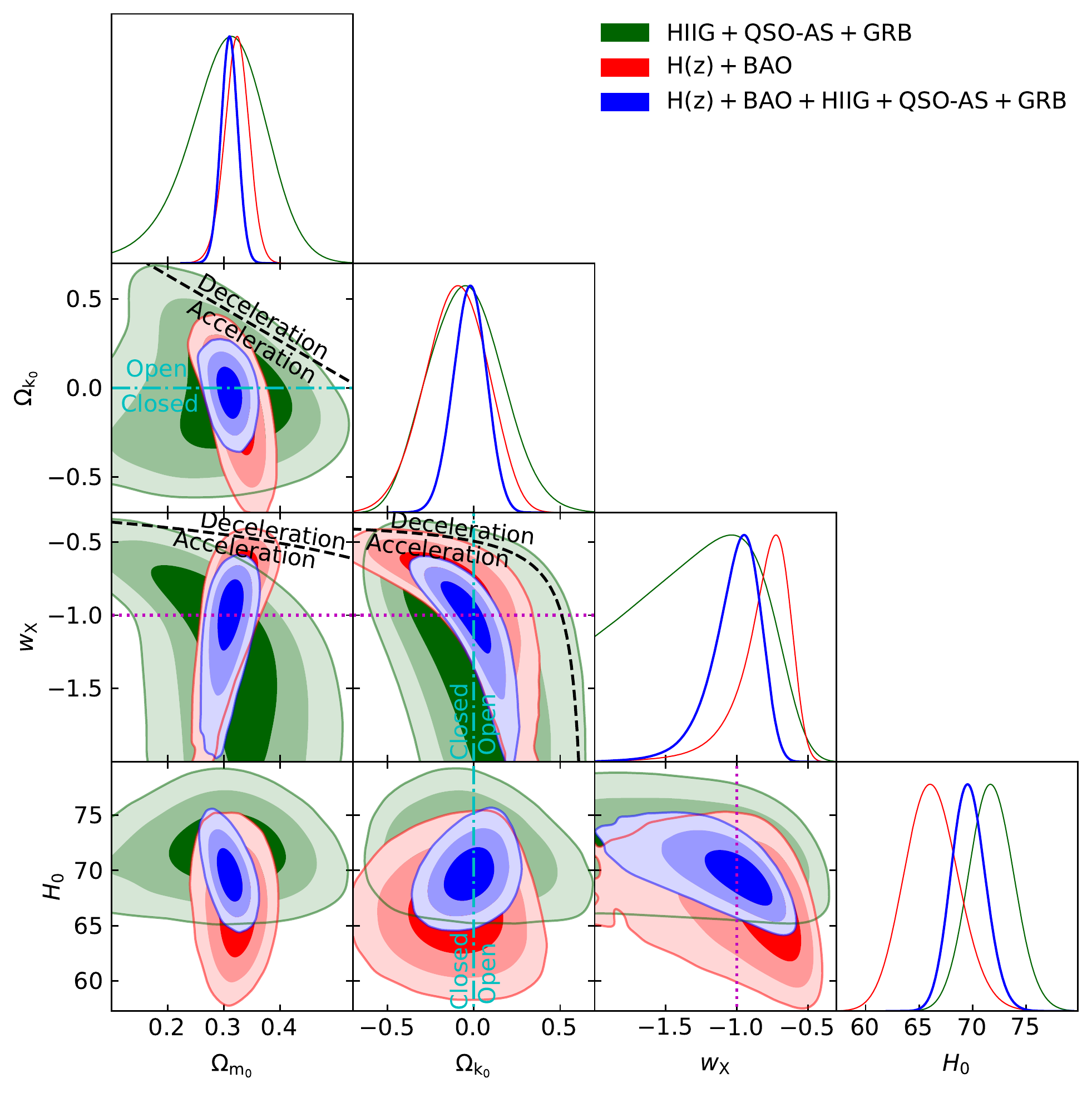}}\\
\caption{Same as Fig. \ref{fig4} (non-flat XCDM) but for different combinations of data.}
\label{fig10}
\end{figure*}

\begin{figure*}
\centering
  \subfloat[All parameters]{%
    \includegraphics[width=3.5in,height=3.5in]{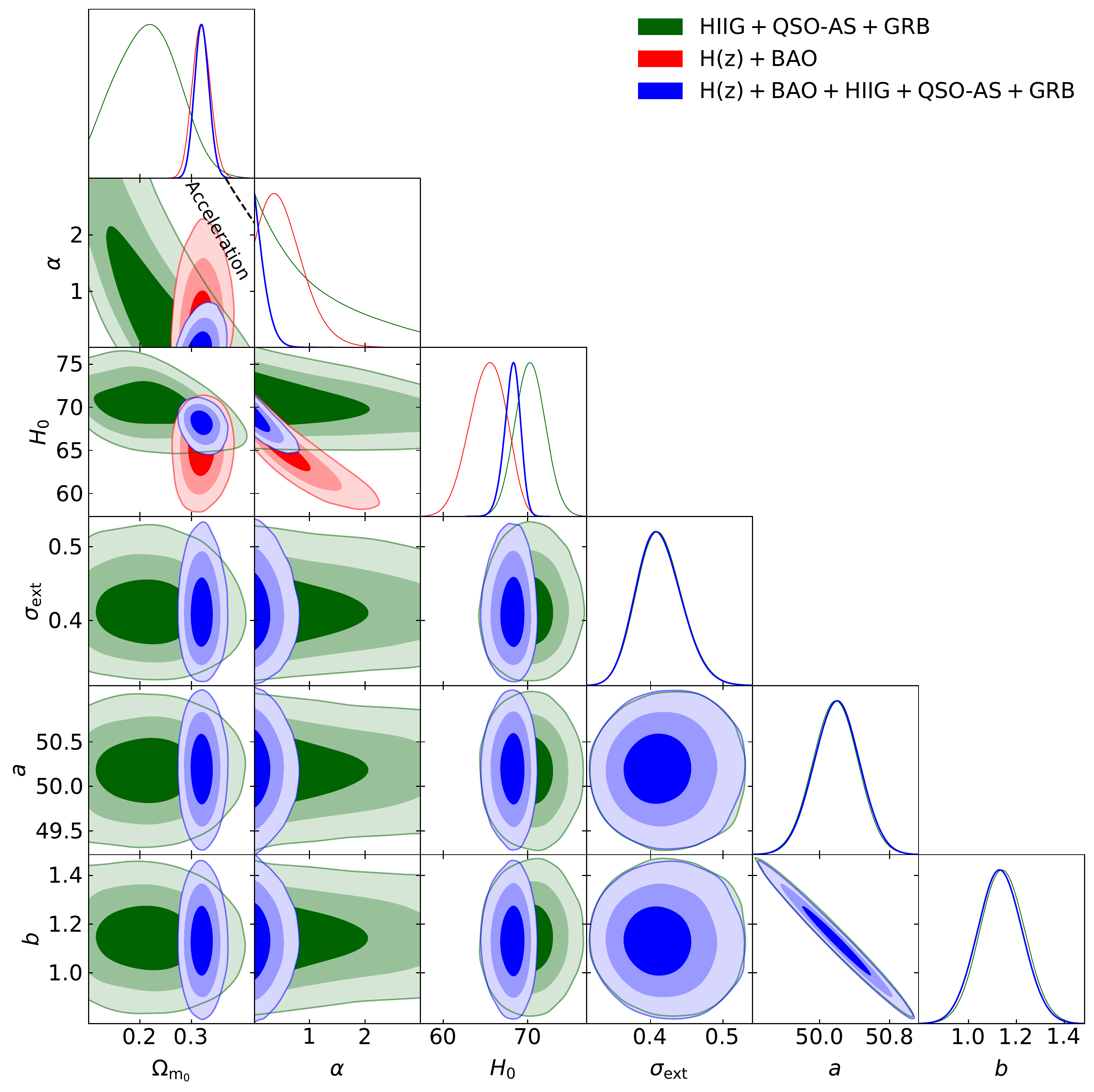}}
  \subfloat[Cosmological parameters zoom in]{%
    \includegraphics[width=3.5in,height=3.5in]{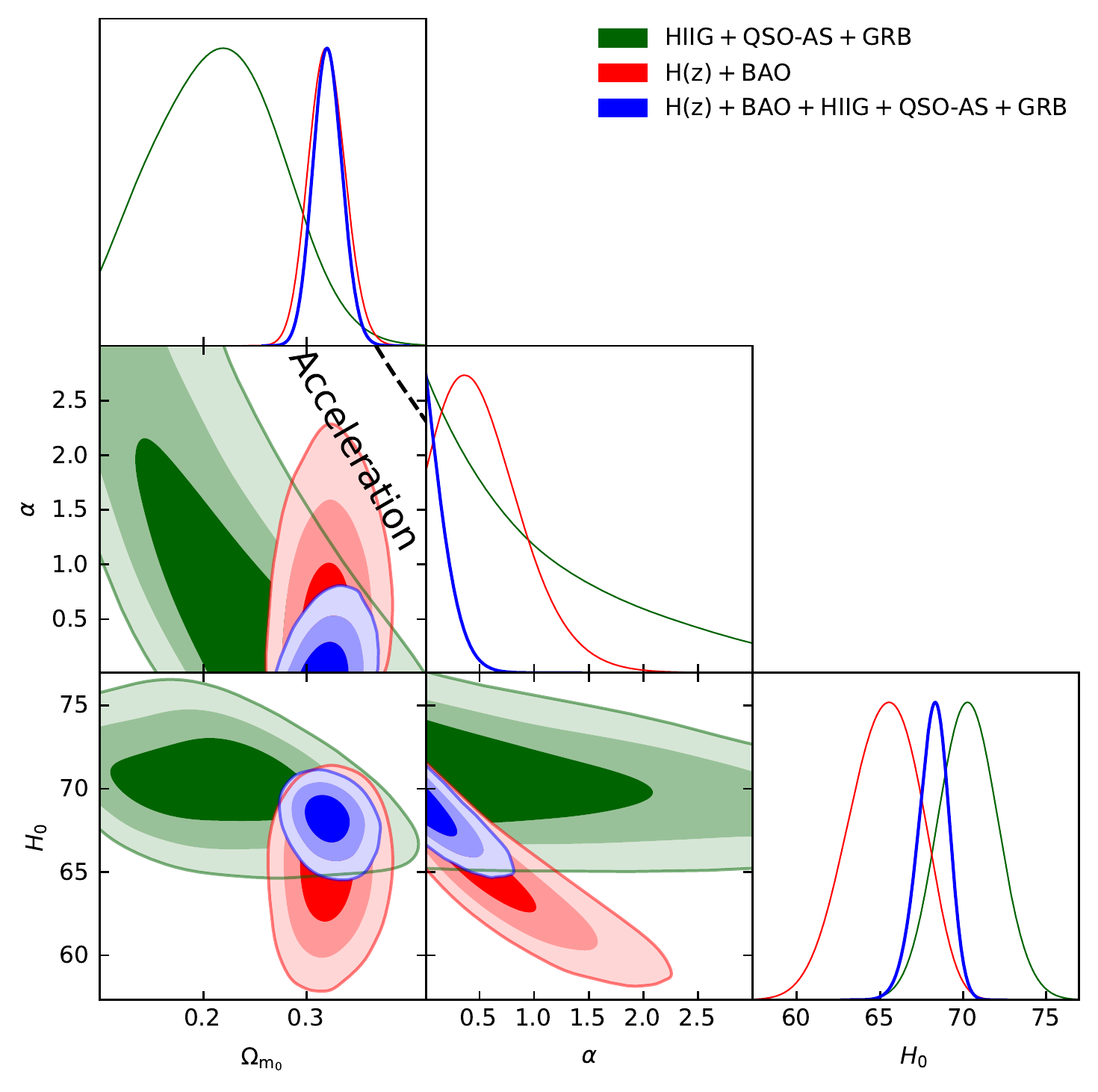}}\\
\caption{Same as Fig. \ref{fig5} (flat \pcdm) but for different combinations of data.}
\label{fig11}
\end{figure*}

\begin{figure*}
\centering
  \subfloat[All parameters]{%
    \includegraphics[width=3.5in,height=3.5in]{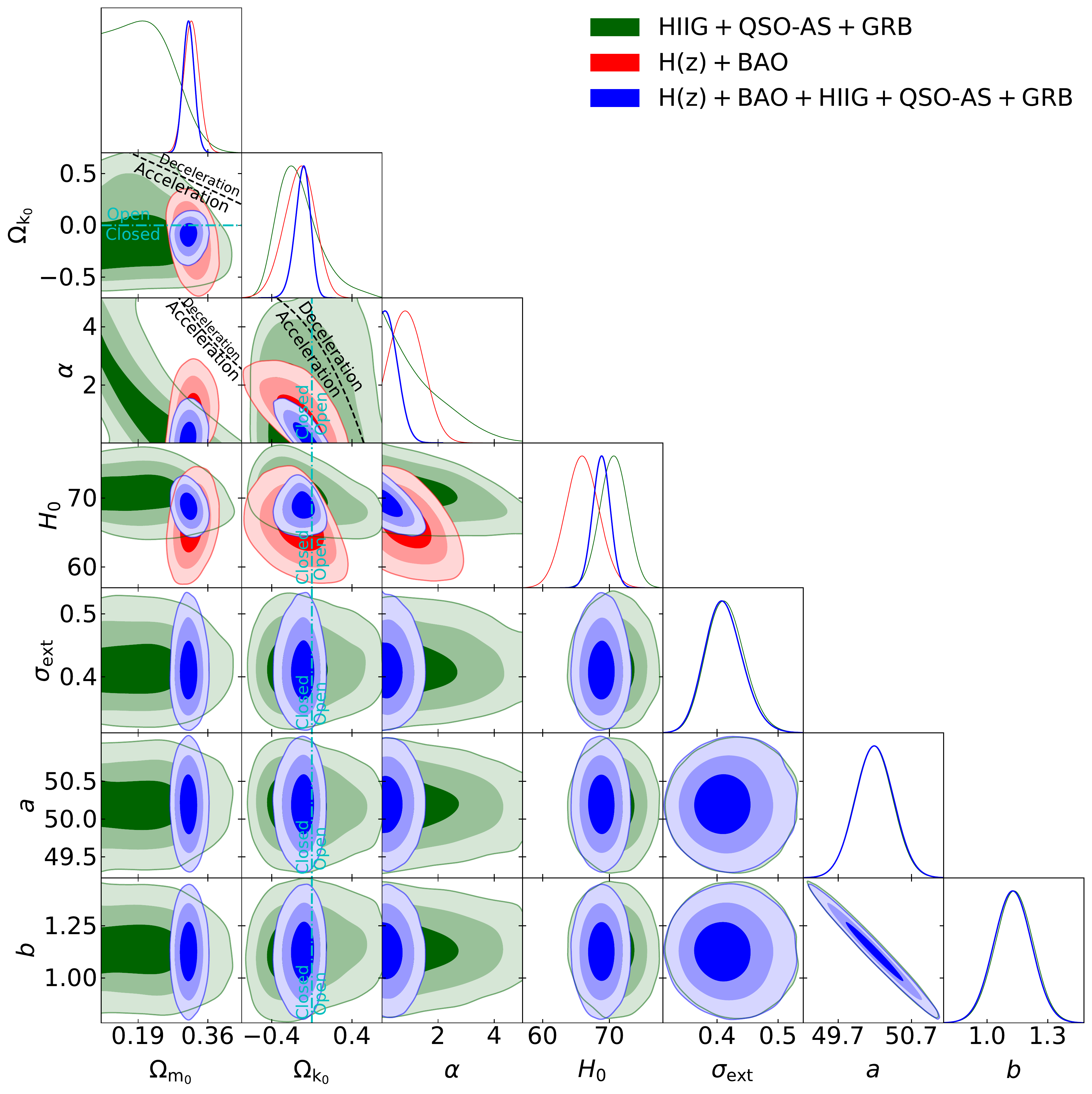}}
  \subfloat[Cosmological parameters zoom in]{%
    \includegraphics[width=3.5in,height=3.5in]{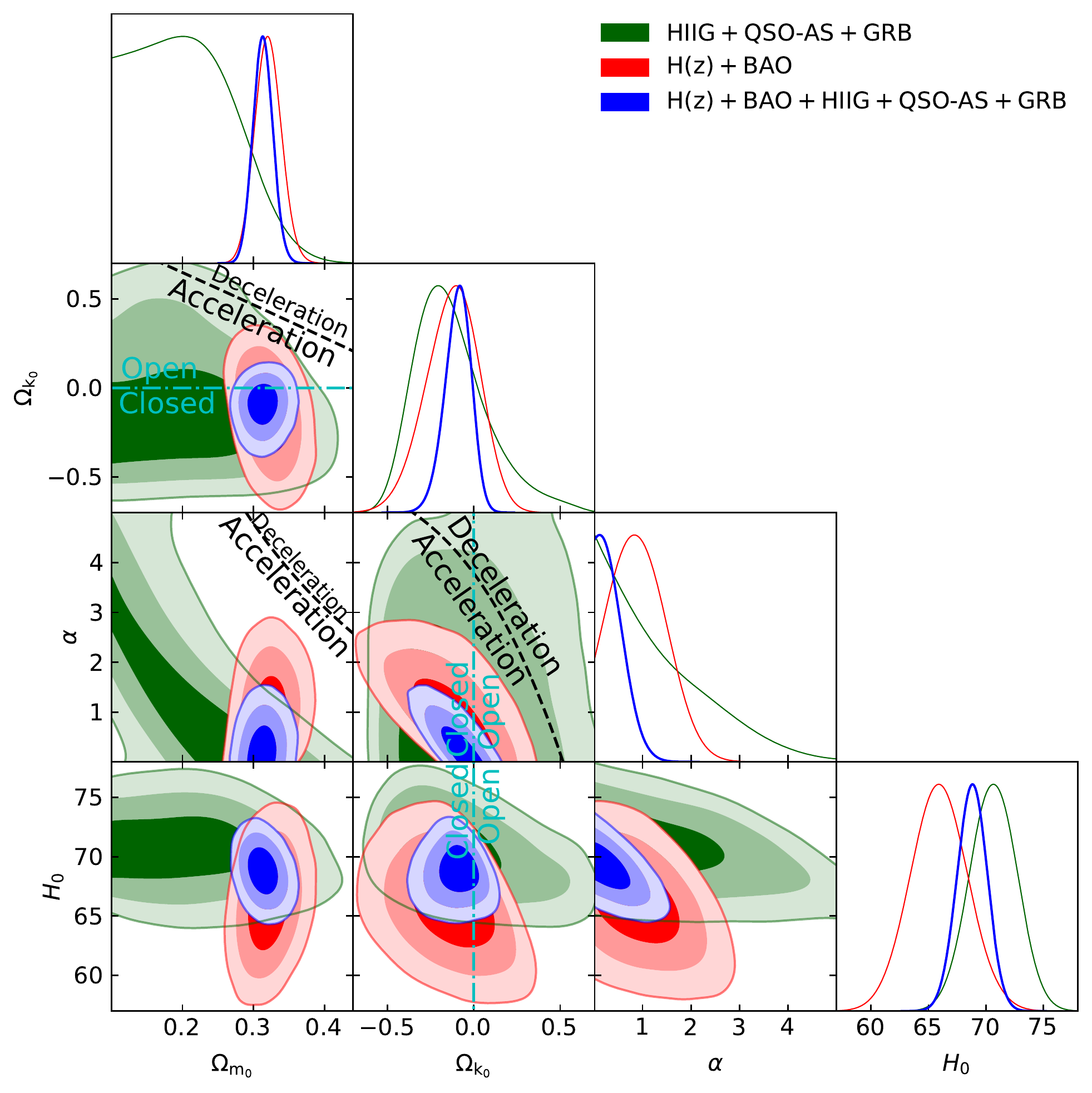}}\\
\caption{Same as Fig. \ref{fig6} (non-flat \pcdm) but for different combinations of data.}
\label{fig12}
\end{figure*}

The 1D probability distributions and 2D confidence regions of the cosmological and Amati relation parameters from the HQASG data are in Figs. \ref{fig1}--\ref{fig6}, in blue, Figs. \ref{fig7}--\ref{fig12}, in green, and panels (a) of Figs. \ref{fig01}--\ref{fig04}, in red. The best-fitting results and uncertainties are in Tables \ref{tab:BFP} and \ref{tab:1d_BFP}.

We find that the HQASG data combination favors currently-accelerating cosmological expansion.

The fit to the HQASG data produces best-fitting values of \om\ that lie between $0.205^{+0.044}_{-0.094}$ (non-flat \pcdm) at the low end, and $0.322^{+0.062}_{-0.044}$ (flat XCDM) at the high end. This range is smaller than the ranges within which \om\ falls when it is determined from the \hiig, QSO-AS, and GRB data individually, but the low and high ends of the range are still somewhat mutually inconsistent, being 2.66$\sigma$ away from each other. This is a consequence of the low \om\ value for non-flat \pcdm; the \om\ values for \lcdm\ and XCDM are quite consistent with the recent estimate of \cite{planck2018b}. In contrast, the best-fitting values of $H_0$ that we measure from the HQASG data are mutually very consistent (within $0.65\sigma$), with $H_0=70.30\pm1.68$ \hunit (flat \pcdm) at the low end of the range and $H_0=72.00^{+1.99}_{-1.98}$ \hunit (flat XCDM) at the high end of the range. These measurements are $0.83\sigma$ (flat XCDM) and $1.70\sigma$ (flat \pcdm) lower than the local Hubble constant measurement of $H_0 = 74.03 \pm 1.42$ \hunit \citep{riess_etal_2019}, and $0.70\sigma$ (flat \pcdm) and $1.16\sigma$ (flat XCDM) higher than the median statistics estimate of $H_0=68 \pm 2.8$ \hunit \citep{chenratmed}.\footnote{Other local expansion rate determinations have slightly lower central values with slightly larger error bars \citep{rigault_etal_2015,zhangetal2017,Dhawan,FernandezArenas,freedman_etal_2019,freedman_etal_2020,rameez_sarkar_2019,Breuvaletal_2020, Efstathiou_2020, Khetan_et_al_2020}. Our $H_0$ measurements are consistent with earlier median statistics estimates \citep{gott_etal_2001,chen_etal_2003} and with other recent $H_0$ determinations \citep{chen_etal_2017,DES_2018,Gomez-ValentAmendola2018, planck2018b, zhang_2018,dominguez_etal_2019,martinelli_tutusaus_2019,Cuceu_2019,zeng_yan_2019,schoneberg_etal_2019,lin_ishak_2019, Blum_et_al_2020, Lyu_et_al_2020, Philcox_et_al_2020, Zhang_Huang_2020, Birrer_et_al_2020, Denzel_et_al_2020}.}

In contrast to the \hiig, QSO-AS, and GRB only cases, when fitted to the HQASG data combination the non-flat models mildly favor closed spatial hypersurfaces. For non-flat \lcdm, non-flat XCDM, and non-flat \pcdm, we find $\Omega_{\rm k_0}=-0.093^{+0.092}_{-0.190}$, $\Omega_{\rm k_0}=-0.044^{+0.193}_{-0.217}$, and $\Omega_{\rm k_0}=-0.124^{+0.127}_{-0.253}$, respectively, with the non-flat \lcdm\ model favoring closed spatial hypersurfaces at 1.01$\sigma$.

The fit to the HQASG data combination produces stronger evidence for dark energy dynamics in the flat and non-flat XCDM parametrizations but weaker evidence in the flat and non-flat \pcdm\ models (in comparison to the \hiig\ and QSO-AS only cases) with tighter error bars on the measured values of $w_{\rm X}$ and $\alpha$. For flat (non-flat) XCDM, $w_{\rm X}=-1.379^{+0.361}_{-0.375}$ ($w_{\rm X}=-1.273^{+0.501}_{-0.321}$), with $w_{\rm X}=-1$ being within the 1$\sigma$ range for non-flat XCDM and being 1.05$\sigma$ larger for flat XCDM. For flat (non-flat) \pcdm, $\alpha<2.584$ ($\alpha<3.414$), where both likelihoods peak at $\alpha=0$.

The constraints on the Amati relation parameters from the HQASG data are also model-independent, but with slightly larger central values and smaller error bars for the parameter $a$. A reasonable summary is $\sigma_{\rm ext}=0.413^{+0.026}_{-0.032}$, $a=50.19\pm0.24$, and $b=1.133\pm0.086$.

The HQASG cosmological constraints are largely consistent with those from other data, like the constraints from the $H(z)$ + BAO data used in \cite{Caoetal_2020} and \cite{Khadka_2020c}, that are shown in red in Figs. \ref{fig7}--\ref{fig12}. We note, however, that there is some mild tension between \pcdm\ \om\ values, and between XCDM and \pcdm\ $H_0$ values determined from $H(z)$ + BAO and HQASG data, with the $2.46\sigma$ difference between \om\ values estimated from the two different data combinations in the non-flat \pcdm\ model being the only somewhat troubling difference (see Table \ref{tab:1d_BFP}).

\begin{table*}
\centering
\scriptsize
\begin{threeparttable} 
\caption{Unmarginalized best-fitting parameter values for all models from various combinations of data.}\label{tab:BFP}
\setlength{\tabcolsep}{0.6mm}{
\begin{tabular}{lcccccccccccccccccc}
\toprule
 Model & Data set & $\Omega_{\mathrm{m_0}}$ & $\Omega_{\Lambda}$ & $\Omega_{\mathrm{k_0}}$ & $w_{\mathrm{X}}$ & $\alpha$ & $H_0$\tnote{c} & $\sigma_{\mathrm{ext}}$ & $a$ & $b$ & $\chi^2$ & $\nu$ & $-2\ln\mathcal{L}_{\mathrm{max}}$ & $AIC$ & $BIC$ & $\Delta\chi^2$ & $\Delta AIC$ & $\Delta BIC$ \\
\midrule
Flat \lcdm & GRB & 0.698 & 0.302 & -- & -- & -- & 80.36 & 0.404 & 49.92& 1.113 & 117.98 & 114 & 130.12 & 140.12 & 154.01 & 1.08 & 0.00 & 0.00\\
 & \hiig\ & 0.276 & 0.724 & -- & -- & -- & 71.81 & -- & -- & -- & 410.75 & 151 & 410.75 & 414.75 & 420.81 & 3.15 & 0.00 & 0.00\\
 & QSO-AS & 0.315 & 0.685 & -- & -- & -- & 68.69 & -- & -- & -- & 352.05 & 118 & 352.05 & 356.05 & 361.62 & 1.76 & 0.00 & 0.00\\
 & HQASG\tnote{d} & 0.271 & 0.729 & -- & -- & -- & 71.13 & 0.407 & 50.18 & 1.138 & 879.42 & 387 & 895.05 & 905.05 & 924.91 & 0.12 & 0.00 & 0.00\\
 & $H(z)$ + BAO & 0.314 & 0.686 & -- & -- & -- & 68.53 & -- & -- & -- & 20.82 & 40 & 20.82 & 24.82 & 28.29 & 2.39 & 0.00 & 0.00\\
 & HzBHQASG\tnote{e} & 0.317 & 0.683 & -- & -- & -- & 69.06 & 0.404 & 50.19 & 1.134 & 903.61 & 429 & 917.79 & 927.79 & 948.16 & 4.05 & 0.00 & 0.00\\
\\
Non-flat \lcdm & GRB & 0.691 & 0.203 & 0.106 & -- & -- & 77.03 & 0.402 & 49.96 & 1.115 & 117.37 & 113 & 129.96 & 141.96 & 158.64 & 0.47 & 1.84 & 4.63\\
 & \hiig\ & 0.311 & 1.000 & $-0.311$ & -- & -- & 72.41 & -- & -- & -- & 410.44 & 150 & 410.44 & 416.44 & 425.53 & 2.84 & 1.69 & 4.72\\
 & QSO-AS & 0.266 & 1.000 & $-0.268$ & -- & -- & 74.73 & -- & -- & -- & 351.30 & 117 & 351.30 & 357.30 & 365.66 & 1.01 & 1.25 & 4.04\\
 & HQASG\tnote{d} & 0.291 & 0.876 & $-0.167$ & -- & -- & 72.00 & 0.406 & 50.22 & 1.120 & 879.30 & 386 & 894.02 & 906.02 & 929.85 & 0.00 & 0.97 & 4.94\\
 & $H(z)$ + BAO & 0.308 & 0.643 & 0.049 & -- & -- & 67.52 & -- & -- & -- & 20.52 & 39 & 20.52 & 26.52 & 31.73 & 2.09 & 1.70 & 3.44\\
 & HzBHQASG\tnote{e} & 0.309 & 0.716 & $-0.025$ & -- & -- & 69.77 & 0.402 & 50.17 & 1.141 & 904.47 & 428 & 917.17 & 929.17 & 953.61 & 4.91 & 1.38 & 5.45\\
\\
Flat XCDM & GRB & 0.102 & -- & -- & $-0.148$ & -- & 55.30 & 0.400 & 50.22 & 1.117 & 118.28 & 113 & 129.79 & 141.79 & 158.47 & 1.38 & 1.67 & 4.46\\
 & \hiig\ & 0.251 & -- & -- & $-0.899$ & -- & 71.66 & -- & -- & -- & 410.72 & 150 & 410.72 & 416.72 & 425.82 & 3.12 & 1.97 & 5.01\\
 & QSO-AS & 0.267 & -- & -- & $-2.000$ & -- & 81.70 & -- & -- & -- & 351.84 & 117 & 351.84 & 357.84 & 366.20 & 1.55 & 1.79 & 4.58\\
 & HQASG\tnote{d} & 0.320 & -- & -- & $-1.306$ & -- & 72.03 & 0.404 & 50.20 & 1.131 & 880.47 & 386 & 894.27 & 906.27 & 930.10 & 1.17 & 1.22 & 5.19\\
 & $H(z)$ + BAO & 0.319 & -- & -- & $-0.865$ & -- & 65.83 & -- & -- & -- & 19.54 & 39 & 19.54 & 25.54 & 30.76 & 1.11 & 0.72 & 2.47\\
 & HzBHQASG\tnote{e} & 0.313 & -- & -- & $-1.052$ & -- & 69.90 & 0.407 & 50.19 & 1.132 & 902.09 & 428 & 917.55 & 929.55 & 953.99 & 2.53 & 1.76 & 5.83\\
\\
Non-flat XCDM & GRB & 0.695 & -- & 0.556 & $-1.095$ & -- & 57.64 & 0.399 & 50.13 & 1.133 & 118.43 & 112 & 129.73 & 143.73 & 163.19 & 1.53 & 3.61 & 9.18\\
 & \hiig\ & 0.100 & -- & $-0.702$ & $-0.655$ & -- & 72.57 & -- & -- & -- & 407.60 & 149 & 407.60 & 415.60 & 427.72 & 0.00 & 0.85 & 6.91\\
 & QSO-AS & 0.100 & -- & $-0.548$ & $-0.670$ & -- & 74.04 & -- & -- & -- & 350.29 & 116 & 350.29 & 358.29 & 369.44 & 0.00 & 2.24 & 7.82\\
 & HQASG\tnote{d} & 0.300 & -- & $-0.161$ & $-1.027$ & -- & 80.36 & 0.405 & 50.21 & 1.122 & 879.48 & 385 & 894.01 & 908.01 & 935.81 & 0.18 & 2.96 & 10.90\\
 & $H(z)$ + BAO & 0.327 & -- & $-0.159$ & $-0.730$ & -- & 65.97 & -- & -- & -- & 18.43 & 38 & 18.43 & 26.43 & 33.38 & 0.00 & 1.61 & 5.09\\
 & HzBHQASG\tnote{e} & 0.312 & -- & $-0.045$ & $-0.959$ & -- & 69.46 & 0.402 & 50.23 & 1.117 & 904.17 & 427 & 917.07 & 931.07 & 959.58 & 4.61 & 3.28 & 11.42\\
\\
Flat $\phi$CDM & GRB & 0.674 & -- & -- & -- & 2.535 & 84.00 & 0.399 & 49.88 & 1.104 & 119.15 & 113 & 130.14 & 142.14 & 158.82 & 2.25 & 2.02 & 4.81\\
 & \hiig\ & 0.255 & -- & -- & -- & 0.260 & 71.70 & -- & -- & -- & 410.70 & 150 & 410.70 & 416.70 & 425.80 & 3.10 & 1.95 & 4.99\\
 & QSO-AS & 0.319 & -- & -- & -- & 0.012 & 68.47 & -- & -- & -- & 352.05 & 117 & 352.05 & 358.05 & 366.41 & 1.76 & 2.00 & 4.79\\
 & HQASG\tnote{d} & 0.282 & -- & -- & -- & 0.012 & 70.81 & 0.402 & 50.19 & 1.135 & 882.56 & 386 & 895.28 & 907.28 & 931.11 & 3.26 & 2.23 & 6.20\\
 & $H(z)$ + BAO & 0.318 & -- & -- & -- & 0.364 & 66.04 & -- & -- & -- & 19.65 & 39 & 19.65 & 25.65 & 30.86 & 1.22 & 0.83 & 2.57\\
 & HzBHQASG\tnote{e} & 0.316 & -- & -- & -- & 0.013 & 69.15 & 0.405 & 50.24 & 1.114 & 903.52 & 428 & 918.12 & 930.12 & 954.56 & 3.96 & 2.33 & 6.40\\
\\
Non-flat $\phi$CDM & GRB & 0.664 & -- & 0.188 & -- & 4.269 & 59.65 & 0.403 & 50.17 & 1.111 & 116.90 & 112 & 129.93 & 143.93 & 163.39 & 0.00 & 3.81 & 9.38\\
 & \hiig\ & 0.114 & -- & $-0.437$ & -- & 2.680 & 72.14 & -- & -- & -- & 409.91 & 149 & 409.91 & 417.91 & 430.03 & 2.31 & 3.16 & 9.22\\
 & QSO-AS & 0.100 & -- & $-0.433$ & -- & 2.948 & 72.37 & -- & -- & -- & 350.98 & 116 & 350.98 & 358.98 & 370.13 & 0.69 & 2.93 & 8.51\\
 & HQASG\tnote{d} & 0.276 & -- & $-0.185$ & -- & $0.145$ & 72.11 & 0.402 & 50.16 & 1.142 & 881.09 & 385 & 894.24 & 908.24 & 936.03 & 1.79 & 3.19 & 11.12\\
 & $H(z)$ + BAO & 0.321 & -- & $-0.137$ & -- & 0.887 & 66.41 & -- & -- & -- & 18.61 & 39 & 18.61 & 26.61 & 33.56 & 0.18 & 1.79 & 5.27\\
 & HzBHQASG\tnote{e} & 0.310 & -- & $-0.052$ & -- & 0.193 & 69.06 & 0.411 & 50.21 & 1.126 & 899.56 & 427 & 917.26 & 931.26 & 959.77 & 0.00 & 3.47 & 11.61\\
\bottomrule
\end{tabular}}
\begin{tablenotes}[flushleft]
\item [c] \hunit.
\item [d] \hiig\ + QSO-AS + GRB.
\item [e] $H(z)$ + BAO + \hiig\ + QSO-AS + GRB.
\end{tablenotes}
\end{threeparttable}
\end{table*}

\subsection{$H(z)$, BAO, \hiig, QSO-AS, and GRB (HzBHQASG) constraints}
\label{subsec:HzBHQASG}

Given the good mutual consistency between constraints derived from $H(z)$ + BAO data and those derived from HQASG data, in this subsection we determine more restrictive joint constraints from the combined $H(z)$, BAO, \hiig, QSO-AS, and GRB (HzBHQASG) data on the parameters of our six cosmological models.

The 1D probability distributions and 2D confidence regions of the cosmological and Amati relation parameters for all models from the HzBHQASG data are in blue in Figs. \ref{fig7}--\ref{fig12}, and in red in panels (b) of Figs. \ref{fig01}--\ref{fig04}. The best-fitting results and uncertainties are in Tables \ref{tab:BFP} and \ref{tab:1d_BFP}.

The measured values of \om\ here are a little larger, and significantly more restrictively constrained, than the ones in the HQASG cases (except for flat XCDM), being between $0.310\pm0.014$ (non-flat XCDM) and $0.320\pm0.013$ (flat \pcdm). The $H_0$ measurements are a little lower, and more tightly constrained, than in the HQASG cases, and are in better agreement with the lower median statistics estimate of $H_0$ \citep{chenratmed} than the higher local expansion rate measurement of $H_0$ \citep{riess_etal_2019}, being between $68.16^{+1.01}_{-0.80}$ \hunit (flat \pcdm) and $69.85^{+1.42}_{-1.55}$ \hunit (flat XCDM).

For non-flat \lcdm, non-flat XCDM, and non-flat \pcdm, we measure $\Omega_{\rm k_0}=-0.019^{+0.043}_{-0.048}$, $\Omega_{\rm k_0}=-0.024^{+0.092}_{-0.093}$, and $\Omega_{\rm k_0}=-0.094^{+0.082}_{-0.064}$, respectively, where the central values are a little higher (closer to 0) than what was measured in the HQASG cases. The joint constraints are more restrictive, with non-flat \lcdm\ and XCDM within 0.44$\sigma$ and 0.26$\sigma$ of $\Omega_{\rm k_0} = 0$, respectively. The non-flat \pcdm\ model, on the other hand, still favors a closed geometry with an $\Omega_{\rm k_0}$ that is 1.15$\sigma$ away from zero.

The HzBHQASG case has slightly larger measured values and tighter error bars for $w_{\rm X}$ and $\alpha$ than the HQASG case, so there is also not much evidence in support of dark energy dynamics. For flat (non-flat) XCDM, $w_{\rm X}=-1.050^{+0.090}_{-0.081}$ ($w_{\rm X}=-1.019^{+0.202}_{-0.099}$). For flat (non-flat) \pcdm, the $2\sigma$ upper limits are $\alpha<0.418$ ($\alpha<0.905$).

The cosmological model-independent constraints from the HzBHQASG data combination on the parameters of the Amati relation can be summarized as $\sigma_{\rm ext}=0.412^{+0.026}_{-0.032}$, $a=50.19\pm0.24$, and $b=1.132\pm0.085$.

\subsection{Model comparison}
\label{subsec:comparison}

From Table \ref{tab:BFP}, we see that the reduced $\chi^2$ values determined from GRB data alone are around unity for all models (being between 1.03 and 1.06) while those values determined from the $H(z)$ + BAO data combination range from 0.48 to 0.53, with the lower reduced $\chi^2$ here being due to the $H(z)$ data (that probably have overestimated error bars). As discussed in \cite{Ryan_2} and \cite{Caoetal_2020}, the cases that involve \hiig\ and QSO-AS data have a larger reduced $\chi^2$ (between 2.11 and 3.02), which is probably due to underestimated systematic uncertainties in both cases.

Based on the $AIC$ and the $BIC$ (see Table \ref{tab:BFP}), the flat \lcdm\ model remains the most favored model, across all data combinations, among the six models we study.\footnote{Note that based on the $\Delta \chi^2$ results of Table \ref{tab:BFP} non-flat \lcdm\ has the minimum $\chi^2$ in the HQASG case and non-flat XCDM has the minimum $\chi^2$ in the \hiig, QSO-AS, and $H(z)$ + BAO cases, whereas non-flat \pcdm\ has the minimum $\chi^2$ for the GRB and HzBHQASG cases. The $\Delta \chi^2$ values do not, however, penalize a model for having more parameters.} From $\Delta AIC$ and $\Delta BIC$, we find mostly weak or positive evidence against the models we considered, and only in a few cases do we find strong evidence against them. According to $\Delta BIC$, the evidence against non-flat XCDM is strong for the \hiig, QSO-AS, and GRB only cases, and very strong for the HQASG and HzBHQASG cases. Similarly, the evidence against flat \pcdm\ is strong for the HQASG and HzBHQASG cases, and the evidence against non-flat \pcdm\ is strong for the \hiig, QSO-AS, and GRB only cases, and very strong for the HQASG and HzBHQASG cases.

Among these six models, a comparison of the $\Delta BIC$ values from Table \ref{tab:BFP} shows that the most disfavored model is non-flat \pcdm, and that the second most disfavored model is non-flat XCDM. This is especially true when these models are fitted to the HQASG and HzBHQASG data combinations, in which cases non-flat \pcdm\ and non-flat XCDM are very strongly disfavored. These models aren't as strongly disfavored by the $AIC$, however; from a comparison of the $\Delta AIC$ values in Table \ref{tab:BFP}, we see that the evidence against the most disfavored model (non-flat \pcdm) is only positive.

\begin{table*}
\centering
\scriptsize
\begin{threeparttable}
\caption{One-dimensional marginalized best-fitting parameter values and uncertainties ($\pm 1\sigma$ error bars or $2\sigma$ limits) for all models from various combinations of data.}\label{tab:1d_BFP}
\setlength{\tabcolsep}{0.45mm}{
\begin{tabular}{lcccccccccc}
\toprule
 Model & Data set & $\Omega_{\mathrm{m_0}}$ & $\Omega_{\Lambda}$ & $\Omega_{\mathrm{k_0}}$ & $w_{\mathrm{X}}$ & $\alpha$ & $H_0$\tnote{c} & $\sigma_{\mathrm{ext}}$ & $a$ & $b$ \\
\midrule
Flat \lcdm & GRB & $>0.208$ & -- & -- & -- & -- & -- & $0.411^{+0.026}_{-0.032}$ & $50.16\pm0.27$ & $1.123\pm0.085$\\
 & \hiig\ & $0.289^{+0.053}_{-0.071}$ & -- & -- & -- & -- & $71.70\pm1.83$ & -- & -- & -- \\
 & QSO-AS & $0.364^{+0.083}_{-0.150}$ & -- & -- & -- & -- & $67.29^{+4.93}_{-5.07}$ & -- & -- & -- \\
 & HQASG\tnote{e} & $0.277^{+0.034}_{-0.041}$ & -- & -- & -- & -- & $71.03\pm1.67$ & $0.413^{+0.026}_{-0.032}$ & $50.19\pm0.24$ & $1.138\pm0.085$\\
 & $H(z)$ + BAO & $0.315^{+0.015}_{-0.017}$ & -- & -- & -- & -- & $68.55\pm0.87$ & -- & -- & -- \\
 & HzBHQASG\tnote{f} & $0.316\pm0.013$ & -- & -- & -- & -- & $69.05^{+0.62}_{-0.63}$ & $0.412^{+0.026}_{-0.032}$ & $50.19\pm0.23$ & $1.133\pm0.085$\\
\\
Non-flat \lcdm & GRB & $0.463^{+0.226}_{-0.084}$ & $<0.658$\tnote{d} & $-0.007^{+0.251}_{-0.234}$ & -- & -- & -- & $0.412^{+0.026}_{-0.032}$ & $50.17\pm0.28$ & $1.121\pm0.086$\\
 & \hiig\ & $0.275^{+0.081}_{-0.078}$ & $>0.501$\tnote{d} & $0.094^{+0.237}_{-0.363}$ & -- & -- & $71.50^{+1.80}_{-1.81}$ & -- & -- & -- \\
 & QSO-AS & $0.357^{+0.082}_{-0.135}$ & -- & $0.017^{+0.184}_{-0.277}$ & -- & -- & $67.32^{+4.49}_{-5.44}$ & -- & -- & -- \\
 & HQASG\tnote{e} & $0.292\pm0.044$ & $0.801^{+0.191}_{-0.055}$ & $-0.093^{+0.092}_{-0.190}$ & -- & -- & $71.33^{+1.75}_{-1.77}$ & $0.413^{+0.026}_{-0.032}$ & $50.19\pm0.24$ & $1.130\pm0.086$\\
 & $H(z)$ + BAO & $0.309\pm0.016$ & $0.636^{+0.081}_{-0.072}$ & $0.055^{+0.082}_{-0.074}$ & -- & -- & $67.44\pm2.33$ & -- & -- & -- \\
 & HzBHQASG\tnote{f} & $0.311^{+0.012}_{-0.014}$ & $0.708^{+0.053}_{-0.046}$ & $-0.019^{+0.043}_{-0.048}$ & -- & -- & $69.72\pm1.10$ & $0.412^{+0.026}_{-0.032}$ & $50.19\pm0.23$ & $1.132\pm0.085$ \\
\\
Flat XCDM & GRB & $>0.366$\tnote{d} & -- & -- & -- & -- & -- & $0.411^{+0.025}_{-0.032}$ & $50.14\pm0.28$ & $1.119\pm0.085$\\
 & \hiig\ & $0.300^{+0.106}_{-0.083}$ & -- & -- & $-1.180^{+0.560}_{-0.330}$ & -- & $71.85\pm1.96$ & -- & -- & -- \\
 & QSO-AS & $0.349^{+0.090}_{-0.143}$ & -- & -- & $-1.161^{+0.430}_{-0.679}$ & -- & $68.39^{+6.14}_{-8.98}$ & -- & -- & -- \\
 & HQASG\tnote{e} & $0.322^{+0.062}_{-0.044}$ & -- & -- & $-1.379^{+0.361}_{-0.375}$ & -- & $72.00^{+1.99}_{-1.98}$ & $0.412^{+0.026}_{-0.032}$ & $50.20\pm0.24$ & $1.130\pm0.085$\\
 & $H(z)$ + BAO & $0.319^{+0.016}_{-0.017}$ & -- & -- & $-0.888^{+0.126}_{-0.098}$ & -- & $66.26^{+2.32}_{-2.63}$ & -- & -- & -- \\
 & HzBHQASG\tnote{f} & $0.313^{+0.014}_{-0.015}$ & -- & -- & $-1.050^{+0.090}_{-0.081}$ & -- & $69.85^{+1.42}_{-1.55}$ & $0.412^{+0.026}_{-0.032}$ & $50.19\pm0.24$ & $1.134\pm0.085$ \\
\\
Non-flat XCDM & GRB & $>0.386$\tnote{d} & -- & $0.121^{+0.464}_{-0.275}$ & $>-1.218$\tnote{d} & -- & -- & $0.411^{+0.026}_{-0.032}$ & $50.12\pm0.28$ & $1.122\pm0.087$\\
 & \hiig\ & $0.275^{+0.084}_{-0.125}$ & -- & $0.011^{+0.457}_{-0.460}$ & $-1.125^{+0.537}_{-0.321}$ & -- & $71.71^{+2.07}_{-2.08}$ & -- & -- & -- \\
 & QSO-AS & $0.359^{+0.111}_{-0.174}$ & -- & $0.115^{+0.466}_{-0.293}$ & $-1.030^{+0.593}_{-0.548}$ & -- & $65.92^{+4.54}_{-9.54}$ & -- & -- & -- \\
 & HQASG\tnote{e} & $0.303^{+0.073}_{-0.058}$ & -- & $-0.044^{+0.193}_{-0.217}$ & $-1.273^{+0.501}_{-0.321}$ & -- & $71.77\pm2.02$ & $0.413^{+0.026}_{-0.031}$ & $50.20\pm0.24$ & $1.129\pm0.085$\\
 & $H(z)$ + BAO & $0.323^{+0.021}_{-0.020}$ & -- & $-0.105^{+0.187}_{-0.162}$ & $-0.818^{+0.212}_{-0.071}$ & -- & $66.20^{+2.29}_{-2.55}$ & -- & -- & -- \\
 & HzBHQASG\tnote{f} & $0.310\pm0.014$ & -- & $-0.024^{+0.092}_{-0.093}$ & $-1.019^{+0.202}_{-0.099}$ & -- & $69.63^{+1.45}_{-1.62}$ & $0.412^{+0.026}_{-0.031}$ & $50.19\pm0.23$ & $1.132\pm0.085$ \\
\\
Flat $\phi$CDM & GRB & $>0.376$\tnote{d} & -- & -- & -- & -- & -- & $0.411^{+0.025}_{-0.032}$ & $50.13\pm0.28$ & $1.121\pm0.087$\\
 & \hiig\ & $0.210^{+0.043}_{-0.092}$ & -- & -- & -- & $<2.784$ & $71.23^{+1.79}_{-1.80}$ & -- & -- & -- \\
 & QSO-AS & $0.329^{+0.086}_{-0.171}$ & -- & -- & -- & $<2.841$ & $64.42^{+4.47}_{-4.62}$ & -- & -- & -- \\
 & HQASG\tnote{e} & $0.214^{+0.057}_{-0.061}$ & -- & -- & -- & $<2.584$ & $70.30\pm1.68$ & $0.413^{+0.026}_{-0.032}$ & $50.18\pm0.24$ & $1.142\pm0.087$\\
 & $H(z)$ + BAO & $0.319^{+0.016}_{-0.017}$ & -- & -- & -- & $0.550^{+0.169}_{-0.494}$ & $65.25^{+2.25}_{-1.82}$ & -- & -- & -- \\
 & HzBHQASG\tnote{f} & $0.320\pm0.013$ & -- & -- & -- & $<0.418$ & $68.16^{+1.01}_{-0.80}$ & $0.412^{+0.027}_{-0.033}$ & $50.20\pm0.24$ & $1.131\pm0.088$ \\
\\
Non-flat $\phi$CDM & GRB & $>0.189$ & -- & $0.251^{+0.247}_{-0.086}$ & -- & -- & -- & $0.411^{+0.026}_{-0.032}$ & $50.11\pm0.28$ & $1.128\pm0.089$\\
 & \hiig\ & $<0.321$ & -- & $0.291^{+0.348}_{-0.113}$ & -- & $<4.590$ & $70.60^{+1.68}_{-1.84}$ & -- & -- & -- \\
 & QSO-AS & $0.362^{+0.117}_{-0.193}$ & -- & $0.254^{+0.304}_{-0.092}$ & -- & $<4.752$ & $61.91^{+2.83}_{-4.92}$ & -- & -- & -- \\
 & HQASG\tnote{e} & $0.205^{+0.044}_{-0.094}$ & -- & $-0.124^{+0.127}_{-0.253}$ & -- & $<3.414$ & $70.66\pm1.90$ & $0.414^{+0.027}_{-0.033}$ & $50.19\pm0.24$ & $1.134\pm0.088$\\
 & $H(z)$ + BAO & $0.321\pm0.017$ & -- & $-0.126^{+0.157}_{-0.130}$ & -- & $0.938^{+0.439}_{-0.644}$ & $65.93\pm2.33$ & -- & -- & -- \\
 & HzBHQASG\tnote{f} & $0.313\pm0.013$ & -- & $-0.094^{+0.082}_{-0.064}$ & -- & $<0.905$ & $68.79\pm1.22$ & $0.412^{+0.027}_{-0.033}$ & $50.20\pm0.24$ & $1.126\pm0.087$ \\
\bottomrule
\end{tabular}}
\begin{tablenotes}[flushleft]
\item [c] \hunit.
\item [d] This is the 1$\sigma$ limit. The $2\sigma$ limit is set by the prior, and is not shown here.
\item [e] \hiig\ + QSO-AS + GRB.
\item [f] $H(z)$ + BAO + \hiig\ + QSO-AS + GRB.
\end{tablenotes}
\end{threeparttable}
\end{table*}

\section{Conclusion}
\label{sec:conclusion}
We find that cosmological constraints determined from higher-$z$ GRB, \hiig, and QSO-AS data are mutually consistent. It is both reassuring and noteworthy that these higher-$z$ data jointly favor currently-accelerating cosmological expansion, and that their constraints are consistent with the constraints imposed by more widely used and more restrictive $H(z)$ and BAO data. Using a data set consisting of 31 $H(z)$, 11 BAO, 120 QSO-AS, 153 \hiig, and 119 GRB measurements, we jointly constrain the parameters of the GRB Amati relation and of six cosmological models. 

The GRB measurements are of special interest because they reach to $z\sim8.2$ (far beyond the highest $z\sim2.3$ reached by BAO data) and into a much less studied area of redshift space. Current GRB data do not provide very restrictive constraints on cosmological model parameters, but in the near future we expect there to be more GRB observations \citep{Shirokov2020} which should improve the GRB data and provide more restrictive cosmological constraints.

Some of our conclusions do not differ significantly between models and so are model-independent. In particular, for the HzBHQASG data (the full data set excluding QSO-Flux data), we find a fairly restrictive summary value of $\Omega_{\rm m_0}=0.313 \pm 0.013$ that agrees well with many other recent measurements. From these data we also find a fairly restrictive summary value of $H_0=69.3 \pm 1.2$ \hunit\ that is in better agreement with the results of \cite{chenratmed} and \cite{planck2018b} than with the result of \cite{riess_etal_2019}; note that we do not take the $H_0$ tension issue into account (for a review, see \citealp{riess_2019}). The HzBHQASG measurements are consistent with flat \lcdm, but do not rule out mild dark energy dynamics or a little spatial curvature energy density. More and better-quality higher-$z$ GRB, \hiig, QSO, and other data will significantly help to test these extensions of flat \lcdm.

\section*{Acknowledgements}

We thank Adam Riess for his comments and the anonymous referee for useful suggestions. This work was partially funded by Department of Energy grant DE-SC0011840. The computing for this project was performed on the Beocat Research Cluster at Kansas State University, which is funded in part by NSF grants CNS-1006860, EPS-1006860, EPS-0919443, ACI-1440548, CHE-1726332, and NIH P20GM113109.

\section*{Data availability}

The \hiig\ data used in this article were provided to us by the authors of \cite{G-M_2019}. These data will be shared on request to the corresponding author with the permission of the authors of \cite{G-M_2019}.




\bibliographystyle{mnras}
\bibliography{mybibfile} 




\appendix

\section{QSO-Flux}
\label{sec:appendix}

QSOs obey a nonlinear relation between their luminosities in the X-ray and UV bands. Using a sample of 808 QSOs in the redshift range $0.061 \leq z \leq 6.280$, \cite{RisalitiandLusso_2015} confirmed that this relation can be written
\begin{equation}
\label{eq:LX-LUV}
  \log L_X=\beta+\gamma\log L_{UV},
\end{equation}
where $L_X$ and $L_{UV}$ are the X-ray and UV luminosities of the QSOs. To make contact with observations, equation \eqref{eq:LX-LUV} must be expressed in terms of the fluxes $F_X$ and $F_{UV}$ measured at fixed rest-frame wavelengths in the X-ray and UV bands, respectively. With this, equation (\ref{eq:LX-LUV}) becomes
\be
\resizebox{0.47\textwidth}{!}{%
    $\log F_X=\beta+(\gamma-1)\log 4\pi+\gamma\log F_{UV} +2(\gamma-1)\log D_{L}.$%
    }
\ee
Here $D_L$ (defined in equation \ref{eq:DL}) is the luminosity distance, which depends on the parameters of our cosmological models. We also treat the slope $\gamma$ and intercept $\beta$ as free parameters in our cosmological model fits.

For QSO-Flux data, the natural log of its likelihood function is
\begin{equation*}
    \ln\mathcal{L_{\rm QF}}=-\frac{1}{2}\sum^{N}_{i=1}\Bigg[\frac{\big[\log(F^{\rm{obs}}_X)_i-\log(F^{\rm{th}}_X)_i\big]^2}{s_i^2}+\ln(2\pi s_i^2)\Bigg], \label{eq:LH_QF} 
\end{equation*}
where $s^2_i = \sigma^2_i + \delta^2$. Here $\sigma_i$ is the uncertainty in $\log\left(F^{\rm obs}_X\right)_i$, and $\delta$ is the global intrinsic dispersion in the data (including the systematic uncertainties), which we treat as a free parameter in our cosmological model fits. We use the \cite{RisalitiandLusso_2019} compilation of 1598 QSO-Flux measurements in the range $0.036 \leq z\leq 5.1003$. The flat priors of cosmological parameters and the Amati relation parameters are in Sec. \ref{sec:analysis} and, as in \cite{Khadka_2020b}, the flat priors of the parameters $\delta$, $\gamma$, and $\beta$ are non-zero over $0\leq\delta\leq e^{10}$, $-2\leq\gamma\leq2$, and $0\leq\beta\leq11$, respectively. 

\begin{figure*}
\centering
  \subfloat[]{%
    \includegraphics[width=3.5in,height=3.5in]{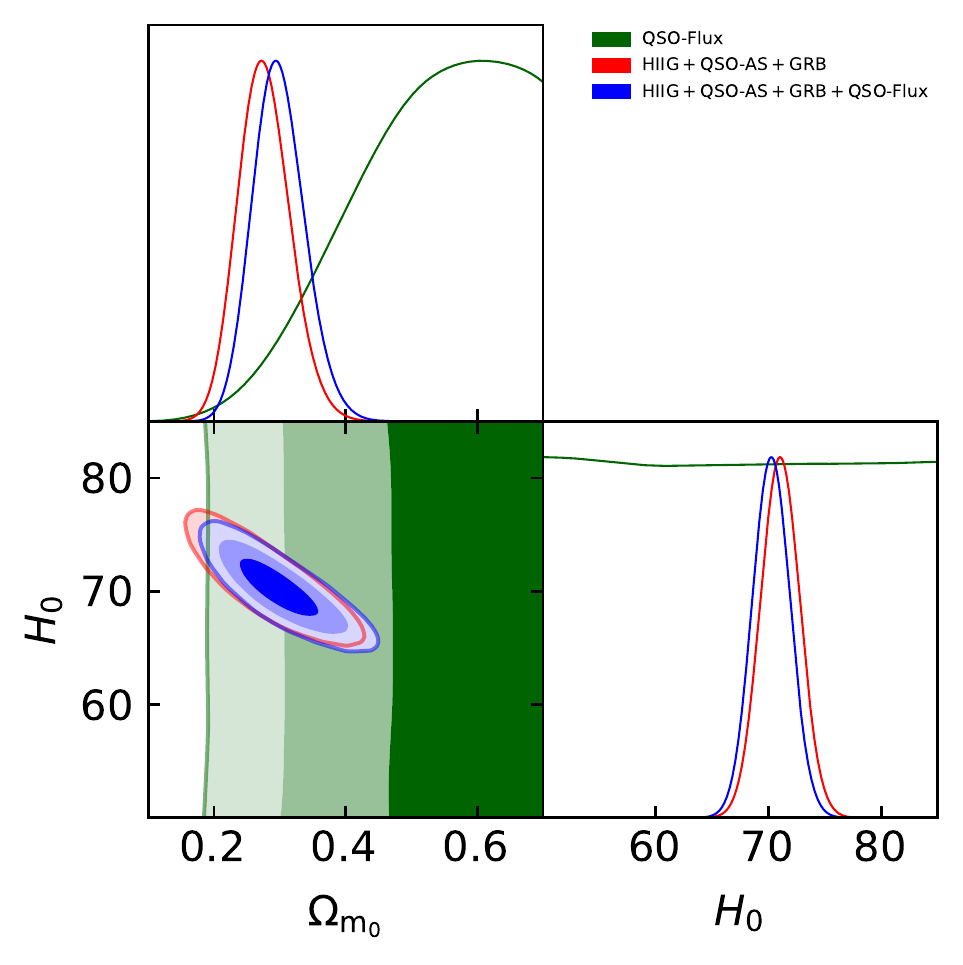}}
  \subfloat[]{%
    \includegraphics[width=3.5in,height=3.5in]{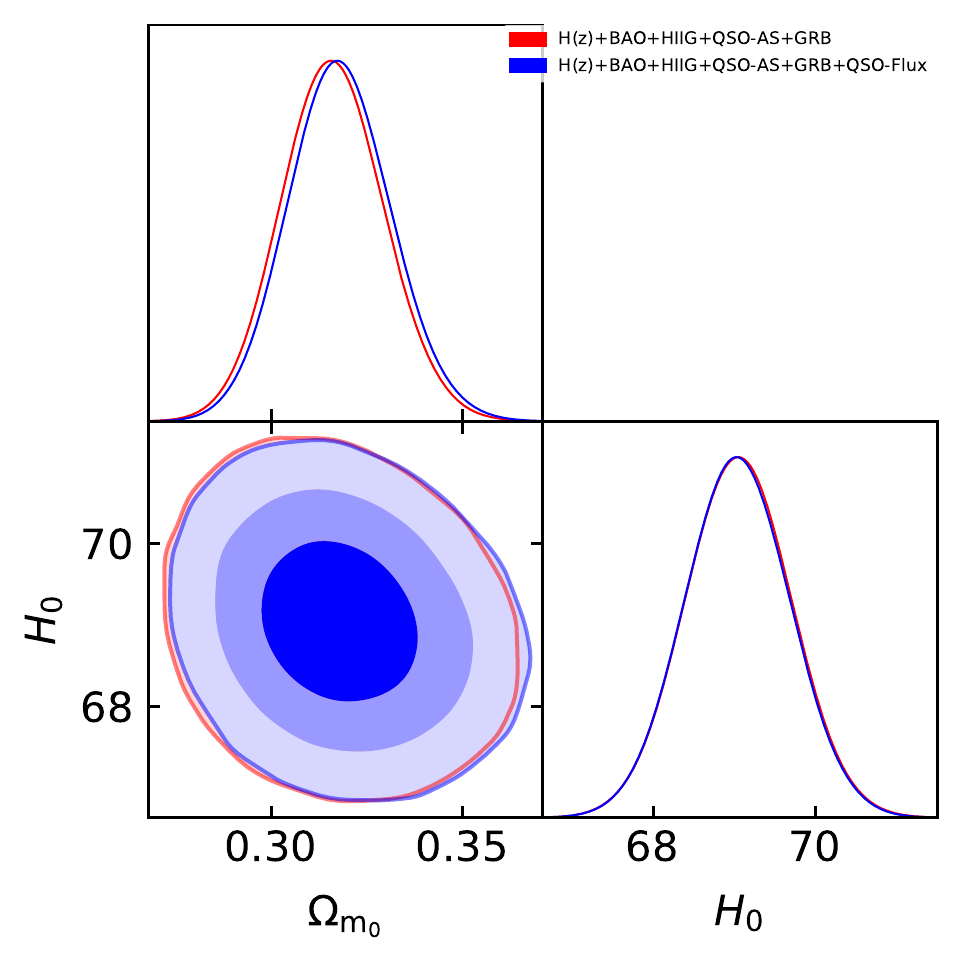}}\\
\caption{Same as Fig. \ref{fig1} (flat \lcdm) but for different combinations of data and showing only cosmological parameters.}
\label{fig01}
\end{figure*}

\begin{figure*}
\centering
  \subfloat[]{%
    \includegraphics[width=3.5in,height=3.5in]{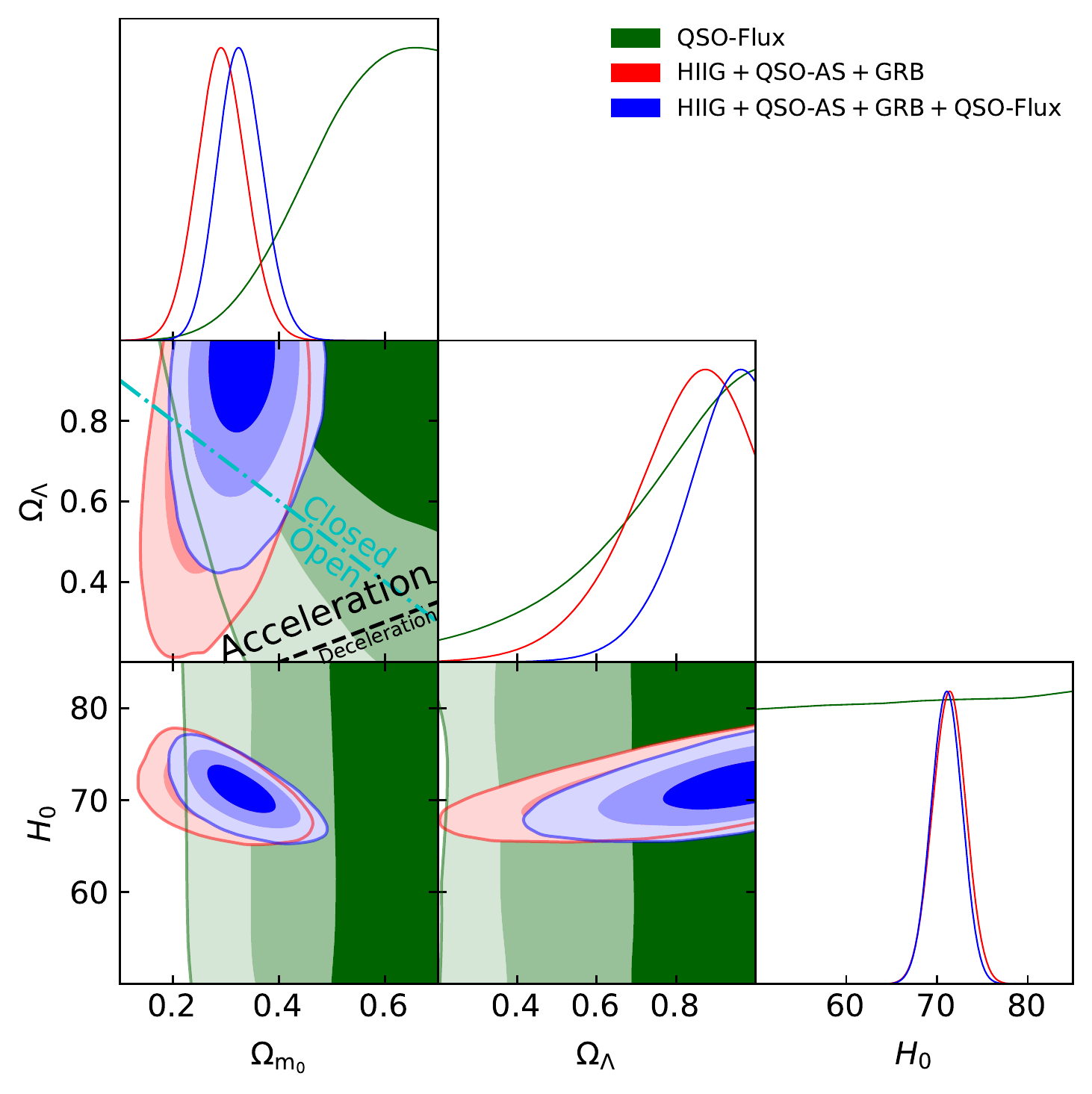}}
  \subfloat[]{%
    \includegraphics[width=3.5in,height=3.5in]{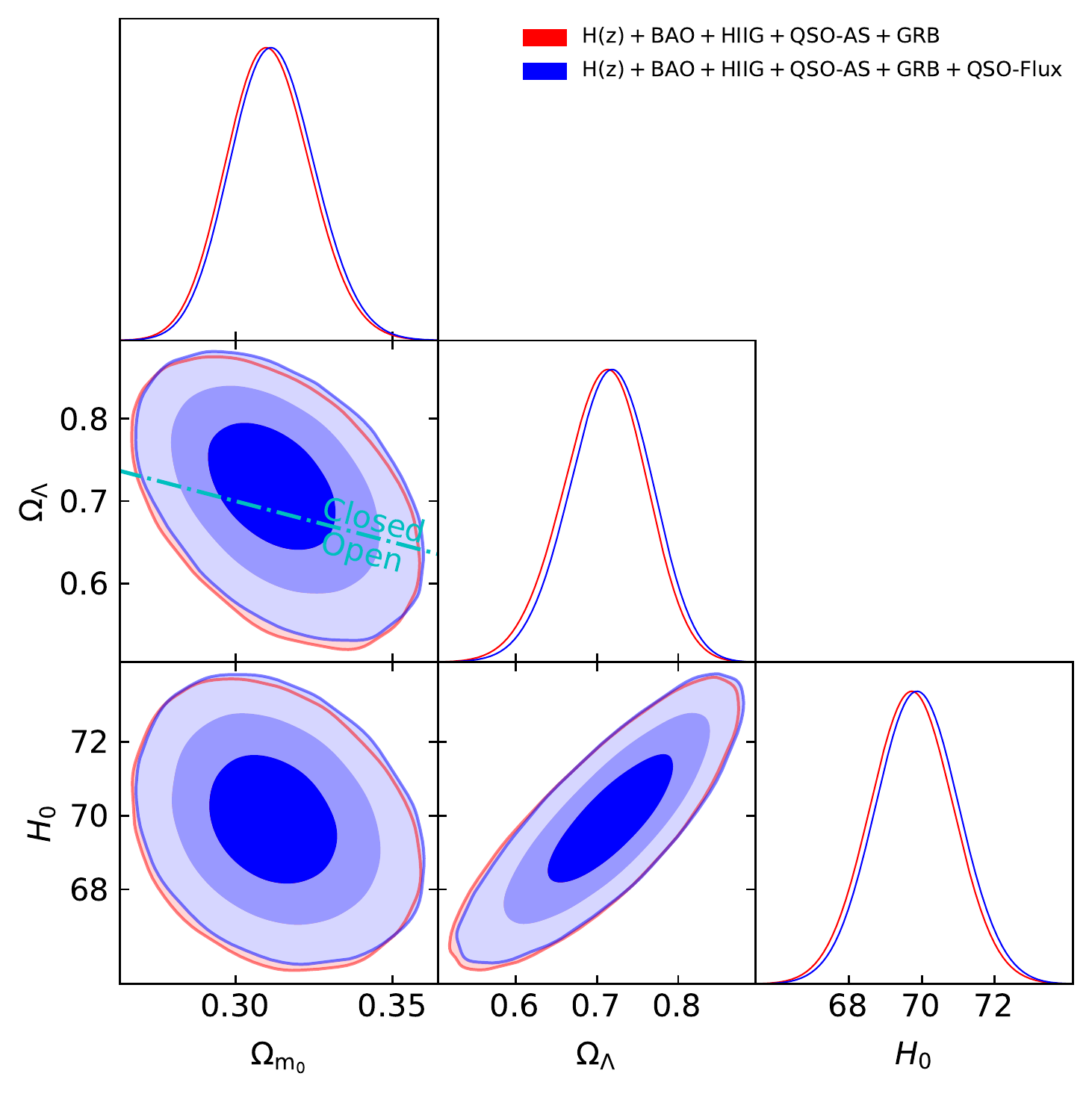}}\\
\caption{Same as Fig. \ref{fig2} (non-flat \lcdm) but for different combinations of data and showing only cosmological parameters.}
\label{fig02}
\end{figure*}

\begin{figure*}
\centering
  \subfloat[]{%
    \includegraphics[width=3.5in,height=3.5in]{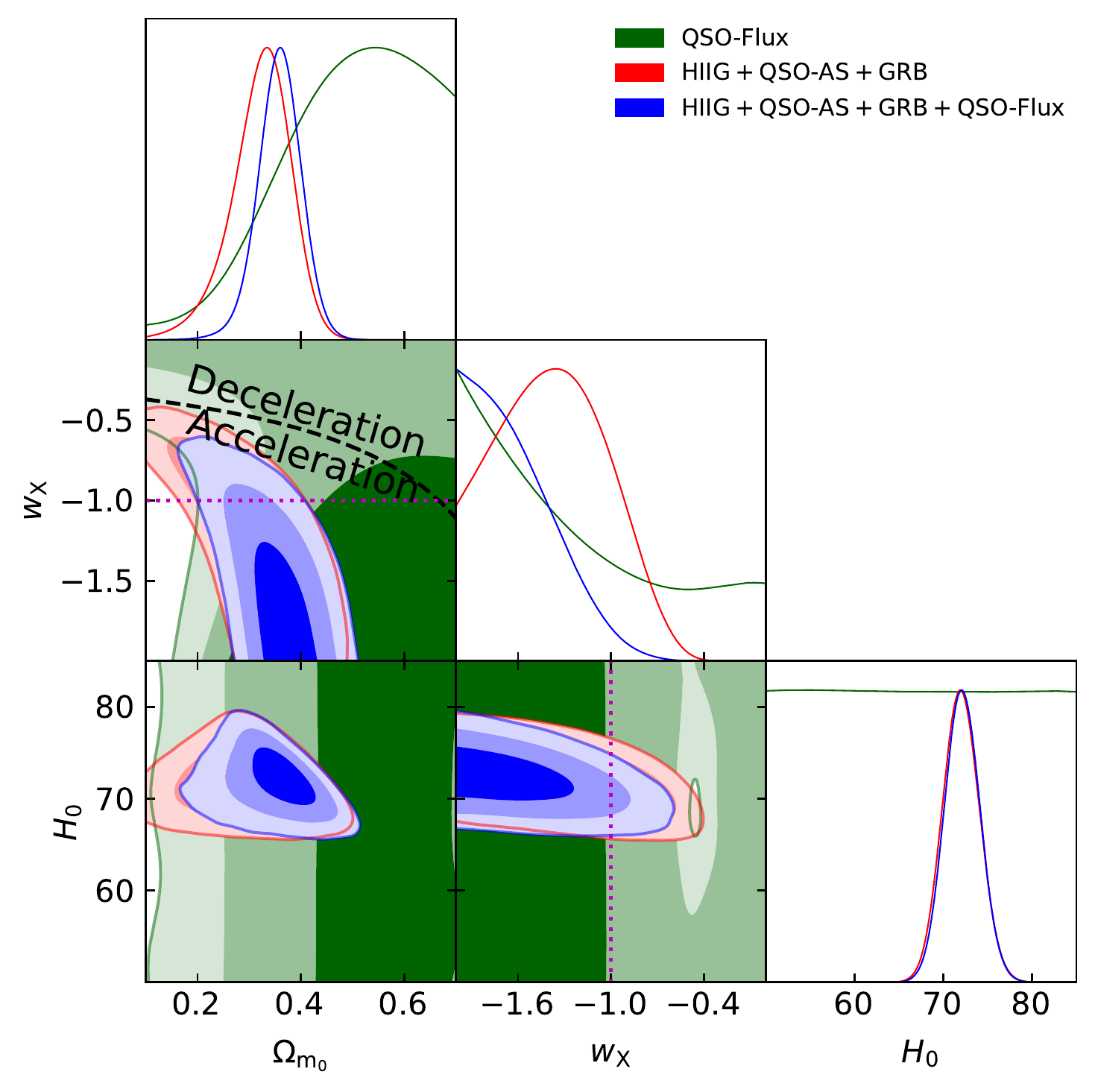}}
  \subfloat[]{%
    \includegraphics[width=3.5in,height=3.5in]{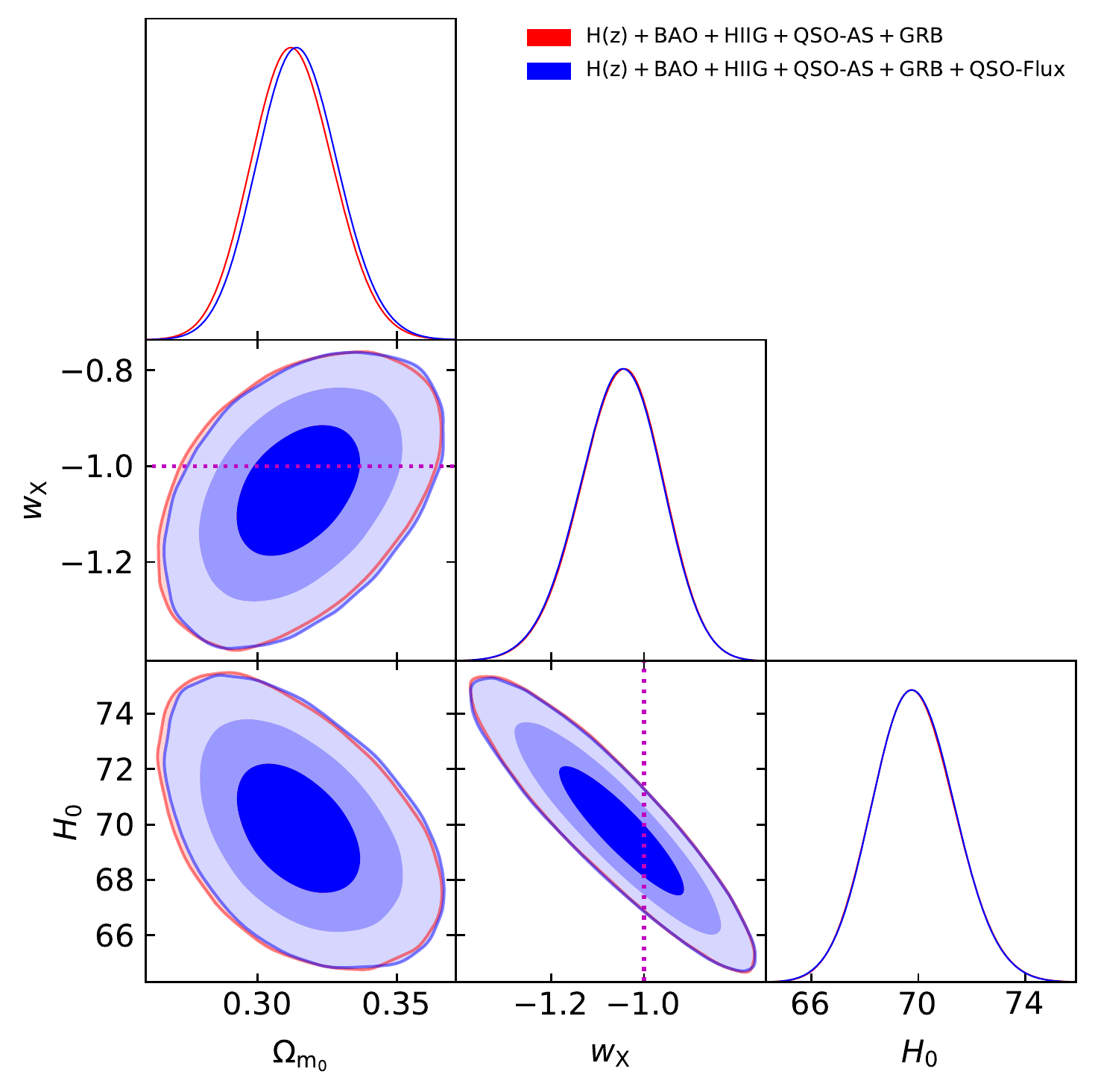}}\\
\caption{Same as Fig. \ref{fig3} (flat XCDM) but for different combinations of data and showing only cosmological parameters.}
\label{fig03}
\end{figure*}

\begin{figure*}
\centering
  \subfloat[]{%
    \includegraphics[width=3.5in,height=3.5in]{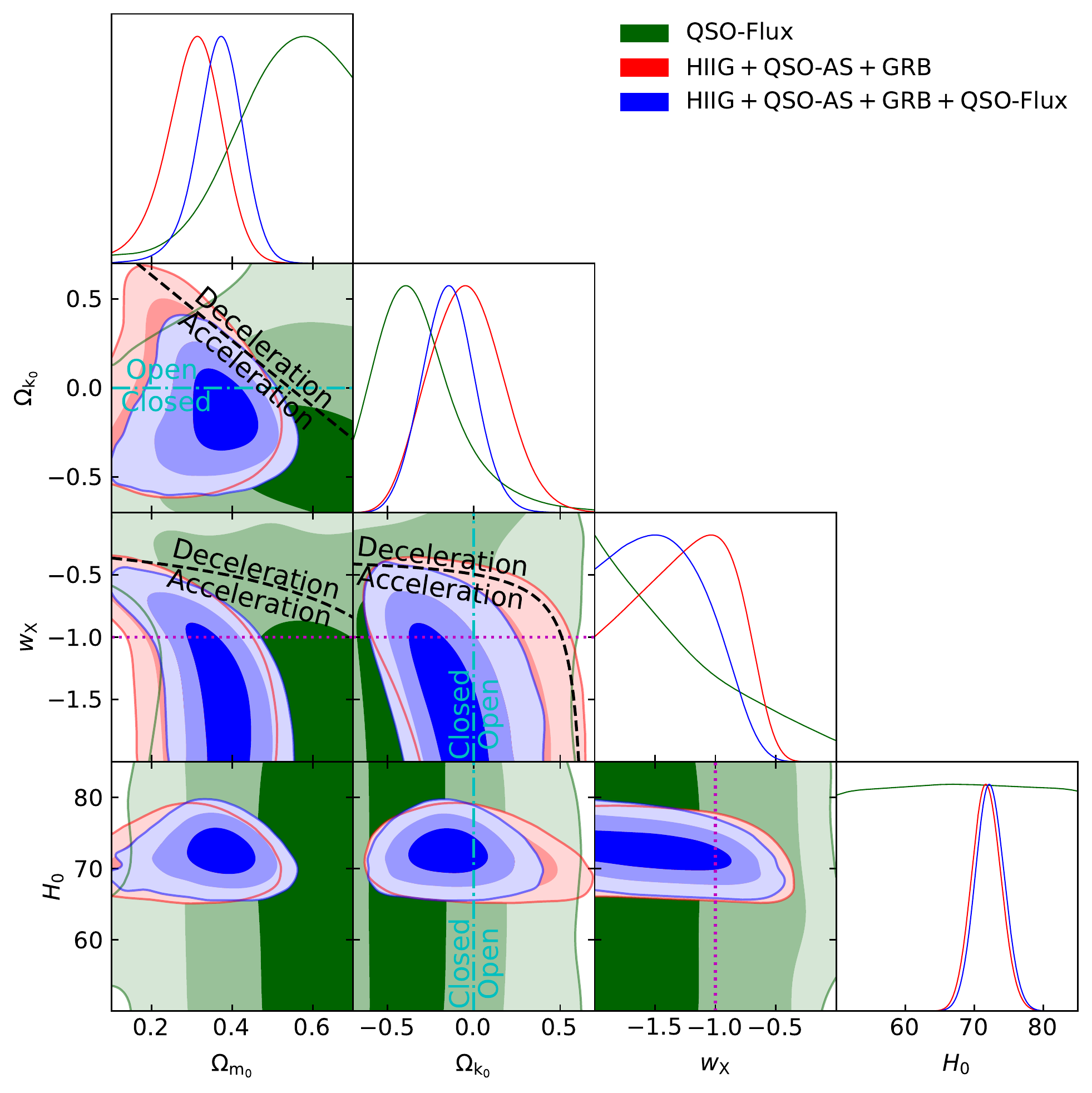}}
  \subfloat[]{%
    \includegraphics[width=3.5in,height=3.5in]{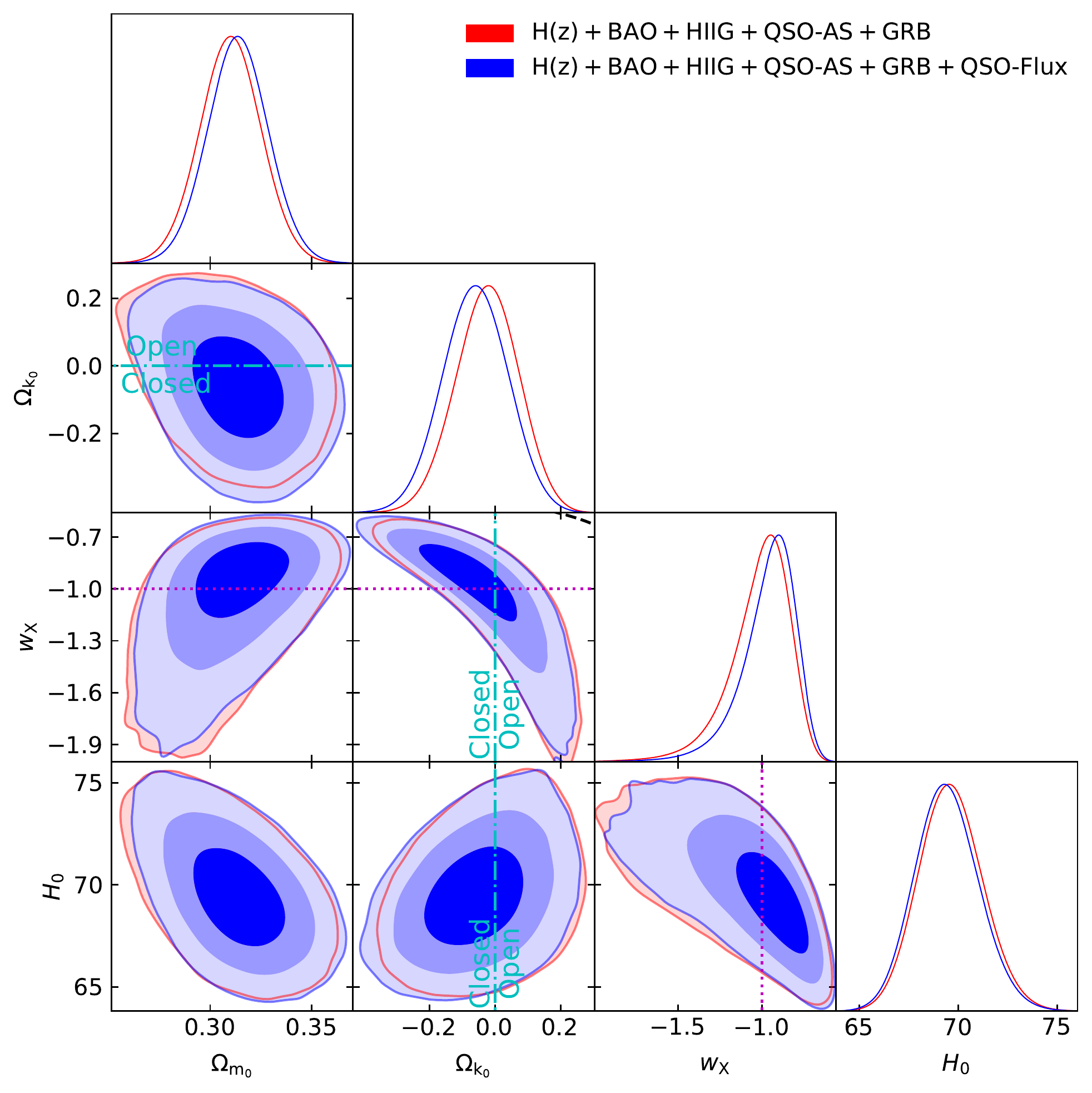}}\\
\caption{Same as Fig. \ref{fig4} (non-flat XCDM) but for different combinations of data and showing only cosmological parameters.}
\label{fig04}
\end{figure*}

As discussed in \cite{Khadka_2020b} the QSO-Flux data alone favors large \om\ values for the physically-motivated flat and non-flat \lcdm\ and \pcdm\ models. \cite{RisalitiandLusso_2019} and \cite{Khadka_2020b} note that this is largely a consequence of the $z\sim 2$--5 QSO data. While these large \om\ values differ from almost all other measurements of \om, the QSO-Flux data have larger error bars and their cosmological constraint contours are not in conflict with those from other data sets. For these reasons we have used the QSO-Flux data, but in this Appendix and not in the main text, and we have not computed QSO-Flux data results for the \pcdm\ cases (these being computationally demanding). We briefly summarize our constraints, listed in Tables \ref{tab:BFP2} and \ref{tab:1d_BFP2} and shown in Figs. \ref{fig01}--\ref{fig04}, below.

\subsection{QSO-Flux constraints}

Except for flat \lcdm, the constraints on \om\ in the QSO-Flux only case are 2$\sigma$ larger than those in the combined HzBHQASQFG case (see Sec. \ref{sec:A3}). QSO-Flux data cannot constrain $\alpha$, nor can they constrain $H_0$ (for the same reason that GRB data cannot constrain this parameter; see Section \ref{subsec:GRB}). QSO-Flux data set upper limits on $w_{\rm X}$ for flat and non-flat XCDM, with $w_{\rm X}=-1$ within the 1$\sigma$ range.

\subsection{\hiig, QSO-AS, QSO-Flux, and GRB (HQASQFG) constraints}

When adding QSO-Flux to HQASG data, the joint constraints favor larger \om\ and lower \ok. In non-flat $\Lambda$CDM closed geometry is favored at 3.24$\sigma$. The $H_0$ constraints are only mildly affected by the addition of the QSO-Flux data. The constraint on $w_{\rm X}$ changes from $-1.379^{+0.361}_{-0.375}$ in the HQASG case to $<-1.100$ (2$\sigma$ limit) in the HQASQFG case for flat XCDM, while for non-flat XCDM, the constraint on $w_{\rm X}$ in the HQASQFG case is 0.40$\sigma$ lower than that in the HQASG case and is 1.80$\sigma$ away from $w_{\rm X}=-1$.

\subsection{$H(z)$, BAO, \hiig, QSO-AS, QSO-Flux, and GRB (HzBHQASQFG) constraints}
\label{sec:A3}

When adding QSO-Flux to the HzBHQASG combination, the \om\ central values are only slightly larger because the $H(z)$ + BAO data dominate this compilation. The joint-constraint central $\Omega_{k0}$ values are lower, and consistent with flat geometry, while the constraints on $H_0$ from this combination are almost unaltered. The constraints on $w_{\rm X}$ are 0.02$\sigma$ lower and 0.23$\sigma$ higher for flat and non-flat XCDM, respectively, both being consistent with $w_{\rm X}=-1$ within 1$\sigma$.

\subsection{Model comparison}

From Table \ref{tab:BFP2}, we see that the reduced $\chi^2$ of the QSO-Flux case for all models is near unity ($\sim1.01$) and that the reduced $\chi^2$ of cases that include QSO-Flux is brought down to $\sim1.24$--1.26 for all models. Based on the $BIC$ (see Table \ref{tab:BFP2}), flat \lcdm\ is the most favored model, while based on the $AIC$, non-flat XCDM, flat XCDM, and flat \lcdm\ are the most favored models for the QSO-Flux, HQASQFG, and HzBHQASQFG combinations, respectively.\footnote{Note that based on the $\Delta \chi^2$ results of Table \ref{tab:BFP2} flat \lcdm\ has the minimum $\chi^2$ in the QSO-Flux, HQASQFG, and HzBHQASQFG cases.} From $\Delta AIC$ and $\Delta BIC$, we find mostly weak or positive evidence against the models, and only in a few cases do we find strong evidence against our models. According to $\Delta BIC$, the evidence against non-flat XCDM is strong for the QSO-Flux data, and very strong for the HQASQFG and HzBHQASQFG data, and the evidence against non-flat \lcdm\ is strong for the HzBHQASQFG data. According to $\Delta AIC$, the evidence against flat XCDM is strong for the HzBHQASQFG data.

\begin{sidewaystable*}
\centering
\footnotesize
\begin{threeparttable}
\caption{Unmarginalized best-fitting parameter values for all models from various combinations of data.}\label{tab:BFP2}
\setlength{\tabcolsep}{1.0mm}{
\begin{tabular}{lccccccccccccccccccccc}
\toprule
 Model & Data set & $\Omega_{\mathrm{m_0}}$ & $\Omega_{\Lambda}$ & $\Omega_{\mathrm{k_0}}$ & $w_{\mathrm{X}}$ & $\alpha$ & $H_0$\tnote{c} & $\sigma_{\mathrm{ext}}$ & $a$ & $b$ & $\delta$ & $\gamma$ & $\beta$ & $\chi^2$ & $\nu$ & $-2\ln\mathcal{L}_{\mathrm{max}}$ & $AIC$ & $BIC$ & $\Delta\chi^2$ & $\Delta AIC$ & $\Delta BIC$ \\
\midrule
Flat \lcdm & QSO-Flux & 0.315 & 0.685 & -- & -- & -- & 68.69 & -- & -- & -- & -- & -- & -- & 1603.28 & 1593 & -50.13 & -40.13 & -13.24 & 0.00 & 1.62 & 0.00\\
 & HQASG\tnote{d} & 0.271 & 0.729 & -- & -- & -- & 71.13 & 0.407 & 50.18 & 1.138 & -- & -- & -- & 879.42 & 387 & 895.05 & 905.05 & 924.91 & 0.12 & 0.00 & 0.00\\
 & HQASQFG\tnote{e} & 0.305 & 0.695 & -- & -- & -- & 70.01 & 0.399 & 50.20 & 1.132 & 0.231 & 0.639 & 7.083 & 2480.01 & 1982 & 848.53 & 864.53 & 909.29 & 0.00 & 2.58 & 0.00\\
 & HzBHQASG\tnote{f} & 0.317 & 0.683 & -- & -- & -- & 69.06 & 0.404 & 50.19 & 1.134 & -- & -- & -- & 903.61 & 429 & 917.79 & 927.79 & 948.16 & 1.52 & 0.00 & 0.00\\
 & HzBHQASQFG\tnote{g} & 0.317 & 0.683 & -- & -- & -- & 69.06 & 0.399 & 50.23 & 1.119 & 0.232 & 0.637 & 7.144 & 2499.87 & 2024 & 870.31 & 886.31 & 931.25 & 0.00 & 0.00 & 0.00\\
\\
Non-flat \lcdm & QSO-Flux & 0.540 & 0.985 & $-0.525$ & -- & -- & 75.75 & -- & -- & -- & 0.230 & 0.611 & 7.888 & 1603.83 & 1592 & -53.25 & -41.25 & -8.99 & 0.55 & 0.50 & 4.25\\
 & HQASG\tnote{d} & 0.291 & 0.876 & $-0.167$ & -- & -- & 72.00 & 0.406 & 50.22 & 1.120 & -- & -- & -- & 879.30 & 386 & 894.02 & 906.02 & 929.85 & 0.00 & 0.97 & 4.94\\
 & HQASQFG\tnote{e} & 0.325 & 0.944 & $-0.269$ & -- & -- & 71.49 & 0.404 & 50.21 & 1.116 & 0.230 & 0.632 & 7.304 & 2486.97 & 1981 & 844.38 & 862.38 & 912.75 & 6.96 & 0.43 & 3.46\\
 & HzBHQASG\tnote{f} & 0.309 & 0.716 & $-0.025$ & -- & -- & 69.77 & 0.402 & 50.17 & 1.141 & -- & -- & -- & 904.47 & 428 & 917.17 & 929.17 & 953.61 & 2.38 & 1.38 & 5.45\\
 & HzBHQASQFG\tnote{g} & 0.309 & 0.709 & $-0.018$ & -- & -- & 69.59 & 0.412 & 50.21 & 1.128 & 0.231 & 0.637 & 7.151 & 2503.43 & 2023 & 869.71 & 887.71 & 938.26 & 3.56 & 1.40 & 7.01 \\
\\
Flat XCDM & QSO-Flux & 0.477 & -- & -- & $-1.988$ & -- & 60.86 & -- & -- & -- & 0.230 & 0.625 & 7.530 & 1604.18 & 1592 & -52.13 & -40.13 & -7.88 & 0.90 & 1.62 & 5.36\\
 & HQASG\tnote{d} & 0.320 & -- & -- & $-1.306$ & -- & 72.03 & 0.404 & 50.20 & 1.131 & -- & -- & -- & 880.47 & 386 & 894.27 & 906.27 & 930.10 & 1.17 & 1.22 & 5.19\\
 & HQASQFG\tnote{e} & 0.370 & -- & -- & $-1.980$ & -- & 73.66 & 0.399 & 50.18 & 1.129 & 0.231 & 0.632 & 7.301 & 2485.59 & 1981 & 843.95 & 861.95 & 912.31 & 5.58 & 0.00 & 3.02\\
 & HzBHQASG\tnote{f} & 0.313 & -- & -- & $-1.052$ & -- & 69.90 & 0.407 & 50.19 & 1.132 & -- & -- & -- & 902.09 & 428 & 917.55 & 929.55 & 953.99 & 0.00 & 1.76 & 5.83\\
 & HzBHQASQFG\tnote{g} & 0.313 & -- & -- & $-1.046$ & -- & 69.84 & 0.401 & 50.18 & 1.134 & 0.231 & 0.635 & 7.215 & 2506.25 & 2023 & 870.16 & 888.16 & 938.71 & 6.38 & 1.85 & 7.46\\
\\
Non-flat XCDM & QSO-Flux & 0.507 & -- & $-0.376$ & $-1.996$ & -- & 75.28 & -- & -- & -- & 0.229 & 0.611 & 7.934 & 1614.59 & 1591 & -55.75 & -41.75 & -4.12 & 11.31 & 0.00 & 9.12\\
 & HQASG\tnote{d} & 0.300 & -- & $-0.161$ & $-1.027$ & -- & 80.36 & 0.405 & 50.21 & 1.122 & -- & -- & -- & 879.48 & 385 & 894.01 & 908.01 & 935.81 & 0.18 & 2.96 & 10.90\\
 & HQASQFG\tnote{e} & 0.395 & -- & $-0.138$ & $-1.639$ & -- & 73.48 & 0.411 & 50.21 & 1.112 & 0.230 & 0.627 & 7.441 & 2486.77 & 1980 & 843.65 & 863.65 & 919.61 & 6.76 & 1.70 & 10.32\\
 & HzBHQASG\tnote{f} & 0.312 & -- & $-0.045$ & $-0.959$ & -- & 69.46 & 0.402 & 50.23 & 1.117 & -- & -- & -- & 904.17 & 427 & 917.07 & 931.07 & 959.58 & 2.08 & 3.28 & 11.42\\
 & HzBHQASQFG\tnote{g} & 0.316 & -- & $-0.089$ & $-0.891$ & -- & 69.05 & 0.410 & 50.23 & 1.111 & 0.230 & 0.633 & 7.247 & 2516.49 & 2022 & 869.25 & 889.25 & 945.41 & 16.62 & 2.94 & 14.16\\
\bottomrule
\end{tabular}}
\begin{tablenotes}[flushleft]
\item [c] \hunit.
\item [d] \hiig\ + QSO-AS + GRB.
\item [e] \hiig\ + QSO-AS + GRB + QSO-Flux.
\item [f] $H(z)$ + BAO + \hiig\ + QSO-AS + GRB.
\item [g] $H(z)$ + BAO + \hiig\ + QSO-AS + GRB + QSO-Flux.
\end{tablenotes}
\end{threeparttable}
\end{sidewaystable*}

\begin{sidewaystable*}
\centering
\scriptsize
\begin{threeparttable}
\caption{One-dimensional marginalized best-fitting parameter values and uncertainties ($\pm 1\sigma$ error bars or $2\sigma$ limits) for all models from various combinations of data.}\label{tab:1d_BFP2}
\setlength{\tabcolsep}{1.3mm}{
\begin{tabular}{lccccccccccccc}
\toprule
 Model & Data set & $\Omega_{\mathrm{m_0}}$ & $\Omega_{\Lambda}$ & $\Omega_{\mathrm{k_0}}$ & $w_{\mathrm{X}}$ & $\alpha$ & $H_0$\tnote{c} & $\sigma_{\mathrm{ext}}$ & $a$ & $b$ & $\delta$ & $\gamma$ & $\beta$ \\
\midrule
Flat \lcdm & QSO-Flux & $>0.313$ & -- & -- & -- & -- & -- & -- & -- & -- & $0.231\pm0.004$ & $0.626\pm0.011$ & $7.469\pm0.321$ \\
 & HQASG\tnote{d} & $0.277^{+0.034}_{-0.041}$ & -- & -- & -- & -- & $71.03\pm1.67$ & $0.413^{+0.026}_{-0.032}$ & $50.19\pm0.24$ & $1.138\pm0.085$ & -- & -- & --\\
 & HQASQFG\tnote{e} & $0.299^{+0.036}_{-0.043}$ & -- & -- & -- & -- & $70.25^{+1.60}_{-1.61}$ & $0.412^{+0.027}_{-0.032}$ & $50.18\pm0.24$ & $1.136\pm0.085$ & $0.231^{+0.005}_{-0.004}$ & $0.639^{+0.009}_{-0.010}$ & $7.091^{+0.281}_{-0.279}$\\
 & HzBHQASG\tnote{f} & $0.316\pm0.013$ & -- & -- & -- & -- & $69.05^{+0.62}_{-0.63}$ & $0.412^{+0.026}_{-0.032}$ & $50.19\pm0.23$ & $1.133\pm0.085$ & -- & -- & --\\
 & HzBHQASQFG\tnote{g} & $0.318\pm0.013$ & -- & -- & -- & -- & $69.03\pm0.62$ & $0.412^{+0.026}_{-0.032}$ & $50.19\pm0.23$ & $1.133\pm0.084$ & $0.231\pm0.004$ & $0.637\pm0.009$ & $7.146\pm0.268$\\
\\
Non-flat \lcdm & QSO-Flux & $>0.353$ & $>0.357$ & $-0.303^{+0.131}_{-0.252}$ & -- & -- & -- & -- & -- & -- & $0.231^{+0.004}_{-0.005}$ & $0.618\pm0.012$ & $7.709\pm0.366$ \\
 & HQASG\tnote{d} & $0.292\pm0.044$ & $0.801^{+0.191}_{-0.055}$ & $-0.093^{+0.092}_{-0.190}$ & -- & -- & $71.33^{+1.75}_{-1.77}$ & $0.413^{+0.026}_{-0.032}$ & $50.19\pm0.24$ & $1.130\pm0.086$ & -- & -- & --\\
 & HQASQFG\tnote{e} & $0.327^{+0.039}_{-0.043}$ & $>0.691$ & $-0.204^{+0.063}_{-0.125}$ & -- & -- & $71.07\pm1.64$ & $0.413^{+0.027}_{-0.032}$ & $50.20\pm0.24$ & $1.120\pm0.086$ & $0.231\pm0.004$ & $0.632\pm0.010$ & $7.291^{+0.306}_{-0.305}$\\
 & HzBHQASG\tnote{f} & $0.311^{+0.012}_{-0.014}$ & $0.708^{+0.053}_{-0.046}$ & $-0.019^{+0.043}_{-0.048}$ & -- & -- & $69.72\pm1.10$ & $0.412^{+0.026}_{-0.032}$ & $50.19\pm0.23$ & $1.132\pm0.085$ & -- & -- & --\\
 & HzBHQASQFG\tnote{g} & $0.312^{+0.012}_{-0.013}$ & $0.716^{+0.052}_{-0.046}$ & $-0.028\pm0.045$ & -- & -- & $69.88\pm1.10$ & $0.412^{+0.025}_{-0.032}$ & $50.19\pm0.23$ & $1.131\pm0.084$ & $0.231\pm0.004$ & $0.637\pm0.009$ & $7.144\pm0.270$\\
\\
Flat XCDM & QSO-Flux & $0.496^{+0.192}_{-0.069}$ & -- & -- & $<-1.042$\tnote{h} & -- & -- & -- & -- & -- & $0.231\pm0.004$ & $0.624\pm0.011$ & $7.508\pm0.326$ \\
 & HQASG\tnote{d} & $0.322^{+0.062}_{-0.044}$ & -- & -- & $-1.379^{+0.361}_{-0.375}$ & -- & $72.00^{+1.99}_{-1.98}$ & $0.412^{+0.026}_{-0.032}$ & $50.20\pm0.24$ & $1.130\pm0.085$ & -- & -- & --\\
 & HQASQFG\tnote{e} & $0.358^{+0.040}_{-0.038}$ & -- & -- & $<-1.100$ & -- & $72.14\pm1.91$ & $0.411^{+0.026}_{-0.031}$ & $50.20\pm0.23$ & $1.125\pm0.084$ & $0.231\pm0.004$ & $0.633^{+0.009}_{-0.010}$ & $7.268^{+0.287}_{-0.288}$\\
 & HzBHQASG\tnote{f} & $0.313^{+0.014}_{-0.015}$ & -- & -- & $-1.050^{+0.090}_{-0.081}$ & -- & $69.85^{+1.42}_{-1.55}$ & $0.412^{+0.026}_{-0.032}$ & $50.19\pm0.24$ & $1.134\pm0.085$ & -- & -- & -- \\
 & HzBHQASQFG\tnote{g} & $0.315^{+0.013}_{-0.015}$ & -- & -- & $-1.052^{+0.091}_{-0.081}$ & -- & $69.85\pm1.48$ & $0.413^{+0.026}_{-0.032}$ & $50.19\pm0.24$ & $1.133\pm0.086$ & $0.231\pm0.004$ & $0.637\pm0.009$ & $7.135^{+0.270}_{-0.271}$\\
\\
Non-flat XCDM & QSO-Flux & $0.515^{+0.184}_{-0.050}$ & -- & $-0.310^{+0.137}_{-0.289}$ & $<-0.294$ & -- & -- & -- & -- & -- & $0.231^{+0.004}_{-0.005}$ & $0.615\pm0.013$ & $7.817^{+0.398}_{-0.400}$ \\
 & HQASG\tnote{d} & $0.303^{+0.073}_{-0.058}$ & -- & $-0.044^{+0.193}_{-0.217}$ & $-1.273^{+0.501}_{-0.321}$ & -- & $71.77\pm2.02$ & $0.413^{+0.026}_{-0.031}$ & $50.20\pm0.24$ & $1.129\pm0.085$ & -- & -- & --\\
 & HQASQFG\tnote{e} & $0.367^{+0.059}_{-0.048}$ & -- & $-0.146^{+0.143}_{-0.147}$ & $-1.433^{+0.241}_{-0.493}$ & -- & $72.27^{+2.01}_{-1.99}$ & $0.413^{+0.026}_{-0.032}$ & $50.21\pm0.24$ & $1.116\pm0.085$ & $0.231\pm0.004$ & $0.629^{+0.011}_{-0.010}$ & $7.382\pm0.321$\\
 & HzBHQASG\tnote{f} & $0.310\pm0.014$ & -- & $-0.024^{+0.092}_{-0.093}$ & $-1.019^{+0.202}_{-0.099}$ & -- & $69.63^{+1.45}_{-1.62}$ & $0.412^{+0.026}_{-0.031}$ & $50.19\pm0.23$ & $1.132\pm0.085$ & -- & -- & -- \\
 & HzBHQASQFG\tnote{g} & $0.314^{+0.014}_{-0.015}$ & -- & $-0.060^{+0.096}_{-0.095}$ & $-0.968^{+0.184}_{-0.087}$ & -- & $69.43^{+1.43}_{-1.63}$ & $0.412^{+0.026}_{-0.032}$ & $50.19\pm0.24$ & $1.130\pm0.085$ & $0.231\pm0.004$ & $0.636^{+0.009}_{-0.010}$ & $7.182^{+0.278}_{-0.281}$\\
\bottomrule
\end{tabular}}
\begin{tablenotes}[flushleft]
\item [c] \hunit.
\item [d] \hiig\ + QSO-AS + GRB.
\item [e] \hiig\ + QSO-AS + GRB + QSO-Flux.
\item [f] $H(z)$ + BAO + \hiig\ + QSO-AS + GRB.
\item [g] $H(z)$ + BAO + \hiig\ + QSO-AS + GRB + QSO-Flux.
\item [h] This is the 1$\sigma$ limit. The $2\sigma$ limit is set by the prior, and is not shown here.
\end{tablenotes}
\end{threeparttable}
\end{sidewaystable*}



\bsp	
\label{lastpage}
\end{document}